\documentclass[11pt]{article}
\usepackage[mathscr]{eucal}
\usepackage{amssymb,latexsym}
\usepackage{verbatim}
\usepackage{amsmath}
\usepackage{amsthm}
\usepackage{enumerate}
\usepackage{authblk}
\usepackage{color}
\usepackage{multicol}
\usepackage{tikz}
\usepackage{caption}
\usepackage{cite}
\usepackage{hyperref}
\usepackage{soul}

\usepackage{physics}
\usepackage[]{graphicx}
\usepackage{cancel}
\usepackage{amsfonts}
\usepackage{mathtools}
\usepackage{bm}
\usepackage{enumitem}
\usepackage[normalem]{ulem}

\newtheorem{thm}{Theorem}[section]
\newtheorem{lem}[thm]{Lemma}
\newtheorem{Def}[thm]{Definition}
\newtheorem{prop}[thm]{Proposition}

\newtheorem{cor}[thm]{Corollary}

\hypersetup{
    colorlinks=true,
    linkcolor=blue,
    filecolor=magenta,      
    urlcolor=cyan,
    citecolor=purple,
}

\renewcommand\l{\lambda}

\newcommand\wt{\widetilde}

\newcommand\s{\sigma}

\newcommand\e{\varepsilon}
\renewcommand\b{\beta}

\renewcommand\l{\lambda}
\newcommand\g{\gamma}

\renewcommand\a{\alpha}

\newcommand\beq{\begin{equation}}
\newcommand\eeq{\end{equation}}
\newcommand\ben{\begin{enumerate}}
\newcommand\een{\end{enumerate}}
\newcommand\bit{\begin{itemize}}
\newcommand\eit{\end{itemize}}

\newcommand{\R}{\mathbb R}

\newcommand{\ov}{\overline}

\newcommand{\half}{\frac{1}{2}}

\newcommand{\ext}{\text{{\rm ext}}}

\newcommand{\pd}{\partial}

\newcommand{\mc}{\mathcal}

\def\undertilde#1{\mathord{\vtop{\ialign{##\crcr
   $\hfil\displaystyle{#1}\hfil$\crcr\noalign{\kern1.5pt\nointerlineskip}
   $\hfil\tilde{}\hfil$\crcr\noalign{\kern1.5pt}}}}}

\newcounter{mnotecount}

\begin{document}

\pagenumbering{roman}

\begin{titlepage}

\baselineskip=15.5pt \thispagestyle{empty}

%%% TITLE %%%
\begin{center}
    {\fontsize{20}{20}\selectfont \bfseries On the initial singularity and extendibility of flat quasi-de Sitter spacetimes}
\end{center}

\vspace{0.1cm}

%%% AUTHORS %%%
\begin{center}
    {\fontsize{12}{18}\selectfont Ghazal Geshnizjani$^{1,2}$, Eric Ling$^{3,4}$ and Jerome Quintin$^{2,1,4}$}
\end{center}

%%% AFFILIATIONS %%%
\begin{center}
    \vskip8pt
    \textsl{$^1$ Perimeter Institute for Theoretical Physics,\\ 31 Caroline Street North, Waterloo, Ontario N2L 2Y5, Canada}\\
    \vskip4pt
    \textsl{$^2$ Department of Applied Mathematics and Waterloo Centre for Astrophysics,\\ University of Waterloo, 200 University Avenue West, Waterloo, Ontario N2L 3G1, Canada}\\
    \vskip4pt
    \textsl{$^3$ Copenhagen Centre for Geometry and Topology (GeoTop), Department of Mathematical Sciences, University of Copenhagen, DK-2100 Copenhagen, Denmark}\\
    \vskip4pt
    \textsl{$^4$ Fields Institute for Research in Mathematical Sciences,\\ University of Toronto, 222 College Street, Toronto, Ontario M5T 3J1, Canada}
\end{center}

\vspace{1.2cm}

%%% ABSTRACT %%%
\hrule
\vspace{0.3cm}
\noindent {\bf Abstract}\\[0.1cm]
Inflationary spacetimes have been argued to be past geodesically incomplete in many situations. However, whether the geodesic incompleteness implies the existence of an initial spacetime curvature singularity or whether the spacetime may be extended (potentially into another phase of the universe) is generally unknown. Both questions have important physical implications. In this paper, we take a closer look at the geometrical structure of inflationary spacetimes and investigate these very questions. We first classify which past inflationary histories have a scalar curvature singularity and which might be extendible and/or non-singular in homogeneous and isotropic cosmology with flat spatial sections. Then, we derive rigorous extendibility criteria of various regularity classes for quasi-de Sitter spacetimes that evolve from infinite proper time in the past. Finally, we show that beyond homogeneity and isotropy, special continuous extensions respecting the Einstein field equations with a perfect fluid must have the equation of state of a de Sitter universe asymptotically. An interpretation of our results is that past-eternal inflationary scenarios are most likely physically singular, except in situations with very special initial conditions.
\vskip10pt
\hrule
\vskip10pt

\end{titlepage}

\thispagestyle{empty}

%%% TABLE OF CONTENTS %%%
\setcounter{page}{2}
\tableofcontents
\newpage
\pagenumbering{arabic}
\setcounter{page}{1}
\clearpage

%%\pagenumbering{arabic}
%%\thispagestyle{empty}
%%\setcounter{page}{1}
%%\tableofcontents

%%%%%%%%%%%%%%%%%%%%%%%%%%%%%%%%%%%%%%%%%%%%%%%%%%%
%%%%%%%%%%%%%%%%%%%%%%%%%%%%%%%%%%%%%%%%%%%%%%%%%%%

\section{Introduction}

Inflation is a theory of the primordial universe, which can nicely solve the problems of standard big bang cosmology (monopole problem, horizon problem, flatness problem)
and provide a framework for generating the seeds of large-scale structures (see, e.g., \cite{Guth:1980zm,Linde:1981mu,Mukhanov:1981xt,Guth:1982ec,Bardeen:1983qw,Brandenberger:1999sw}). There are many inflationary models that can be in agreement with observations (nearly scale-invariant curvature power spectrum with slight red tilt, small tensor-to-scalar ratio, small non-Gaussianities, etc.; see, e.g., \cite{Bean:2008ga, Martin:2013tda,Planck:2018jri}). However, inflation is not a theory `of the big bang,' i.e., it does not necessarily tell us anything about the very beginning of the universe (if there was one).

In fact, it is known that if certain energy conditions are respected (together with additional assumptions), then the universe is past geodesically incomplete \cite{Hawking:1966sx,Hawking:1966jv,Hawking:1967ju,Hawking:1970zqf} (theorems more specific to inflation are also found in \cite{Borde:1993xh,Borde:1994ye,Borde:1994ai,Borde:1996pt,GalLingSingThm}). Moreover, it has been shown that a universe that has always been expanding (in an appropriately defined average sense) is similarly past geodesically incomplete \cite{Borde:2001nh}, thus demonstrating that any past-eternal expanding cosmology including past-eternal inflation is also past geodesically incomplete regardless of energy conditions. Yet, geodesic incompleteness does not tell us anything about the nature of the spacetime singularities (in particular, it does not always imply the presence of curvature singularities), a well-known drawback of any geodesic incompleteness theorems (nevertheless usually called `singularity theorems'; see, e.g., \cite{Senovilla:2014gza} for a review and a discussion). In the context of an accelerating homogeneous and isotropic universe, de Sitter (dS) spacetime is a prototypical example: the flat Friedmann-Lema\^{i}tre-Robertson-Walker (FLRW) patch (also known as the planar coordinates \cite{Spradlin:2001pw} of dS) only covers half of the whole (global) dS spacetime. As such, null geodesics appear past incomplete within the flat FLRW patch, yet the whole maximal spacetime is non-singular. This known fact, reviewed later in this paper and elsewhere, can be the basis for constructing past-eternal geodesically complete inflationary scenarios (e.g., \cite{Aguirre:2001ks,Aguirre:2003ck,Aguirre:2007gy}; see \cite{Vilenkin:2013rza,Vilenkin:2013tua} for related aspects).

There are very few concrete generic results regarding the actual presence and strength of a singularity to the past of inflation. It is generally unknown whether the past geodesic incompleteness of inflation always translates into a `big bang-like' initial singularity or whether the spacetime is actually extendible beyond its past boundary. A first step in this direction was given by \cite{Yoshida:2018ndv} (further developed and applied in \cite{Numasawa:2019juw,Nomura:2021lzz,Nishii:2021ylb,Nomura:2022vcj,Harada:2021yul}), where the nature of the singularity in quasi-de Sitter spacetimes was explored. Building on mathematical results relating spacetime extendibility to parallelly propagated curvature singularities \cite{Clarke:1973,Ellis:1974,Ellis:1977,Clarke:1982,Clarke:1994cw}, it was shown in \cite{Yoshida:2018ndv,Nomura:2021lzz} that the universe had to approach dS sufficiently quickly in order to avoid parallelly propagated curvature singularities.

The goal of this paper is to examine the geometrical structure of inflationary spacetimes and determine when singularities are present prior to inflation versus when an inflationary spacetime is potentially non-singular and extendible. 
In particular, we will investigate sufficient conditions for determining metric extendibility. 
Along the way, we will expand on the results obtained in \cite{Yoshida:2018ndv,Nomura:2021lzz} as well as put them on more rigorous grounds.
A related important mathematical question is to pinpoint when spacetimes are inextendible with a continuous metric, i.e., $C^0$ inextendible. However, this is a notoriously difficult question to answer and only partial results are known (see \cite{Galloway:2016bej,Galloway:2017qkr,Sbierski:2015nta,Sbierski:2017ezr,Sbierski:2020toa,Graf:2017xym, Minguzzi:2019dpz}). In the cosmological setting, it is known that FLRW spacetimes with finite particle horizons are $C^{0,1}$ inextendible \cite{Sbierski:2020toa}. Also, some $C^0$-inextendibility results can be achieved within the class of spherically symmetric spacetimes \cite{Galloway:2016bej}.  For a study of pre-big bang geometric extensions that still contain a singularity, see \cite{Klein_2017}.

As it will be physically and mathematically justified in Section \ref{sec:3cases}, the framework of the current paper is set within situations that are close enough to the flat patch of dS. Specifically, we will mostly explore spacetimes that are flat past-asymptotically dS and derive conditions on their extendibility. There will be many parallels with the works of \cite{Galloway:2016bej,Ling:2018jzl,Ling:2018tih,Ling:2022uzs,Ling:2022kjk} on the topic of metric extendibility of Milne-like spacetimes, which are homogeneous and isotropic universes with \emph{negative} (as opposed to \emph{zero} here) constant spatial curvature.

The structure of the paper is as follows: in Section \ref{sec:3cases}, we further introduce and motivate the questions of interest in this study. Specifically, we classify what are the possible past histories of inflationary cosmology within a flat FLRW spacetime. We define various classes of inflation and show which ones inevitably have an initial scalar curvature singularity and which ones might not. Cosmologies that are inflating from the infinite past (past eternal) fit in the latter class and motivate us to seek specific conditions permitting spacetime extendibility. We define a generic subclass of such spacetimes, dubbed `flat past-asymptotically de Sitter'. Section \ref{sec:metricext} is devoted to showing sufficient conditions for metric extendibility beyond the past boundary of past-eternal inflationary spacetimes. To do so, we introduce a new set of coordinates that allows us to explicitly write down a spacetime extension. We derive asymptotic criteria on the scale factor and/or the Hubble parameter (and derivatives thereof when needed) that yield continuous ($C^0$), continuously differentiable ($C^1$), twice-continuously differentiable ($C^2$), etc.~(up to smooth; $C^\infty$) extendibility. We also comment on the notion of scalar curvature singularities versus parallelly propagated singularities in those spacetimes. In Section \ref{sec:confEmbedd}, we present an alternative set of coordinates in which past-asymptotically de Sitter spacetimes are continuously extendible. This is done by showing that such spacetimes can be conformally embedded into the Einstein static universe, just like exact dS. This also serves for the result of Section \ref{sec:inhomoaniso}: assuming the Einstein equations with a perfect fluid, we can prove that, under certain extendibility assumptions that do \emph{not} require homogeneity and isotropy, the initial equation of state of matter must be that of a positive cosmological constant. (Recall dS solves the Einstein equations with a positive cosmological constant.) After presenting some applications to inflationary cosmology, we summarize and discuss our results in Section \ref{sec:conclusions}, especially with regard to physical applications, implications, and interpretations.

\section{Classification of the past of inflation}\label{sec:3cases}

\subsection{Preliminaries}\label{sec:firstPreliminaries}

The first sections of this work deal with homogeneous and isotropic (FLRW) universes with flat spatial sections. The $3+1$-dimensional manifold is $M=I\times\mathbb{R}^3$, where $I\subseteq\mathbb{R}$ is an open interval (of time). The metric $g$ may be written in various ways, e.g.,
\begin{equation}
    g=-\dd t^2+a(t)^2h_{\mathbb{E}}=-\dd t^2+a(t)^2\left(\dd r^2+r^2\dd\Omega^2\right)\,,\label{eq:gFLRW}
\end{equation}
where $h_{\mathbb{E}}:=\dd x^2+\dd y^2+\dd z^2$ is the usual 3-dimensional Euclidean metric, where the Cartesian coordinates $(x,y,z)$ are related to the spherical coordinates $(r,\theta,\phi)$ in the usual way. In the above, $\dd\Omega^2=\dd\theta^2+\sin^2\theta\,\dd\phi^2$ is the round metric on $\mathbb{S}^2$. The scale factor $a(t):I\to\mathbb{R}_{>0}$ is a smooth, positive, dimensionless function of the physical time $t$, and it describes the evolution of the size of spatial sections.\footnote{The convention in cosmology is to set the present-time value of the scale factor to $a_0=1$. Its value at other times relates to the physical redshift $z$ of electromagnetic waves through the relation $1/a(t)=1+z$.} The Hubble parameter $H(t):I\to\mathbb{R}$, defined by $H(t):=\dot a(t)/a(t)$ with a dot denoting a derivative with respect to $t$, characterizes the expansion or contraction rate of the spatial sections.

Throughout Section \ref{sec:3cases}, we denote the infimum of $I$ by $t_\mathrm{ini}$, which thus corresponds to the `initial time' of comoving observers in this coordinate system. Either $t_\mathrm{ini}>-\infty$ or $t_\mathrm{ini}=-\infty$. In the former case, we will assume that $a(t)\searrow 0$ as $t\to t_\mathrm{ini}^+$, i.e., we assume
\begin{equation}
    t_\mathrm{ini}>-\infty \quad \implies \quad \lim_{t\to t_\mathrm{ini}^+}a(t)=0^+\,,\label{eq:assumtinifiniteato0}
\end{equation}
since otherwise one could easily extend the spacetime continuously to times $t < t_\mathrm{ini}$ and we wish to ignore such situations. In the latter case ($t_\mathrm{ini}=-\infty$), many things can occur, e.g., $a(t)$ could go to $0$ in the infinite past, it could go to infinity (the universe started infinitely large in the asymptotic past as in bouncing scenarios), or it could go to a constant (the universe started close to Minkowski space in the asymptotic past as in loitering scenarios). Bouncing and loitering scenarios are the subject of vast literature, which we do not review here (for some examples and reviews, see, e.g., \cite{Novello:2008ra,Cai:2014bea,Battefeld:2014uga,Brandenberger:2016vhg,Boruah:2018pvq}). For the rest of this paper (from Section \ref{sec:metricext} onwards), we will be mostly concerned with the situation where $a(t)\searrow 0$ as $t\to-\infty$.

\subsection{Inflationary classification}\label{sec:inflClass}

\begin{figure}[h]
\centering
\includegraphics[scale=0.5]{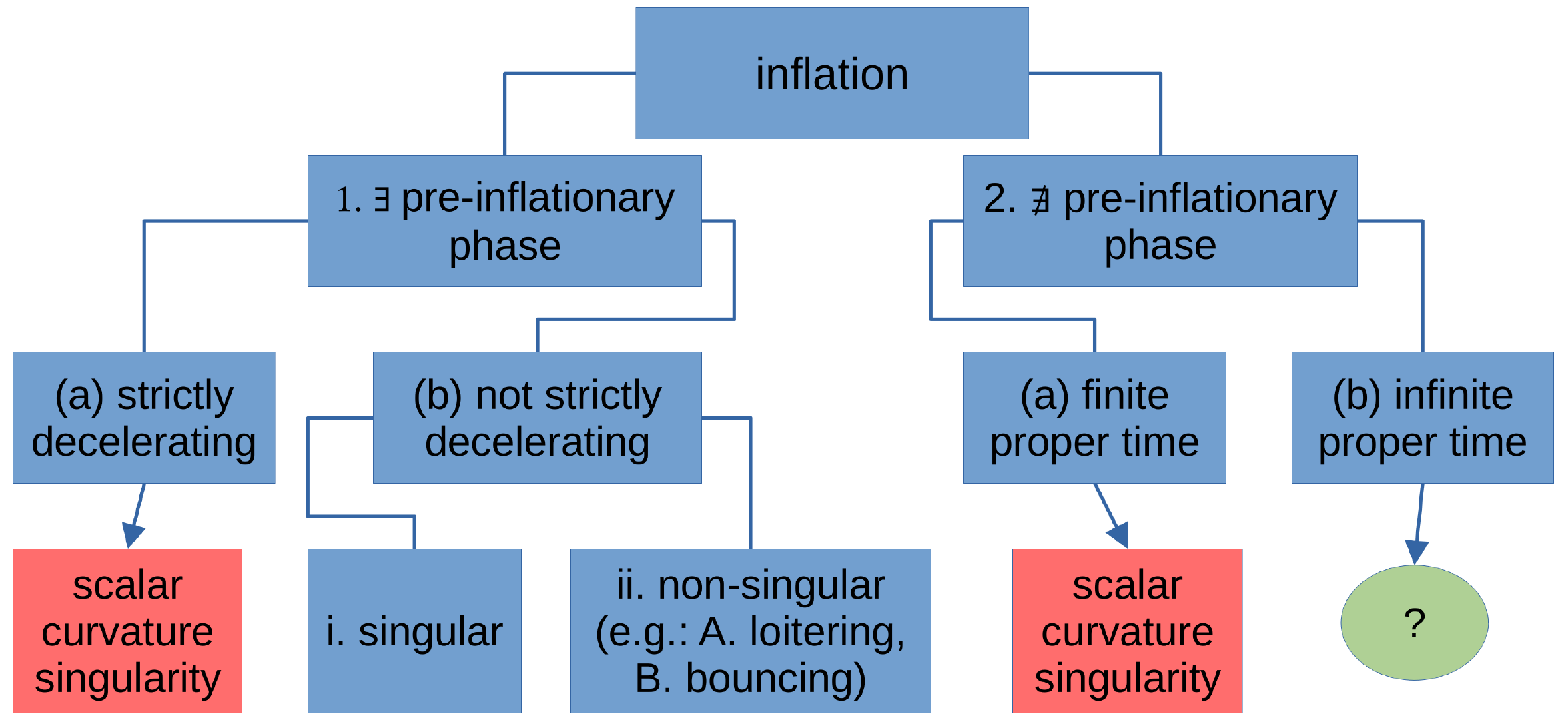}
\caption{{\small Classification of (the past of) inflationary cosmology within flat FLRW described in the text. Note that `$\nexists$ pre-inflationary phase' is meant to apply only within the spacetime $(M,g)$. Indeed, the spacetime might well be extendible beyond the boundary that lives at infinite proper time in Case \ref{itm:case2b}. This is the subject of the next sections.}}
\label{fig:flow}
\end{figure}

One can define inflation in many ways, but for the purpose of this work, we define it to be a period of time during which the universe is expanding at an accelerating rate. Therefore, we say that, for some interval $I_\mathrm{inf}\subset I$,
\begin{equation}
    \dot a(t)>0\qquad\textrm{and}\qquad\ddot a(t)>0
    \label{eq:defInf}
\end{equation}
for all $t\in I_\mathrm{inf}$.
Then, we can classify (the past of) inflationary scenarios into two general cases, each containing various subcases:
\begin{enumerate}
    \item Inflation started at some finite time in the past when the universe had a finite size. Therefore, the universe had already undergone some evolution prior to inflation (we say that there was a pre-inflationary phase). This corresponds to saying that there are starting and ending times, say $t_1,t_2\in\mathbb{R}$ respectively, during which the universe is inflating, so $I_\mathrm{inf}=[t_1,t_2]$. Note that the universe had a finite size at the initial inflationary time (i.e., $0<a(t_1)<\infty$) and finite expansion rate (i.e., $0<H(t_1)<\infty$). Prior to inflation, two general situations could occur, so we define two subcases:
    \begin{enumerate}
        \item\label{itm:case1a} The universe was always decelerating prior to inflation, i.e., $\ddot a(t)\leq 0$ for all $t\in I_{\textrm{pre-inf}}:=(t_\mathrm{ini},t_1)$. As we will prove shortly, this situation implies that
        \begin{equation}
            \lim_{t\to t_\mathrm{ini}^+}a(t)=0^+\,,\nonumber
        \end{equation}
        at which point scalar curvature invariants diverge, hence there is an initial scalar curvature singularity. A depiction is given in Figure \ref{fig:1a}.

\begin{figure}[h]
\centering
\includegraphics[width=0.35\textwidth]{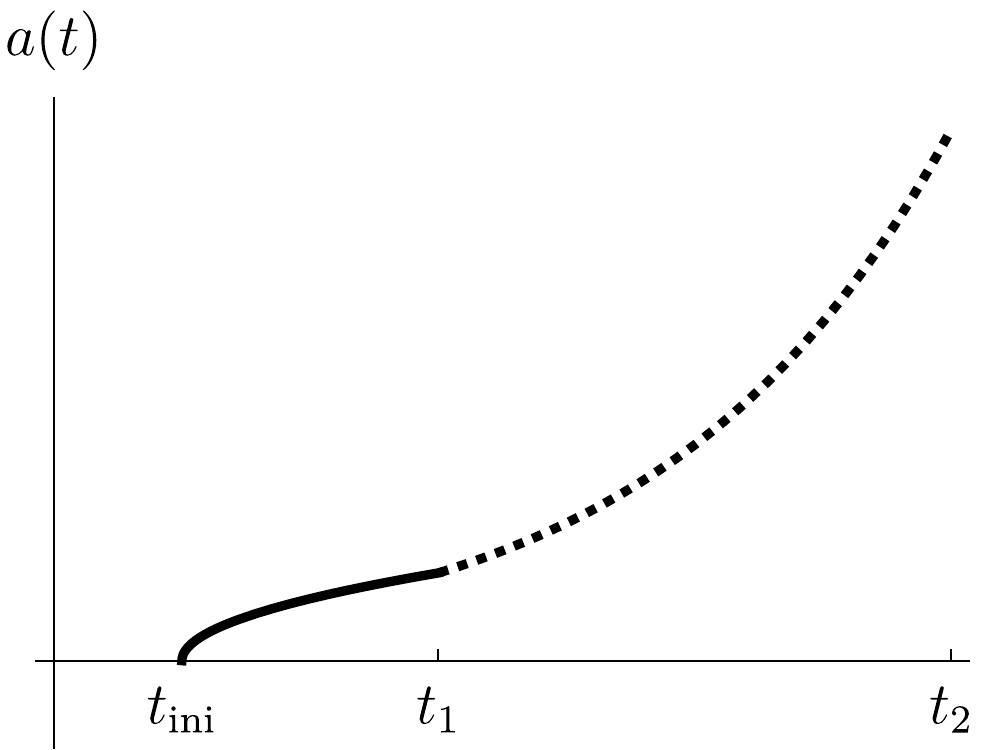}
\caption{{\small Sketch of an example of scale factor evolution in Classification \ref{itm:case1a}. The universe is inflating ($\dot a(t)>0$ and $\ddot a(t)>0$) for the interval $[t_1,t_2]$, but it is decelerating in a pre-inflationary phase from $t_\mathrm{ini}$ to $t_1$. A singularity is reached at $t_\mathrm{ini}$.}}
\label{fig:1a}
\end{figure}
        
        \item\label{itm:case1b} Alternatively, the universe was not always decelerating prior to inflation. In such a situation, many things can happen, for instance:
        \begin{enumerate}
            \item\label{itm:case1bi} there could still be an initial singularity to the past; or
            \item\label{itm:case1bii} \begin{enumerate}
                \item\label{itm:case1biiA} the universe could slowly approach a loitering phase with constant scale factor (i.e., it is past-asymptotically Minkowski), in which case it may be non-singular and geodesically complete (a depiction is given in the left panel of Figure \ref{fig:1b});
                \item\label{itm:case1biiB} or the universe could undergo a non-singular bounce, before which the universe started large and was contracting. A depiction is given in the right panel of Figure \ref{fig:1b}.

\begin{figure}[h]
\centering
\includegraphics[width=0.35\textwidth]{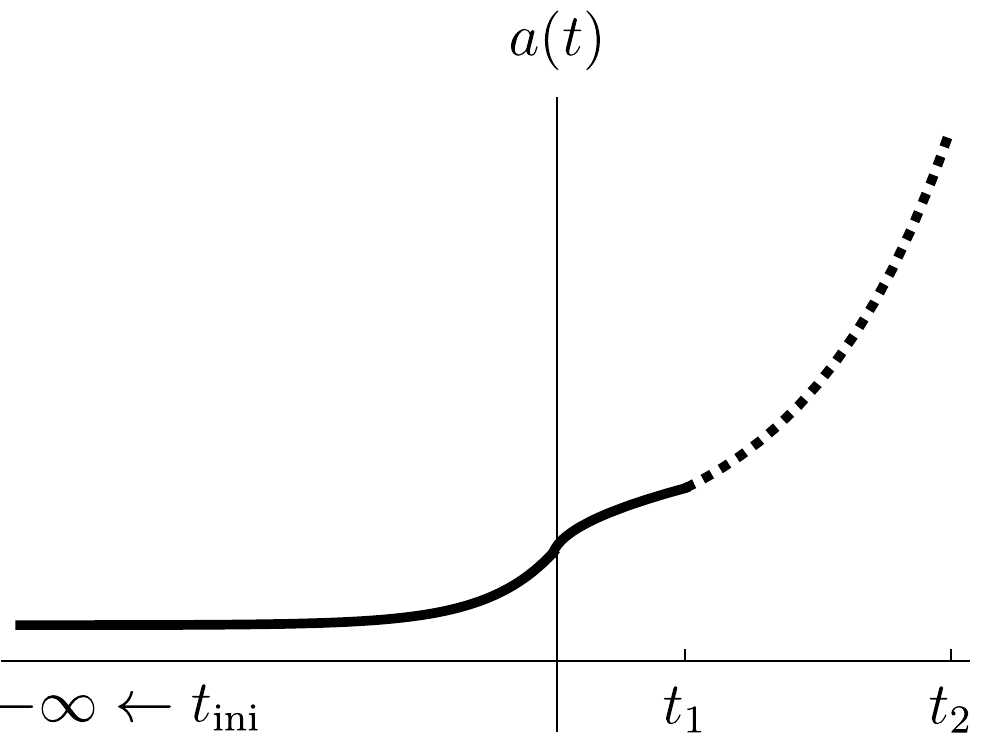}
\hspace*{0.1\textwidth}
\includegraphics[width=0.35\textwidth]{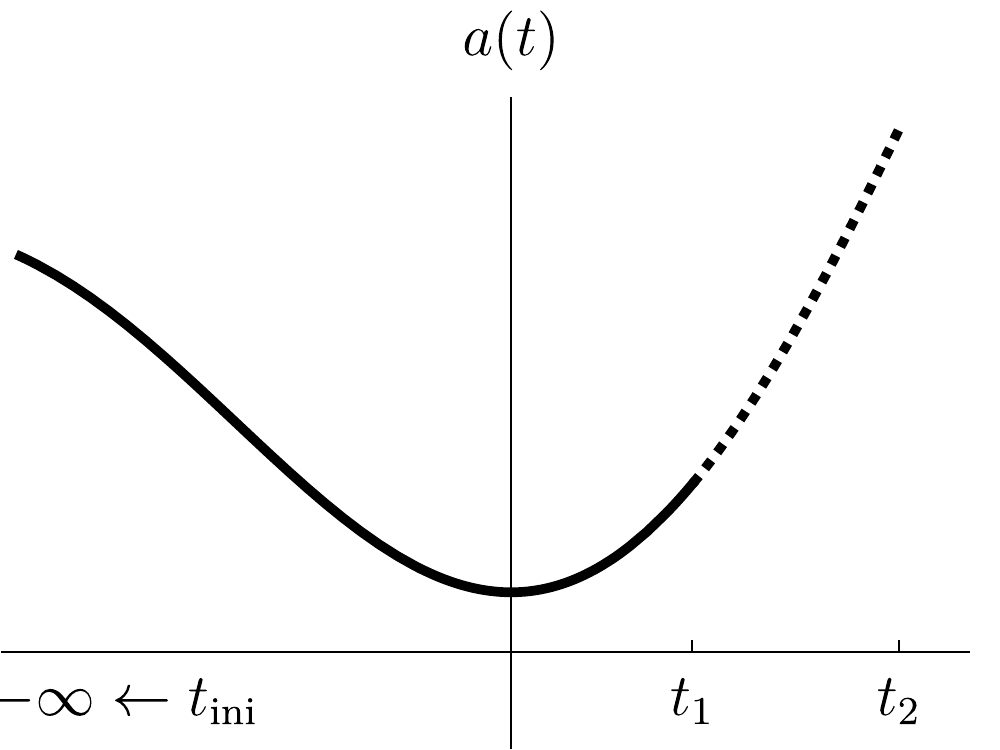}
\caption{{\small \textit{Left:} Sketch of an example of scale factor evolution in Classification \ref{itm:case1biiA}, where the universe is inflating from $t_1$ to $t_2$, but before that, $a(t)$ both decelerates and accelerates. In particular, the universe approaches a `loitering phase' in the past, i.e., the scale factor goes to a positive constant as time approaches $-\infty$. \textit{Right:} Sketch of an example of scale factor evolution in Classification \ref{itm:case1biiB}, where the universe is again inflating from $t_1$ to $t_2$ and both accelerating and decelerating before that. In particular, the universe undergoes a non-singular bounce, i.e., a transition from contraction to expansion about a non-zero minimum for $a(t)$. If the scale factor grows to infinity in the past or reaches a constant, the initial time $t_\mathrm{ini}$ is pushed to $-\infty$.}}
\label{fig:1b}
\end{figure}
                
            \end{enumerate}
        \end{enumerate}
        These latter situations (loitering and bouncing) often have to invoke some modifications to general relativity, matter violating the standard energy conditions (the null energy condition in particular), or quantum gravity. While we do not review them here, it is nevertheless important to stress that there are possible non-singular, geodesically complete pre-inflationary scenarios.
    \end{enumerate}
    \item\label{itm:case2} The second possibility is that inflation happened all the way `to the beginning', at which point $a(t)$ approaches $0$. Here, the `beginning' might be at finite time or at infinite time in the past, so we distinguish additional subcases:
    \begin{enumerate}
        \item\label{itm:case2a} $I_\mathrm{inf}=(t_\mathrm{ini},t_2]$, $t_\mathrm{ini}>-\infty$, and
        \begin{equation}
            \lim_{t\to t_\mathrm{ini}^+}a(t)=0^+\,.\nonumber
        \end{equation}
        A depiction is given in the left panel of Figure \ref{fig:2}.
        \item\label{itm:case2b} Alternatively, $I_\mathrm{inf}=(-\infty,t_2]$ and
        \begin{equation}
            \lim_{t\to -\infty}a(t)=0^+\,.\nonumber
        \end{equation}
        A depiction is given in the right panel of Figure \ref{fig:2}.

\begin{figure}[h]
\centering
\includegraphics[width=0.35\textwidth]{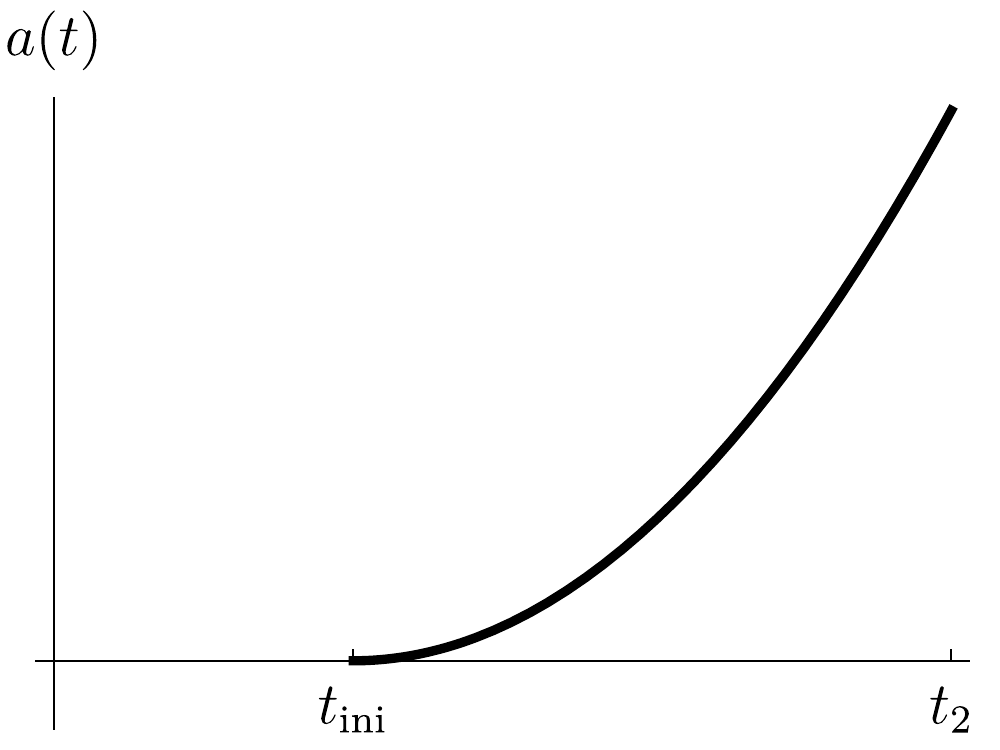}
\hspace*{0.1\textwidth}
\includegraphics[width=0.35\textwidth]{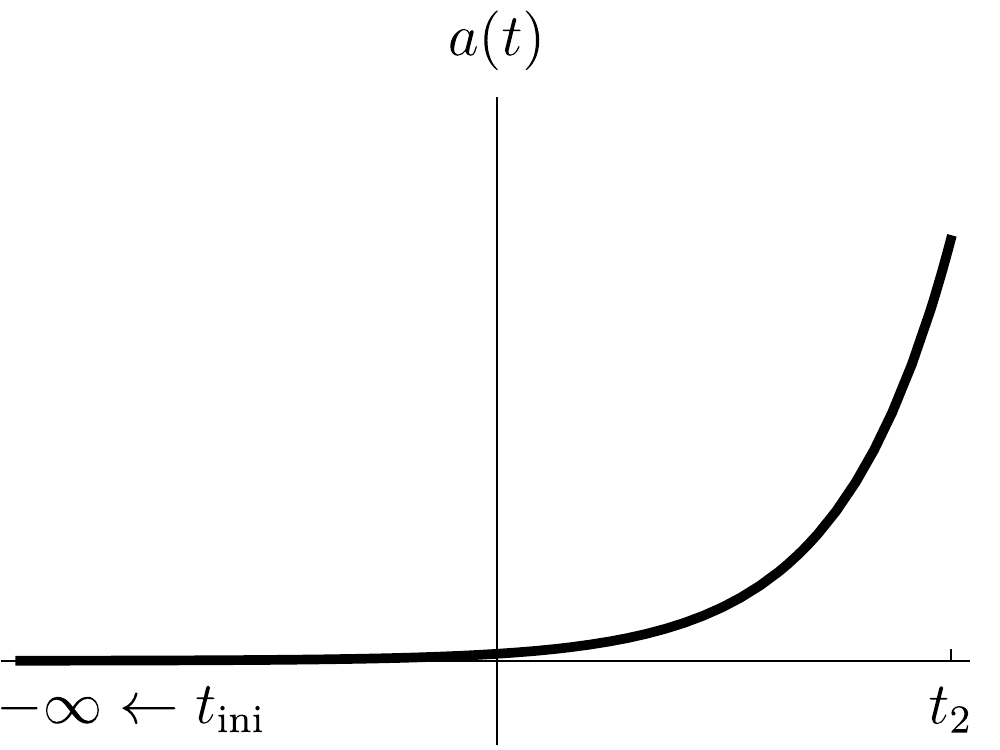}
\caption{{\small \textit{Left:} Example of scale factor evolution in Classification \ref{itm:case2a}. The universe is inflating from a finite initial time $t_\mathrm{ini}$, which corresponds to a singularity. \textit{Right:} Example of scale factor evolution in Classification \ref{itm:case2b}, where the universe is inflating all the way to $t_\mathrm{ini}=-\infty$.}}
\label{fig:2}
\end{figure}

    \end{enumerate}
    We will shortly show that Case \ref{itm:case2a} implies that there is an initial scalar curvature singularity at $t_\mathrm{ini}$. Correspondingly, this will motivate us to consider Case \ref{itm:case2b} more closely for the rest of this paper.
\end{enumerate}
A diagrammatic summary of the different cases is given in Figure \ref{fig:flow}.

\medskip
\medskip

Let us repeat the statement of Case \ref{itm:case1a} defined above:
\begin{thm}\label{thm:pre-inf-sing}
    Let $\ddot a(t)\leq 0$ for all $t\in I_{\mathrm{pre}\textrm{-}\mathrm{inf}}:=(t_\mathrm{ini},t_1)$, where $t_1$ denotes the time at which inflation would start (so the second time derivative of $a(t)$ would change sign from that point onward). Then, $t_\mathrm{ini}>-\infty$, hence $a(t)\searrow 0$ as $t\to t_\mathrm{ini}^+$, and at least two scalar curvature invariants diverge in the same limit. Correspondingly, the spacetime has an initial scalar curvature singularity in the limit to $t_\mathrm{ini}$.
\end{thm}

\proof
By integration, $\ddot a(t)\leq 0$ implies $\dot a(t)\geq\dot a(t_1)$ for all $t\in I_{\textrm{pre-inf}}$, and the fact that $\ddot a(t)\leq 0$ also implies $\dot a(t)$ is a
decreasing function of $t$ in the interval $I_{\textrm{pre-inf}}$. From our definition of inflation in equation \eqref{eq:defInf}, we have $\dot a(t_1)>0$, so
\begin{equation}
    \dot a(t)\geq\dot a(t_1)>0
    \label{eq:dotaineq}
\end{equation}
for all $t\in I_{\textrm{pre-inf}}$.
Integrating this inequality, we have, always for all $t\in I_{\textrm{pre-inf}}$,
\begin{align}
    \int_t^{t_1}\dd\tilde t\,\dot a(\tilde t)\geq\int_t^{t_1}\dd\tilde t\,\dot{a}(t_1)>0 &\implies a(t_1)-a(t)\geq\dot{a}(t_1)(t_1-t)>0\nonumber\\
    &\implies a(t)\leq a(t_1)-\dot a(t_1)(t_1-t)\,.\label{eq:aUpBound}
\end{align}
Let $H_1:=\dot a(t_1)/a(t_1)$, which is a finite and positive number --- it is the Hubble expansion rate at the beginning of inflation.
Then we can define $t_\mathrm{out}:=t_1-H_1^{-1}$, $-\infty<t_\mathrm{out}<t_1$, which would suggest $a(t_\mathrm{out})\leq 0$ according to \eqref{eq:aUpBound}. However, from our definition of the scale factor in \ref{sec:firstPreliminaries}, $a(t)>0$ for all $t\in I$. Since $a(t)$ is considered to be smooth all the way to $\inf I=t_\mathrm{ini}$, it must be that $-\infty<t_\mathrm{out}\leq t_\mathrm{ini}<t_1$, and thus by the assumption \eqref{eq:assumtinifiniteato0},
\begin{equation}
    \lim_{t\to t_\mathrm{ini}^+>-\infty}a(t)=0^+\,.\label{eq:abblim}
\end{equation}
Consequently, the pre-inflationary phase can only have a finite duration with $I_\textrm{pre-inf}=(t_\mathrm{ini},t_1)$, and this is the first phase of the universe since $t_\mathrm{ini}=\inf I=\inf I_\textrm{pre-inf}$ in that case.
Then from \eqref{eq:dotaineq} and from the fact that $\dot a(t)$ is a
decreasing function of $t$, we can tell that $\dot a(t)$ reaches a positive limit (either finite or infinite) as $t\to t_\mathrm{ini}^+$. Therefore, recalling $H:=\dot a/a$, equation \eqref{eq:abblim} implies
\begin{equation}
    \lim_{t\to t_\mathrm{ini}^+}H(t)=+\infty\,.\label{eq:Hbblim}
\end{equation}
At this point, it is important to recall a few linearly independent curvature invariants in a flat FLRW spacetime:
\begin{subequations}\label{eq:lqci}
\begin{align}
    R&=6(2H^2+\dot H)\,,\\
    R_{\mu\nu}R^{\mu\nu}&=3\big((3H^2+2\dot H)^2+3H^4\big)\,,\\
    R_{\mu\nu\rho\sigma}R^{\mu\nu\rho\sigma}&=12\big((H^2+\dot H)^2+H^4\big)\,,
\end{align}
\end{subequations}
are the Ricci scalar, the contraction of the Ricci tensor with itself, and the contraction of the Riemann tensor with itself (also known as the Kretschmann scalar), respectively. We can finally make the observation that at least two of them will diverge as $t\to t_\mathrm{ini}^+$ because of \eqref{eq:Hbblim}. Indeed, $R_{\mu\nu}R^{\mu\nu}$ and the Kretschmann scalar are non-negative functions of $H^2$ and $\dot H$, so they will diverge when $H$ diverges as $t\to t_\mathrm{ini}^+$, regardless of what $\dot H$ does. Correspondingly, there is an initial scalar curvature singularity as $t\to t_\mathrm{ini}^+$.
\qed

\medskip
\medskip

\noindent\textit{Remark.}
If one assumed general relativity and the strong energy condition, deceleration ($\ddot a(t)\leq 0$) would follow and thus the above theorem. Whenever the strong energy condition is satisfied, it is not unsurprising for the spacetime to have an inevitable initial singularity. The previous theorem, though, applies irrespectively of the gravitational field equations at play: only a geometric `convergence' condition (deceleration) is assumed, albeit the theorem is restricted to flat FLRW spacetimes.

\medskip
\medskip

Let us then present results pertaining to the Case \ref{itm:case2} defined earlier:
\begin{lem}\label{lem:Htoconsttminfty}
    Suppose $H(t)\to H_\Lambda\in\mathbb{R}_{\geq 0}$ as $t\to t_\mathrm{ini}^+$, where we recall $t_\mathrm{ini}$ is the infimum of the domain of $a(t)$. Then, $t_\mathrm{ini}=-\infty$.
\end{lem}

\proof
The proof goes by contradiction: suppose $t_\mathrm{ini}>-\infty$. Then, by the assumption \eqref{eq:assumtinifiniteato0}, we have $a(t)\searrow 0$ as $t\to t_\mathrm{ini}^+$.
Then, by definition of $\lim_{t\to t_\mathrm{ini}^+}H(t)=H_\Lambda$, there exists a $t_\mathrm{L}>t_\mathrm{ini}$ such that, for all $t\in(t_\mathrm{ini},t_\mathrm{L})$, $0<|H(t)-H_\Lambda|<\e$ for any $\e\in\mathbb{R}_{>0}$. Upon integrating from $t$ to $t_\mathrm{L}$, we thus have, at any time $t\in(t_\mathrm{ini},t_\mathrm{L})$,
\begin{align}
    &-\e(t_\mathrm{L}-t)<\int_t^{t_\mathrm{L}}\dd\tilde t\,(H(\tilde t)-H_\Lambda)<\e(t_\mathrm{L}-t)\nonumber\\
    \implies & (\e-H_\Lambda)(t-t_\mathrm{L})<\ln\!\big(a(t_\mathrm{L})\big)-\ln\!\big(a(t)\big)<(\e+H_\Lambda)(t_\mathrm{L}-t)\nonumber\\
    \implies & a(t_\mathrm{L})e^{(\e+H_\Lambda)(t-t_\mathrm{L})}<a(t)<a(t_\mathrm{L})e^{(H_\Lambda-\e)(t-t_\mathrm{L})}\,.\label{eq:alimitsHconst}
\end{align}
Therefore, $a(t)$ is bounded below by a decaying exponential of $t$ as time decreases. In particular, since \eqref{eq:alimitsHconst} applies for all $t\in(t_\mathrm{ini},t_\mathrm{L})$ and for any $\e>0$, we can take the limit of \eqref{eq:alimitsHconst} as $t\to t_\mathrm{ini}^+$, where we recall $t_\mathrm{ini}>-\infty$, and we find that $\lim_{t\to t_\mathrm{ini}^+}a(t)$ is bounded below by a positive number, $a(t_\mathrm{L})e^{(\e+H_\Lambda)(t_\mathrm{ini}-t_\mathrm{L})}$, in contradiction with $\lim_{t\to t_\mathrm{ini}^+}a(t)=0^+$.
\qed

\medskip
\medskip

\noindent\emph{Remarks.}
Since the previous proof holds for any $\e>0$, in particular it must be true for a small $0<\e<H_\Lambda$ whenever $H_\Lambda>0$. Thus, in such a situation, $a(t)$ is actually bounded in between two decaying exponentials of $t$ as time decreases according to \eqref{eq:alimitsHconst}, and hence $a(t)$ itself decays to zero.

Another observation is that the contrapositive of the above lemma tells us that whenever $t_\mathrm{ini}>-\infty$, $H(t)$ cannot approach a non-negative constant as $t\to t_\mathrm{ini}^+$. This will be useful shortly.

\medskip
\medskip

\begin{prop}\label{prop:cases2a}
    Suppose the universe reaches zero volume in finite time, i.e., $a(t)\searrow 0$ as $t\to t_\mathrm{ini}^+$ with $\inf I=t_\mathrm{ini}>-\infty$, and suppose the universe is initially expanding in some neighborhood of $t_\mathrm{ini}$, i.e., $\dot a(t)>0$ for all $t\in U:=(t_\mathrm{ini},t_\mathrm{ini}+\delta)$ for some fixed $\delta\in\mathbb{R}_{>0}$. Then, $H(t)$ is unbounded in the neighborhood $U$ of $t_\mathrm{ini}$.
\end{prop}

\proof
The proof goes by contradiction: suppose $H(t)$ is bounded in the neighborhood $U$ of $t_\mathrm{ini}$. This means that there exists $H_\mathrm{max}\in\mathbb{R}_{>0}$ such that $0<H(t)<H_\mathrm{max}$ for all $t\in U$. The lower bound comes from the fact that $H(t)$ must be positive in $U$, because $a(t)$ is always positive and $\dot a(t)$ is positive in $U$. Writing $H:=\dot a/a=\dd\ln a/\dd t$, we find $0<\dd\ln a<H_\mathrm{max}\dd t$ for all $t\in U$, and so, integrating from an arbitrary $t\in U$ to $t_\mathrm{ini}+\delta$ yields
\begin{equation}
    0<\ln\!\big(a(t_\mathrm{ini}+\delta)\big)-\ln\!\big(a(t)\big)<H_\mathrm{max}(t_\mathrm{ini}+\delta-t)\,.\nonumber
\end{equation}
In particular, this must hold as $t\to t_\mathrm{ini}^+$, in which case $a(t)\searrow 0$, but since $a(t_\mathrm{ini}+\delta)>0$, this would then imply $0<+\infty<H_\mathrm{max}\delta$, hence the contradiction.
\qed

\medskip
\medskip

\begin{thm}\label{thm:cases2}
    If inflation happens all the way to the beginning of the universe, i.e., $a(t)\searrow 0$ as $t\to t_\mathrm{ini}^+$, where $\inf I=\inf I_\mathrm{inf}=t_\mathrm{ini}\in\mathbb{R}\cup\{-\infty\}$, and $\dot a(t)>0$ (so $H(t)>0$) and $\ddot a(t)>0$ for all $t\in(t_\mathrm{ini},t_2]$, then either
    \begin{enumerate}[label=(\roman*)]
        \item $t_\mathrm{ini}=-\infty$ and $H(t)$ has a finite limit as $t\to -\infty$;
        \item $t_\mathrm{ini}=-\infty$ and $H(t)$ does not have a finite limit as $t\to -\infty$;
        \item $t_\mathrm{ini}>-\infty$ and $H(t)$ does not have a finite limit as $t\to t_\mathrm{ini}^+$.
    \end{enumerate}
    In case (iii), the spacetime has an initial scalar curvature singularity.
\end{thm}

\proof
According to the contrapositive of Lemma \ref{lem:Htoconsttminfty}, if $t_\mathrm{ini}>-\infty$, then $H(t)\nrightarrow H_\Lambda\in\mathbb{R}_{\geq 0}$. By the hypothesis that $H(t)>0$ for all $t\in(t_\mathrm{ini},t_2]$, it must therefore be that $H(t)$ does not have a finite limit (i.e., it cannot converge) as $t\to t_\mathrm{ini}^+$ --- this corresponds to case (iii) of the theorem (or \ref{itm:case2a} in the classification defined earlier). The other alternative (Classification \ref{itm:case2b}) is that $t_\mathrm{ini}=-\infty$, in which case either $H(t)$ converges to a finite positive number [case (i)] or it does not [case (ii)].

In case (iii), Proposition \ref{prop:cases2a} tells us that $H(t)$ is actually unbounded in a neighborhood of $t_\mathrm{ini}$. This means that either $H(t)$ diverges to $+\infty$ or $H(t)$ oscillates without bound as $t\to t_\mathrm{ini}^+$. In both cases, we can use the same argument as in the proof of Theorem \ref{thm:pre-inf-sing} to say that a scalar curvature singularity is reached as $t\to t_\mathrm{ini}^+$ (in particular, the Kretschmann scalar diverges, i.e., it either goes to $\infty$ or oscillates without bound).
\qed

\medskip
\medskip

\noindent\emph{Remark.}
Note that there is no analogous statement to Proposition \ref{prop:cases2a} when $t_\mathrm{ini}=-\infty$. In fact, it is straightforward to construct a counter-example: if $a(t)=\exp(t+\frac{1}{2}\sin t)$, then $a\searrow 0$ as $t\to-\infty$ and $\ddot a(t)$ is everywhere positive, but while $H(t)=1+\frac{1}{2}\cos t$ does not reach a limit as $t\to-\infty$, it is nevertheless bounded within $1/2$ and $3/2$. Therefore, case (ii) of the above theorem does not necessarily imply a `blow up' of $H(t)$ and thus a scalar curvature singularity.

\medskip
\medskip

\noindent\emph{Examples.}
The previous theorem tells us that if we are in the Classification \ref{itm:case2a} of inflation, then there is an initial big bang-like scalar curvature singularity at a finite time in the past.
A prototypical example would be a scale factor that is asymptotically a power-law,
\begin{equation}
	a(t)\stackrel{t\to 0^+}{=}a_0t^q+\ldots\,,\qquad q\in\mathbb{R}_{>1}\,,\ a_0\in\mathbb{R}_{>0}\,,\nonumber
\end{equation}
for which $R$, $R_{\mu\nu}R^{\mu\nu}$, and $R_{\mu\nu\rho\sigma}R^{\mu\nu\rho\sigma}$ are all divergent as $t\to 0^+$. Here, without loss of generality, we set $t_\mathrm{ini}=0$. Another example is a function of the form $a(t)\propto\mathrm{exp}(-1/t^\alpha)$ for $\alpha\in\mathbb{R}_{>0}$. Yet another would be $a(t)\propto\mathrm{exp}[A(\ln t)^\gamma]$ with $A\in\mathbb{R}_{>0}$ and $\gamma\in\{3,5,7,9,\ldots\}$. The latter is a subclass of what is known as `logamediate' inflation in the literature \cite{Barrow:1996bd,Barrow:2007zr}.

\medskip
\medskip

Let us end this subsection by commenting on the fact that we specialized the discussion to \emph{flat} FLRW. This is crucial since many statements do not necessarily apply with either positively or negatively curved spatial sections. It is known that inflation `dilutes' spatial curvature, hence assuming flatness during and after inflation is well justified. Prior to inflation (or at the very onset of inflation), all three cases --- flat, closed, open --- are possible. We choose to focus on flat spatial sections. The issue of extendibility in negatively curved FLRW spacetimes has been studied quite extensively in \cite{Galloway:2016bej,Ling:2018jzl,Ling:2018tih,Ling:2022uzs,Ling:2022kjk,Nomura:2021lzz}. With positive spatial curvature, some investigations exist as well \cite{Nomura:2021lzz}; in this context, there are examples of pre-inflationary non-singular bounces \cite{Bramberger:2019zez,Anabalon:2019equ}, and the issue of setting up the quantum initial conditions has been tackled in \cite{Letey:2022hdp}.

\subsection{Past-asymptotically de Sitter}\label{sec:padS}

To summarize what we have shown so far, either inflation in flat FLRW did not occur all the way to the past, in which case it is singular [cases \ref{itm:case1a} and \ref{itm:case1bi}] or potentially geodesically complete due to some modifications to general relativity or quantum gravity effects [example cases \ref{itm:case1biiA} and \ref{itm:case1biiB}], or inflation happened all the way to the past, in which case either there is a curvature singularity at finite time [Case \ref{itm:case2a}] or the beginning of inflation is at $t\to-\infty$ [Case \ref{itm:case2b}].

Therefore, this very last case is the only inflationary scenario in flat FLRW that might evade the singularity problem within the validity of general relativity and is the topic of interest for the rest of this paper. In this context, the existing result --- often referred to as the Borde-Guth-Vilenkin theorem \cite{Borde:2001nh} in the cosmological literature\footnote{Also reviewed in \cite{Yoshida:2018ndv}; see \cite{Lesnefsky:2022fen} for a discussion of its drawbacks though.} --- states that if $H(t)>0$ for all $t\in(-\infty,t_2]$, then null geodesics are past incomplete. Yet, this does not always imply the presence of a singularity as $t\to-\infty$, and the spacetime might actually be extendible beyond the past boundary at $t_\mathrm{ini}=-\infty$. As alluded to in the introduction, flat de Sitter is a nice example of this feature, but \emph{exact} dS is a very special solution that cannot represent our universe. A necessary feature of all the inflationary scenarios is that $H(t)$ is not constant in time, allowing quantum fluctuations to generate the seeds of our universe's large-scale structures consistent with observations. Therefore, it is more relevant to examine the singularity for `quasi-de Sitter' universes. In particular the next natural step is to study those that approach dS in the infinite past, as to fit in the Classification \ref{itm:case2b}. A similar premise was used in \cite{Yoshida:2018ndv}, but we shall make it more precise in the following definition.

\medskip
\medskip

\begin{Def}\label{def:padS}
    \emph{A flat FLRW spacetime as defined in Section \ref{sec:firstPreliminaries} is said to be flat\footnote{Since we will always be concerned with flat spatial sections, the adjective `flat' shall often be omitted and simply implied.} \emph{past-asymptotically de Sitter} if
    \begin{equation}
        H(t)\stackrel{t\to-\infty}=H_\Lambda+o(1)\,,\qquad\dot H(t)\stackrel{t\to-\infty}{=}0+o(1)\,,\nonumber
    \end{equation}
    where $H_\Lambda\in\mathbb{R}_{>0}$. Note that the asymptotic little-$o$ notation is equivalent to saying that
    \begin{equation}
        \lim_{t\to-\infty}H(t)=H_\Lambda\,,\qquad\lim_{t\to-\infty}\dot H(t)=0\,.\nonumber
    \end{equation}}
\end{Def}

\medskip
\medskip

\noindent\emph{Remark.}
Exact de Sitter is a solution to the Einstein equations with a positive cosmological constant $\Lambda\in\mathbb{R}_{>0}$. In flat FLRW, the corresponding Hubble parameter is a constant given by $H_\Lambda=\pm\sqrt{\Lambda/3}$, and here we are choosing the expanding branch with positive constant Hubble parameter. This is the motivation for the above notation in past-asymptotically dS.

\medskip
\medskip

In the following Lemma, we show that the definition of past-asymptotically de Sitter could have been $H(t)\to H_\Lambda$ and the limit of $\dot H$ exists (and is finite) as $t\to-\infty$.

\medskip

\begin{lem}\label{lem:HtohdotHtoconst}
    Suppose a flat FLRW spacetime as defined in Section \ref{sec:firstPreliminaries} is such that
    \begin{equation}
        H(t)\stackrel{t\to-\infty}=H_\Lambda+o(1)\,,\qquad\dot H(t)\stackrel{t\to-\infty}{=}c+o(1)\,,\nonumber
    \end{equation}
    where $H_\Lambda\in\mathbb{R}_{>0}$ and $c \in\mathbb{R}$. Then the spacetime is past-asymptotically de Sitter.
\end{lem}
\proof
We note that 
\begin{equation*}
    \lim_{t\to-\infty}\frac{\ddot{a}}{a}= \lim_{t\to-\infty}(\dot{H}+H^2)= c+H_\Lambda^2\,,
\end{equation*}
leading to 
\begin{equation}
        \lim_{t\to-\infty}\ddot{a}=\lim_{t\to-\infty} \frac{\ddot{a}}{a}\times a= \lim_{t\to-\infty} \frac{\ddot{a}}{a}\times \lim_{t\to-\infty} a= (c+H_\Lambda^2)\times 0=0\,.\nonumber
    \end{equation}
    Note that we are using $a(t) \searrow 0$ as $t \to -\infty$. This follows since $H_\Lambda > 0$; see the remark after Lemma \ref{lem:Htoconsttminfty}.
    At the same time, we have 
    \begin{equation}
        \lim_{t\to-\infty}\dot{a}=\lim_{t\to-\infty} \frac{\dot{a}}{a}\times a= \lim_{t\to-\infty} \frac{\dot{a}}{a}\times \lim_{t\to-\infty} a= H_\Lambda\times 0=0\,.\nonumber
    \end{equation}
    Therefore, applying L'H\^{o}pital's rule results in 
       \begin{equation}
         \lim_{t\to-\infty}H=\lim_{t\to-\infty} \frac{\dot{a}}{a}= \lim_{t\to-\infty} \frac{\ddot{a}}{\dot{a}}= \lim_{t\to-\infty} \frac{a}{\dot{a}}\times \frac{\ddot{a}}{a}=\frac{1}{H_\Lambda}(c+H_\Lambda^2)\,,\nonumber
    \end{equation}
    which can hold only if $c=0$ since $\lim_{t\to-\infty}H= H_\Lambda$. 
\qed

\medskip
\medskip

In the following Lemma, we show another alternative notion of past-asymptotically de Sitter.

\medskip

\begin{lem}\label{lem:Hto1dotHto0}
    Suppose there exists a smooth function $f\colon I\to\mathbb{R}$ such that
    \begin{equation}
        a(t)=e^{H_\Lambda t}+f(t)\,,\qquad\dot a(t)=H_\Lambda e^{H_\Lambda t}+\dot f(t)\,,\qquad\ddot a(t)=H_\Lambda^2e^{H_\Lambda t}+\ddot f(t)\,,\nonumber
    \end{equation}
    and such that $f(t),\dot f(t),\ddot f(t)$ are asymptotically $o(e^{H_\Lambda t})$ as $t\to -\infty$, i.e.,
    \begin{equation}
        \lim_{t\to-\infty}\frac{f(t)}{e^{H_\Lambda t}}=\lim_{t\to-\infty}\frac{\dot f(t)}{e^{H_\Lambda t}}=\lim_{t\to-\infty}\frac{\ddot f(t)}{e^{H_\Lambda t}}=0\,.\nonumber
    \end{equation}
    Then the spacetime is past-asymptotically de Sitter.
\end{lem}

\proof
We note that
\begin{equation}
	H(t)-H_\Lambda=\frac{\dot a(t)}{a(t)}-H_\Lambda=\frac{H_\Lambda e^{H_\Lambda t}+\dot f(t)}{e^{H_\Lambda t}+f(t)}-H_\Lambda=\frac{e^{-H_\Lambda t}(\dot f(t)-H_\Lambda f(t))}{1+e^{-H_\Lambda t}f(t)}\stackrel{t\to-\infty}{\longrightarrow}0\,,\nonumber
\end{equation}
and
\begin{align}
    \dot H(t)&=\frac{\ddot a(t)}{a(t)}-\frac{\dot a(t)^2}{a(t)^2}=\frac{(H_\Lambda^2e^{H_\Lambda t}+\ddot f(t))(e^{H_\Lambda t}+f(t))-(H_\Lambda e^{H_\Lambda t}+\dot f(t))^2}{(e^{H_\Lambda t}+f(t))^2}\nonumber\\
    &=\frac{e^{-H_\Lambda t}(H_\Lambda^2f(t)-2H_\Lambda\dot f(t)+\ddot f(t))+e^{-2H_\Lambda t}(f(t)\ddot f(t)-\dot f(t)^2)}{(1+e^{-H_\Lambda t}f(t))^2}\stackrel{t\to-\infty}{\longrightarrow}0\,.\nonumber
\end{align}
\qed

\medskip
\medskip

\noindent\emph{Remarks.}
Note that the converse of the previous lemma does not generally hold. A counter-example could simply be $a(t)=(1-t)e^{H_\Lambda t}$, which respects $H\to H_\Lambda$ and $\dot H\to 0$ as $t\to-\infty$, but which is not asymptotically of the form $a(t)=e^{H_\Lambda t}+o(e^{H_\Lambda t})$. Therefore, a scale factor that is past-asymptotically of exponential form is not equivalent to our Definition \ref{def:padS} of a past-asymptotically dS spacetime. Nevertheless, we shall explore both premises where the scale factor is past-asymptotically of exponential form and where the spacetime is past-asymptotically dS.

In the hypothesis of the previous lemma, note that we could have written $a(t)\stackrel{t\to-\infty}{=}a_0e^{H_\Lambda t}+o(e^{H_\Lambda t})$, $\dot a(t)\stackrel{t\to-\infty}{=}a_0H_\Lambda e^{H_\Lambda t}+o(e^{H_\Lambda t})$, etc., for some constant $a_0\in\mathbb{R}_{>0}$, but
$a_0$ is a physically irrelevant dimensionless quantity in flat FLRW, so we can
set it to unity.

As another remark, one could replace the hypothesis of Lemma \ref{lem:Hto1dotHto0} by just $\ddot a(t)\stackrel{t\to-\infty}{=}e^t+o(e^t)$ and the conditions that $a(t)\searrow 0$ and $\dot a(t)\searrow 0$ as $t\to-\infty$. Indeed, if $(\ddot a(t)-e^t)/e^t\to 0$ as $t\to-\infty$, then by L'H\^{o}pital's rule since $\dot a(t)\searrow 0$, we must have $(\dot a(t)-e^t)/e^t\to 0$ [so $\dot a(t)\stackrel{t\to-\infty}{=}e^t+o(e^t)$], and applying L'H\^{o}pital's rule once more, together with $a(t)\searrow 0$, implies $(a(t)-e^t)/e^t\to 0$ [so $a(t)\stackrel{t\to-\infty}{=}e^t+o(e^t)$]. Therefore, the asymptotic conditions on $a(t)$ and $\dot a(t)$ are implied by the asymptotic condition on $\ddot a(t)$ in this context.

\medskip
\medskip

Another motivation for Definition \ref{def:padS} is that if the spacetime approaches dS in the asymptotic past, we expect to recover the non-singular nature of exact dS, i.e., there should not be any scalar curvature singularity on the past boundary. To be precise, we conjecture the following:

\medskip

    \textit{Any scalar contractions of the Riemann tensor, Ricci tensor, and Ricci scalar are finite as $t\to-\infty$ in a past-asymptotically de Sitter spacetime.}

\medskip

\noindent We do not provide a rigorous proof of this statement, but as we will see and argue later in Section \ref{sec:curv}, any scalar curvature invariants constructed out of the product of $n$ Riemann tensors contracted with the metric tensor and its inverse (thus possibly involving the Ricci tensor and Ricci scalar) in flat FLRW seems to consist of a linear combination of $\{H^{2n},H^{2(n-1)}\dot H,H^{2(n-2)}\dot H^2,\,\cdots,H^2\dot H^{n-1},\dot H^n\}$. Therefore, it is straightforward to see from the definition of past-asymptotically dS that any such scalar curvature invariants should have a finite limit as $t\to-\infty$.

While past-asymptotically dS spacetimes should not have $C^2$ scalar curvature singularities according to the previous statement (where $C^2$ refers to the differentiability class of the metric tensor), this does not guarantee metric extendibility beyond the spacetime boundary at $t_\mathrm{ini}=-\infty$. Conditions for metric extendibility will be the subject of the next section. We can already mention that other kinds of singularity such as parallelly propagated singularities can still be present as $t\to -\infty$ in past-asymptotically dS universes, and those may be an obstruction to spacetime extendibility (cf.~\cite{Yoshida:2018ndv,Nomura:2021lzz} and references therein).

\medskip
\medskip

To end this section, let us mention that one might want to consider other leading-order functions for $a(t)$, $\dot{a}(t)$, and $\ddot a(t)$ in the limit $t\to-\infty$ than those in the hypothesis of Lemma \ref{lem:Hto1dotHto0}. However, ignoring (pathological) cases where $H(t)$ oscillates without reaching a limit or where $H(t)$ is unbounded (in which case there is a scalar curvature singularity to the past), the only possible situations as $t\to-\infty$ are $H(t)\to H_\Lambda>0$ or $H(t)\searrow 0$ (when $H(t)>0$ at least in a neighborhood of $t_\mathrm{ini}=-\infty$). The case $H(t)\to H_\Lambda>0$ as $t\to-\infty$ is the motivation for our definition of past-asymptotically dS spacetime, while the case $H(t)\searrow 0$ would correspond to something like a past-asymptotically Minkowski spacetime. We are only interested in the former case in the present work. In such a situation, as mentioned in the remark after Lemma \ref{lem:Htoconsttminfty}, $a(t)$ must be bounded in between two decaying exponentials,
hence additional justification for looking at the asymptotic functional expressions for $a(t)$ and its derivatives given in the hypothesis of Lemma \ref{lem:Hto1dotHto0}.

As a couple of examples of asymptotic functions that are \emph{not} of interest in the current work, one could have $a(t)=1/|t|^\alpha$ ($\alpha\in\mathbb{R}_{>0}$), which goes to 0 slower than $e^t$ as $t\to-\infty$, or $a(t)=\exp(-\exp(|t|^\alpha))$ ($\alpha\in(0,1]$), which goes to 0 faster than $e^t$ as $t\to-\infty$. Both examples respect the conditions of \eqref{eq:defInf} for inflation with $I_\mathrm{inf}=(-\infty,0)$, but by direct evaluation, one finds that $H(t)\to\infty$ as $t\to-\infty$ in both cases. A couple more examples could be $a(t)=\exp(-|t|^\alpha)$ and $a(t)=\exp(-(\ln|t|)^\alpha)$ ($\alpha\in\mathbb{R}_{>0}$ in both cases), again respecting \eqref{eq:defInf} with $I_\mathrm{inf}=(-\infty,0)$, but in those cases one finds $H(t)\searrow 0$ as $t\to-\infty$. Past-asymptotically Minkowski spacetimes could be the topic of a different paper.

\section{Metric extendibility}\label{sec:metricext}

After formally defining spacetime extendibility (Section \ref{sec: further prelim}) and presenting how exact dS may be continously extended as a warm-up (Section \ref{flat de Sitter sec}), we derive a $C^0$-extendibility theorem in Section \ref{sec:C0}, which either requires an asymptotic condition on the scale factor alone, $a(t)=e^{ht}+o(e^{ht})$ as $t\to-\infty$ for some $h>0$, or the condition that $H\to h>0$ as $t\to-\infty$. Asymptotic conditions on $a(t)$ and its derivatives lead to higher-regularity theorems in Section \ref{sec:Ckfull}, and we finally discuss curvature singularities in Section \ref{sec:curv}.

\subsection{Further preliminaries}\label{sec: further prelim}

Our conventions will follow \cite[Sec.~2]{Ling:2018tih}, which we briefly review. 

Let $k \geq 0$ be an integer or $\infty$. A $C^k$ \emph{spacetime} $(M,g)$ is a $C^{k +1}$  four-dimensional manifold $M$ (connected, Hausdorff, second countable) equipped with a $C^k$ Lorentzian metric tensor $g$ (i.e., its components $g_{\mu\nu}$ are $C^k$ functions in any coordinate system) and a time orientation induced by some $C^0$ timelike vector field. A \emph{future-directed timelike curve} $\g \colon [a,b] \to M$ is a piecewise $C^1$ curve such that $\dot\g(s):=\dd\gamma/\dd s$ is future-directed timelike for all $s \in [a,b]$, including its break points and endpoints (understood as one-sided derivatives). \emph{Past-directed timelike curves} are defined time-dually. The \emph{timelike future} of a point $p \in M$ with respect to an open set $U \subset M$ containing $p$ is denoted by $I^+(p,U)$ and is defined by the set of points $q \in U$ such that there is a future-directed timelike curve $\g \colon [a,b] \to U$ from $p$ to $q$. The timelike past $I^-$ is defined time-dually. If we need to emphasize the metric $g$, then we will write $I^\pm_g$.

\medskip
\begin{Def}\label{def: extension}
\emph{
Suppose $(M,g)$ is a $C^\infty$ spacetime and $(M_\ext, g_\ext)$ is a $C^k$ spacetime. Then we say that $(M_\ext, g_\ext)$ is a \emph{$C^k$ extension of $(M,g)$} if there is an isometric embedding $(M,g) \hookrightarrow (M_\ext, g_\ext)$ such that $M \subset M_\ext$ is a proper open subset. Note that we are identifying $M$ with its image under the embedding.}
\end{Def}

\medskip

\begin{Def}\label{def: future and past boundaries}
\emph{
Let $(M_\ext, g_\ext)$ be a $C^k$ extension of a $C^\infty$ spacetime $(M,g)$. The topological boundary of $M$ within $M_\ext$ is denoted by $\pd M$. A future-directed timelike curve $\g \colon [a,b] \to M_\ext$ is called a \emph{future-terminating timelike curve} for a point $p \in \pd M$ provided $\g(b) = p$ and $\g\big([a,b)\big) \subset M$. \emph{Past-terminating timelike curves} are defined time-dually. The \emph{future} and \emph{past boundaries} of $M$ within $M_\ext$ are 
\begin{align*}
\pd^+M \,&=\, \{p \in \pd M \mid \text{there is a future-terminating timelike curve for $p$}\}\,,\\
 \pd^-M \,&=\, \{p \in \pd M \mid \text{there is a past-terminating timelike curve for $p$}\}\,.
\end{align*}}
\end{Def}

\subsection{Warm-up: exact de Sitter}\label{flat de Sitter sec}

The \emph{flat de Sitter} model is a FLRW spacetime with vanishing spatial curvature and an exponentially growing scale factor. Specifically, the manifold is $M = \R \times \R^3$, and the metric is \eqref{eq:gFLRW} with scale factor $a(t) \propto \exp(\sqrt{\Lambda/3}\,t)$
for some real positive constant $\Lambda$. This is a solution to the Einstein field equations with cosmological constant $\Lambda$. From here on, we set the arbitrary proportionality constant to unity, and for this subsection we also work in units where $\Lambda=3$, such that the scale factor simply reads $a(t)=e^t$.
The time orientation is given by declaring $\pd_t$ to be future directed. 

We define \emph{conformal time}, $\eta$, such that it satisfies $\dd\eta=\dd t/a(t)$, i.e., so that a flat FLRW metric may be expressed as a conformal transformation of Minkowski spacetime, $g=a(\eta)^2(-\dd\eta^2+h_{\mathbb{E}})$. In dS, we may take $\eta(t)=-\int_t^\infty\dd\tilde t/a(\tilde t)=-e^{-t}=-1/a(t)$, and this allows us to express the metric of dS as
\begin{equation}\label{eq:gdSconf}
    g=\frac{1}{\eta^2}\left(-\dd\eta^2+\dd r^2+r^2\dd\Omega^2\right)\,.
\end{equation}
As inspired by \cite{Yoshida:2018ndv}, we then introduce new coordinates $(\l,v)$ as functions of $(t,r)$ via\footnote{By virtue of satisfying $\dd\lambda=a\,\dd t$, $\lambda$ can be interpreted as the affine parameter of null geodesics, which are characterized by $v=\mathrm{constant}$ (as well as constant angular coordinates).} 
\begin{equation}\label{l and v coord}
    \l(t) \,:=\, \int_{-\infty}^t \dd\tilde t\,a(\tilde t) \:\:\:\: \text{ and } \:\:\:\: v(t,r) \,:=\, \eta(t) + r\,,
\end{equation}
so for dS this gives
\begin{equation}\label{l and v coord expl}
    \l(t) \,=\, e^t\,=\,a(t) \:\:\:\: \text{ and } \:\:\:\: v(t,r) \,=\, -e^{-t} + r \,=\, -\frac{1}{\lambda(t)} + r\,.
\end{equation}
Since $v + 1/\lambda = r > 0$, the coordinates $(\l, v)$ are a diffeomorphism from 
\begin{equation}\label{diffeo image}
    \{(t,r) \mid t \in \R \text{ and } r > 0\} \:\: \text{ onto } \:\: U := \{(\lambda, v) \mid \lambda \in (0, \infty) \text{ and } v \in (-1/\lambda, \infty)\}\,.
\end{equation}
Using $\dd \lambda = a \, \dd t$ (equivalently $\dd t = \dd\lambda/a$) and $\dd v = \dd t/a + \dd r$ (equivalently $\dd r = \dd v - \dd\lambda/a^2$),
we have 
\begin{equation*}
    -\dd t^2 + a^2 \dd r^2 \,=\, -\frac{1}{a^2}\dd\lambda^2 + a^2\left(\dd v - \frac{1}{a^2}\dd \lambda \right)^2 \,=\, a^2 \dd v^2 - 2\,\dd\lambda\,\dd v\,.
\end{equation*}
Therefore the metric on $U \times \mathbb{S}^2$ is given by
\begin{equation}\label{metric in l and v}
    g \,=\, -2\,\dd\lambda\,\dd v + a^2 \dd v^2 + a^2r^2 \dd\Omega^2\,.
\end{equation}
In the above expression, $a$ and $r$ are functions of $\lambda$ and $v$. Explicitly, plugging (\ref{l and v coord expl}) into  (\ref{metric in l and v}) specifically gives for dS
\begin{equation}\label{metric in l and v II}
    g \,=\, -2\,\dd\lambda\,\dd v + \l^2 \dd v^2 + (\lambda v + 1)^2\dd\Omega^2\,.
\end{equation}
Note that $t = -\infty$ corresponds to $\lambda = 0$. From equation \eqref{metric in l and v II}, there is no  degeneracy at $\lambda = 0$. Therefore these coordinates can be used to extend the spacetime through $\lambda = 0$. Of course the extension is not unique without any further assumptions.\footnote{Evidently, the extension should still be a solution to whatever field equations are at play. Here we are not prescribing any field equations and keep the discussion purely geometrical, i.e., results apply for arbitrary metrical theories of gravity.}

To illustrate the existence of a $C^0$ extension, note that $U \cup \{(\l, v) \mid  \l \leq 0 \text{ and } v \in \R\}$ is a smooth manifold since $U$ is given by (\ref{diffeo image}). Therefore,
\begin{equation}\label{ext manif}
    M_\ext \,=\, M \cup \big(\{(\l, v) \mid  \l \leq 0 \text{ and } v \in \R\} \times \mathbb{S}^2\big)
\end{equation}
is a smooth manifold which contains $M$ as a proper subset. We define the extended metric via
\begin{equation}\label{metric ext}
 g_\ext \,=\,  \left\{
\begin{array}{ll}
      g & \text{ on } M \\
      -2\,\dd\l\,\dd v + \dd\Omega^2 & \text{ on } \{(\l, v) \mid \l \leq 0 \text{ and } v \in \R\} \times \mathbb{S}^2. \\
\end{array} 
\right.
\end{equation}
We choose the time orientation on $M_\ext$ such that $\pd_\l + \pd_v$ is future directed for $\l \leq 0$. This  time orientation agrees with the original one on $M$.  Then $(M_\ext, g_\ext)$ is a continuous extension of $(M,g)$ as defined in Section \ref{sec: further prelim}. Of course, other choices could be made for the metric in the extension. In particular, there exists $C^\infty$ extensions such as full (global) de Sitter spacetime.

Lastly, we note that $\l = 0$ serves as the past boundary $\pd^-M$ of $M$ for this extension. To see this, fix $v_0 \in \R$ and $\omega_0 \in \mathbb{S}^2$. Define $\g \colon [0,f(v_0)] \to M_\ext$ by
\begin{equation}\label{past boundary coord}
    \l \circ \g(s) \,=\, s \:\:\:\: \text{ and } \:\:\:\: v \circ \g(s) \,=\, v_0 + s\,,
\end{equation}
and $\g(s)$ keeps the fixed value $\omega_0$ on $\mathbb{S}^2$. Here $f(v_0)$ is given by $f(v_0) = 1$ for $v_0 \geq -1/2$ and $f(v_0) = -1/(2v_0)$ for $v_0 \leq -1/2$.
(This choice of $f(v_0)$ is mostly arbitrary --- this example is sufficient to ensure $\gamma$ remains in the manifold, but it is not necessary.)
Then $\g$ has past endpoint, $\g(0)$, at $(0, v_0, \omega_0)$. Since $U$ is given by (\ref{diffeo image}), the choice of $f$ ensures that $\g$ maps into $M_\ext$. Using (\ref{metric in l and v II}), we see that $\g$ is a timelike curve:
\begin{equation*}
    g_\ext(\g', \g')|_{\g(s)} \,=\, g_\ext(\pd_\l  + \pd_v, \pd_\lambda + \pd_v)|_{\g(s)} =\, -2 + \big(\l\circ \g(s)\big)^2=\, -2 + s^2\leq -1\,.
\end{equation*}
Conversely, it is straightforward to see that if $p$ is any point on $\pd^-M$, then $\l(p) = 0$. Hence $\pd^-M$ coincides with the $\l = 0$ hypersurface.

\subsection{Conditions for continuous extendibility}\label{sec:C0}

Now we consider spacetimes that in some sense asymptotically approach flat de Sitter towards the past. The manifold is given by $M = (-\infty, t_{\rm max}) \times \R^3$ for some $t_{\rm max} \in \R$. The metric is flat FLRW as before,
but now the scale factor $a \colon (-\infty, t_{\rm max}) \to (0,\infty)$ is either assumed to satisfy
\begin{equation}
    a(t) \,\stackrel{t\to-\infty}{=}\, e^{h t} + o(e^{h t})\label{eq:aAsymExp}
\end{equation}
or
\begin{equation}
    H(t):=\frac{\dot a(t)}{a(t)}\stackrel{t\to-\infty}{\to}h\label{eq:HAsymConst}
\end{equation}
for some $h\in\mathbb{R}_{>0}$.
(By definition, the former assumption means $\big(a(t) -e^{ht}\big)/e^{ht} \to 0$ as $t \to -\infty$.)
We assume $a(t)$ is smooth so that $(M,g)$ is a smooth spacetime.
Again, the time orientation is given by declaring $\pd_t$ to be future directed.

\medskip
\medskip

\noindent\emph{Remarks.} The positive finite number $h$ plays the role of $H_\Lambda$ in Definition \ref{def:padS}. Note that while the scale factor in \eqref{eq:aAsymExp} is past-asymptotically exponential (so it has the same form as exact dS in the asymptotic past), it does not necessarily qualify as \emph{past-asymptotically de Sitter} as per our Definition \ref{def:padS}. Indeed, according to Lemma \ref{lem:Hto1dotHto0}, we would need knowledge of the first and second derivatives of $a(t)$ to properly call the scale factor past-asymptotically dS. Likewise, $H\to h>0$ is not past-asymptotically dS without proper knowledge of $\dot H$. However, as it will soon become clear, knowledge of the derivatives of $a(t)$ or $H(t)$ is not necessary to derive $C^0$ extendibility results. Still, we will see that the $C^0$-extendibility result does apply to past-asymptotically dS spacetimes.

Let us mention that all the results in this subsection apply without knowledge of the sign of $\ddot a$ [and even of $\dot a$ if the assumption is\footnote{If the assumption is \eqref{eq:HAsymConst}, then we necessarily know that $\dot a\searrow 0$ as $t\to-\infty$, so the universe has to be expanding in a neighborhood of $t=-\infty$. However, this still holds regardless of $\ddot a$, so it may or may not be past-asymptotically inflationary.} \eqref{eq:aAsymExp}]. Therefore, the results can apply to universes that are not necessarily past-asymptotically inflationary --- one could have examples that are cyclic (i.e., undergoing a series of bounces and turnarounds) and only inflationary in an `averaged' sense (e.g., \cite{Ijjas:2019pyf,Kinney:2021imp}). If the universe is past-asymtotically inflationary (as per Classification \ref{itm:case2b}), then the results certainly apply.

\medskip
\medskip

Fix any $t_\mathrm{L} \in (-\infty, t_{\rm max})$; its choice does not matter for what follows. We take our definition of conformal time to be 
\begin{equation}\label{def: conformal time}
    \eta(t)=-\int_t^{t_\mathrm{L}}\frac{\dd\tilde t}{a(\tilde t)}\,.
\end{equation}
Again, we introduce new coordinates $(\l,v)$ as functions of $(t,r)$ via \eqref{l and v coord}. (The asymptotic assumption on $a$ implies that the integral for $\l$ converges.)
Note that $\l$ maps $(-\infty, t_{\rm max})$ onto $(0, \l_{\rm max})$ where $\l_{\rm max} = \int_{-\infty}^{t_{\max}}\dd t\,a(t)$. Since $a$ is positive, the mapping $t\mapsto\lambda$ is invertible. From this it follows that the coordinates $(\l, v)$ are a diffeomorphism from $\{(t,r) \mid t \in (-\infty, t_{\rm max}) \text{ and } r > 0\}$ onto its image. Unlike (\ref{diffeo image}), there is not an easily deduced closed form for the image, however we know that it lies in $\l > 0$. 

\medskip
\medskip
\noindent\emph{Notation remark.} Since $t$ and $\lambda$ are invertible, we can view $\eta$ as a function of $\lambda$ via $\eta \circ t(\lambda)$; we will abuse notation and call this $\eta(\lambda)$. Context can be used to distinguish between $\eta(t)$ and $\eta(\lambda)$. Likewise, if we write $a(\lambda)$, then we mean $a\circ t(\lambda)$. Again, context can be used to distinguish between the two.
\medskip
\medskip

Note that $\eta(\lambda)$ is a strictly increasing function defined on $(0, \lambda_{\rm max})$ and is a diffeomorphism onto its image. Since $v-\eta=r>0$, the coordinates $(\lambda, v)$ are a diffeomorphism from $(-\infty, t_{\rm max}) \times (0,\infty)$ onto
\begin{equation}
 U := \big\{(\lambda, v) \mid \lambda \in (0, \lambda_{\rm max}) \text{ and } v \in \big(\eta(\lambda), \infty\big)\big\}.
\end{equation}
In analogy to \eqref{ext manif} in exact dS, we then wish to choose the extended manifold to be $M_\ext = M \cup (\{(\l, v) \mid \l \leq 0 \text{ and } v \in \R\} \times \mathbb{S}^2)$.
The following lemma ensures that if we attach the lower half plane $\lambda \leq 0$ to $U$, then the resulting space is indeed a manifold.

\medskip

\begin{lem}\label{ext manif lem}
Let $U$ denote the image of the $(\l,v)$ coordinates. For each $v \in \R$, there is a $\lambda > 0$ such that $(0, \lambda) \times (v-\lambda, \infty) \subset U$. Consequently, $U \cup \{(\l, v) \mid  \l \leq 0 \text{ and } v \in \R\}$ is a smooth manifold.
\end{lem}

\proof
We showed that $\eta(t) \to -\infty$ as $t \to -\infty$, so $\eta(\lambda) \to -\infty$ as $\lambda \to 0^+$.  Fix $v \in \R$. There is a $\lambda > 0$ such that $\eta(\lambda) < v -\lambda$. Since $\eta$ is an increasing function, it follows that the vertical line $\{(\l', v-\lambda) \mid \l' \in (0, \lambda) \}$ is contained in $U$. Likewise, for any $v' > v-\lambda$, we also have $v' > \eta(\lambda)$ and so the same vertical line based at $v'$ is also contained in $U$.
\qed

\medskip
\medskip

The next lemma will guarantee that the extended metric is not degenerate on the boundary of the extension.

\medskip

\begin{lem}\label{lem: aeta limit}
Suppose $a(t)$ satisfies \eqref{eq:aAsymExp} or \eqref{eq:HAsymConst}. Then $a(t)\eta(t) \to -1/h$ as $t \to -\infty$.
\end{lem}
\proof
Let us start with the assumption \eqref{eq:aAsymExp}.
Define $\Theta(t) := 1/a(t) - e^{-ht}$. Claim 1: $\Theta(t) = o(e^{-ht})$ as $t \to -\infty$. Indeed,
\[
    \frac{\Theta(t)}{e^{-ht}} \,=\, \frac{1/a(t) - e^{-ht}}{e^{-ht}} \,=\, \frac{e^{ht}}{a(t)} - 1 \:\:\to\:\: 0 \quad \text{ as } \quad t \:\:\to\:\: -\infty\,;
\]
the limit follows since the asymptotic assumption on the scale factor implies $a(t)/e^{ht} \to 1$ as $t \to -\infty$. This proves claim 1.

Claim 2: $e^{ht}\int_t^{t_\mathrm{L}}\dd \tilde{t}\,\Theta(\tilde{t}) \to 0$ as $t \to -\infty$. Fix $\e > 0$. By claim 1, there is a $\delta > 0$ such that $t < \delta$ implies 
\begin{equation}\label{eq: Theta bound}
    -\e e^{-ht} \,<\, \Theta(t) \,<\, \e e^{-ht}\,.
\end{equation}
To prove claim 2, it suffices to show $e^{ht}\int_t^{\delta}\dd \tilde{t}\,\Theta(\tilde{t}) \to 0$ as $t \to -\infty$. To prove this, integrate \eqref{eq: Theta bound} from $t$ to $\delta$:
\[
    -\frac{\e}{h}\big(-e^{-h\delta} + e^{-ht} \big) \,<\, \int_t^{\delta} \dd \tilde{t}\,\Theta(\tilde{t}) \,<\, \frac{\e}{h}\big(-e^{-h\delta} + e^{-ht} \big)\,.
\]
Multiplying by $e^{ht}$, we have
\[
    -\frac{\e}{h} \,<\, -\frac{\e}{h}\big(-e^{h(t-\delta)} + 1 \big) \,<\, e^{ht} \int_t^{\delta} \dd \tilde{t}\,\Theta(\tilde{t}) \,<\, \frac{\e}{h}\big(-e^{h(t-\delta)} + 1 \big) \,<\, \frac{\e}{h}\,.
\]
This proves claim 2. 

Next, we prove the lemma for the case of \eqref{eq:aAsymExp}. Since $1/a(t) = e^{-ht} + \Theta(t)$, we have
\[
    \eta(t) \,=\, -\int_t^{t_\mathrm{L}}\dd \tilde{t}\,\big(e^{-h\tilde{t}} + \Theta(\tilde{t}) \big) \,=\, -\frac{e^{-ht}}{h}\left(1-e^{h(t-t_\mathrm{L})} + e^{ht}\int_t^{t_\mathrm{L}}\dd\tilde{t}\,\Theta(\tilde{t})\right).
\]
Define $F(t) := a(t)/e^{ht} - 1$, so that $a(t) = e^{ht}\big(1 + F(t)\big).$ The asymptotic assumption on $a(t)$ implies $F(t) \to 0$ as $t \to -\infty$. Taking the product of $a(t)\eta(t)$, we find
\[
    a(t)\eta(t) \,=\, -\frac{1 + F(t)}{h}\left(1-e^{h(t-t_\mathrm{L})} + e^{ht}\int_t^{t_\mathrm{L}}\dd\tilde{t}\,\Theta(\tilde{t})\right).
\]
The lemma now follows from claim 2 and the fact that $F(t) \to 0$ as $t \to -\infty$.

If we start with the assumption \eqref{eq:HAsymConst} instead of \eqref{eq:aAsymExp}, then $a(t) \eta(t) \to -1/h$ follows from applying L'H{\^o}pital's rule to $\frac{\eta(t)}{1/a(t)}$,
\[
    \lim_{t\to-\infty}\frac{\eta(t)}{1/a(t)}=\lim_{t\to-\infty}\frac{\dot\eta}{-\dot a/a^2}=-\lim_{t\to-\infty}\frac{1}{H(t)}=-\frac{1}{h}\,,
\]
where we recall conformal time satisfies $\dd\eta=a^{-1}\dd t$.
\qed

\medskip
\medskip

\begin{thm}\label{cont ext thm}
Let $(M,g)$ be the smooth spacetime given by 
\[
    M \,=\, (-\infty, t_{\rm max}) \times \R^3 \quad \text{ and } \quad g\,=\,-\dd t^2 + a^2(t)h_{\mathbb{E}}\,,
\]
time oriented by declaring $\pd_t$ to be future directed. If the scale factor satisfies 
\[
    a(t) \,\stackrel{t\to-\infty}{=}\, e^{ht} + o(e^{ht}) \qquad \text{ for some } \:\: h \in\mathbb{R}_{>0}\,,
\]
or if it satisfies
\[
    H(t):=\frac{\dot a(t)}{a(t)}\stackrel{t\to-\infty}{\to}h \qquad \text{ for some } \:\: h \in\mathbb{R}_{>0}\,,
\]
then a continuous extension $(M_\ext, g_\ext)$ of $(M,g)$ is given by 
\[
    M_\ext \,=\, M \cup M_{\lambda\leq 0}\,,\qquad M_{\lambda\leq 0}:=\{(\l, v) \mid \l \leq 0 \text{ and } v \in \R\} \times \mathbb{S}^2
\]
and 
\[
    g_\ext \,=\,  \left\{
    \begin{array}{ll}
      -2\,\dd\lambda\,\dd v + a(\lambda)^2 \dd v^2 + a(\lambda)^2(v-\eta(\lambda))^2 \dd\Omega^2 & \text{ on } M \\
      -2\,\dd \l\,\dd v + h^{-2}\dd\Omega^2 & \text{ on } M_{\lambda\leq 0}\,. \\
    \end{array} 
    \right.
\]
The $(\l,v)$ coordinates are defined in \eqref{l and v coord}, and $\eta$ is the conformal time given by \eqref{def: conformal time}. The time orientation on $(M_\ext, g_\ext)$ is determined by declaring $\pd_\l + \pd_v$ to be future directed, which agrees with the original time orientation on $M$.  The past boundary $\pd^-M$ is given by the $\l  = 0$ hypersurface which is topologically $\R \times \mathbb{S}^2$.    
\end{thm}

\proof
The same computations leading to \eqref{metric in l and v} in the previous subsection still apply, hence the metric in $(\l, v, \theta, \phi)$ coordinates is again given by \eqref{metric in l and v}.
Since $a(t(\lambda)) \searrow 0$ as $\lambda \to 0^+$, the worrisome term in the metric is $a^2r^2$, where $a(\lambda)r(\l,v)=a(\lambda)\big(v-\eta(\lambda)\big)$.
We want to show that this expression limits to a finite nonzero number as $\lambda \to 0^+$ in order to achieve nondegeneracy of the metric. Since $\lambda \to 0^+$ if and only if $t \to -\infty$, this follows from Lemma \ref{lem: aeta limit}.

Now we show how to obtain a continuous extension. We choose the extended manifold to be $M_\ext = M \cup M_{\l \leq 0}$ --- Lemma \ref{ext manif lem} ensures that this is indeed a manifold. Again in analogy to \eqref{metric ext} in exact dS, the metric $g_\ext$ on $M_\ext$ is taken to be $g$ on $M$ and\footnote{Note that the metric in the extension here is the same as in \eqref{metric ext}, where we had set $h=1$.} $-2\,\dd\l\,\dd v + h^{-2}\dd\Omega^2$ on $\{(\l, v) \mid \l \leq 0\text{ and } v \in \R\} \times \mathbb{S}^2$. Thus $(M_\ext, g_\ext)$ gives the desired continuous extension.

Furthermore, as in exact dS, $\l = 0$ coincides with the past boundary $\pd^-M$ for this extension. One can show this with a similar curve $\g \colon [0, f(v_0)] \to M_\ext$ as defined in \eqref{past boundary coord}, but one has to be a little more careful choosing the function $f(v_0)$; the existence of some $f$ is guaranteed by Lemma \ref{ext manif lem} since we can draw a tiny timelike curve from $v_0$ on $\l = 0$. It will be timelike for a small enough curve since so long as $a(\lambda) < \sqrt{2}$.
\qed

\medskip
\medskip

\begin{cor}\label{cor:padSC0}
	A flat past-asymptotically de Sitter spacetime is continuously extendible.
\end{cor}

\proof
Past-asymptotically dS spacetimes satisfy $H\to h>0$ and $\dot H\to 0$ as $t\to-\infty$, so Theorem \ref{cont ext thm} applies.
\qed

\medskip
\medskip

\begin{figure}
    \centering
    \includegraphics[scale=0.30]{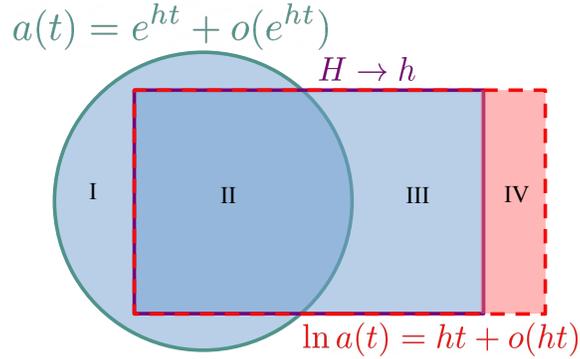}
    \caption{{\small{Venn-like diagram of classes of functions [either $a(t)$ or $H(t)$] characterized by their asymptotic behavior as $t\to -\infty$. We see that $a(t)=e^{ht}+o(e^{ht})$ (green circle) is not equivalent to $H\to h$ (purple rectangle; neither one implies the other without further assumptions), but there is some overlap (called region II in the figure). The condition $H\to h$ implies $\ln(a)=ht+o(ht)$ (dashed red rectangle), hence the purple rectangle lies within the red one, but the converse is not true, leaving region IV. Functions that fall inside the green circle and the purple rectangle (so regions I, II, and III) are $C^0$ extendible. We remain agnostic about functions that fall in region IV. Examples are presented in the text.}}}
    \label{fig:Venn}
\end{figure}

\noindent\textit{Examples.}
Let us consider the following scale factor first:
\begin{equation}
    a(t)=e^t+\sin^2(e^{-3t})e^{2t}\stackrel{t\to-\infty}{=}e^t+o(e^t)\,.\label{eq:C0ButSingularExample}
\end{equation}
Since it respects \eqref{eq:aAsymExp},
the resulting spacetime must be $C^0$ extendible by Theorem \ref{cont ext thm}. However, note that $H=\dot a/a$ does not have a limit as $t\to -\infty$; hence it does not respect \eqref{eq:HAsymConst}. Therefore, this is an example of $C^0$-extendible spacetime that is not past-asymptotically dS.
This falls in region I of the diagram presented in Figure \ref{fig:Venn}.
Conversely, let us consider
\begin{equation}
    a(t)=(1+t^2)e^t\stackrel{t\to-\infty}{\neq}e^t+o(e^t)\,.\nonumber
\end{equation}
This does not satisfy \eqref{eq:aAsymExp},
but a simple calculation shows $H\to 1$ (and $\dot H\to 0$) as $t\to-\infty$, so it does satisfy \eqref{eq:HAsymConst}. Therefore, it still qualifies for $C^0$ extendibility by Theorem \ref{cont ext thm}. This falls in region III of Figure \ref{fig:Venn}.

These examples highlight the fact that conditions on $a(t)$ alone and on $H(t)$ [which involves $\dot a(t)$] are not always equivalent. Of course, in many instances they are equivalent; a simple example could be $a(t)=e^t+e^{2t}$, which falls in the overlapping region II of Figure \ref{fig:Venn}.

There exists a straightforward relation between $H\to h$ and a condition on $a(t)$ if we express the latter as a logarithm. Specifically, we note that $H\to h$ as $t\to-\infty$ implies $\ln a(t)\stackrel{t\to-\infty}{=}ht+o(ht)$. Indeed, using L'H\^opital's rule
\[
	\lim_{t\to-\infty}\frac{\ln a(t)}{ht}=\lim_{t\to-\infty}\frac{H(t)}{h}=1\,.
\]
However, the converse is not true in general.
A counter-example could be $\ln a(t)=t+\sin(t^2)\sqrt{-t}\stackrel{t\to-\infty}{=}t+o(t)$, $t<0$, for which one can verify that $H(t)$ does not have a limit as $t\to-\infty$. Thus this example falls in region IV of Figure \ref{fig:Venn}.

\medskip
\medskip

\noindent\textit{Remarks.}
Note that, contrary to regions I, II, and III of that diagram, we do not know whether functions falling in region IV admit $C^0$ extensions. For the example $\ln a(t)=t+\sin(t^2)\sqrt{-t}$, one can numerically integrate $1/a(t)$ to show that $a\eta$ does not converge as $t\to-\infty$. Therefore, this suggests functions in region IV are generally not extendible within the $(\lambda,v)$ coordinates of this section. Let us stress again that this does not mean general inextendibility since our statements are restricted to a particular coordinate system. In fact, this holds for any functions that do not fit in the classes of functions of Figure \ref{fig:Venn}. Furthermore, even within $(\lambda,v)$ coordinates, we do not know whether regions I, II, and III of Figure \ref{fig:Venn} exhaust all possibilities allowing $C^0$ extensions.

Let us end this subsection with an important observation. As is known but perhaps not always appreciated, the concepts of spacetime extendibility and singularities are not equivalent. We have encountered such an example above: indeed, the scale factor \eqref{eq:C0ButSingularExample} allows a $C^0$ extension of the spacetime, but the divergence of the Hubble parameter in this case implies the presence of a scalar curvature singularity at $t=-\infty$ [e.g., the Ricci and Kretschmann scalar curvature invariants diverge; recall \eqref{eq:lqci}]. We thus stress that while past-asymptotically dS spacetimes are expected to be always free of scalar curvature singularities and while they are always $C^0$ extendible, there certainly are $C^0$-extendible examples that do not fit in that class and that might still be singular.

\subsection{Gaining higher regularity}\label{sec:Ckfull}

We have so far found an asymptotic condition on $a(t)$ or on $H(t)$ that ensures $C^0$ metric extendibility. Naturally, we then wish to ask what asymptotic conditions are needed to find more regular extensions, e.g., an extension that is continuously differentiable once ($C^1$), twice ($C^2$), or more. We address these questions in the following subsections.

\subsubsection{Continuously differentiable once and twice}

From the metric $g = -2\,\dd\lambda\,\dd v + a^2 \dd v^2 + a^2r^2 \dd\Omega^2$, we see that we have to control $a$ and $ar$ (as functions of $\lambda$ and $v$) to gain higher regularity. The $v$-derivatives are well-behaved at $\l = 0$ since $\frac{\pd}{\pd v}a(t(\lambda))=0$ and, recalling $r(\lambda,v)=v-\eta(t(\lambda))$,
\[
    \frac{\pd}{\pd v}\Big(a(t(\lambda))r(\lambda, v)\Big) \,=\, a(t(\lambda)) \stackrel{\lambda\to 0^+}{\longrightarrow}0^+\,.
\]
Also recalling $\dd\lambda=a\,\dd t$, $\dd\eta=\dd t/a$, and $H=\dot a/a$, we have
\begin{equation}\label{eq:dadlambda}
    \frac{\partial}{\partial\lambda}a(t(\lambda))=\frac{\dd a}{\dd t}\frac{\dd t}{\dd\lambda}=\frac{\dot a}{a}=H
\end{equation}
and
\begin{equation}\label{eq:dardlambda}
    \frac{\pd}{\pd \lambda}\Big(a(t(\lambda))r(\lambda, v)\Big) \,=\, \frac{\pd}{\pd \lambda}\Big(a(v-\eta)\Big)=H(v-\eta)-a\frac{\dd\eta}{\dd t}\frac{\dd t}{\dd\lambda}=H(v-\eta)-\frac{1}{a}\,.
\end{equation}
The above calculation shows that the following limits are desirable:
\begin{equation}\label{C1 condition}
    H(t) \to h \:\:\:\: \text{ and } \:\:\:\: \frac{1}{a(t)} + H(t)\eta(t) \to c \in \R \qquad \text{ as } \:\: t \to -\infty\,,
\end{equation}
for some real number $c$. Recall that taking the limit as $t \to -\infty$ is equivalent to taking the limit as $\lambda \to 0^+$.

\medskip
\medskip

\begin{lem}\label{C1 lem}
Suppose both $a(t)=e^{ht} + o\big(e^{(2+\e)ht} \big)$ and $\dot{a}(t)=he^{ht} + o\big(e^{(2+\e)ht} \big)$ as $t \to -\infty$ for some $h,\e \in \R_{>0}$. Then \emph{(\ref{C1 condition})} holds.
\end{lem}

\proof
Choose $t_\mathrm{L1} < 0$ such that $t < t_\mathrm{L1}$ implies $he^{ht}-e^{(2+\e)ht}>0$, $|a(t) - e^{ht}| < e^{(2+\e)ht}$, and  $|\dot{a}(t) - he^{ht}| < e^{(2+\e)ht}$. Then
\[
    0<he^{ht}-e^{(2+\e)ht}<\dot a(t)<he^{ht}+e^{(2+\e)ht}
\]
and
\begin{equation}\label{1/a C^1}
    \frac{1}{e^{ht} + e^{(2+\e)ht}} \,<\, \frac{1}{a(t)} \,<\, \frac{1}{e^{ht} - e^{(2+\e)ht}}\,,
\end{equation}
so $H(t) = \dot{a}(t)/a(t) \to h$ as $t \to -\infty$ follows via the squeeze theorem.

Now we show $1/a(t) + H(t) \eta(t)$ approaches some real number $c$. Note that part of the problem is deducing the value of $c$. 

Set $\Theta(t) := 1/a(t) -e^{-ht}$. Claim:  $\Theta(t) = o(e^{\e ht/2})$ as $t\to-\infty$. 
To prove the claim, note that \eqref{1/a C^1} implies  $\frac{1}{a(t)} < \frac{1}{e^{ht} - e^{(2+\e)ht}} = \frac{e^{-ht}}{1-e^{(1+\e)ht}}= e^{-ht}\left(1 + e^{(1+\e)ht} + e^{2(1+\e)ht} + \dotsb \right)$. Hence 
\[
    \frac{1/a(t) - e^{-ht}}{e^{\e h t/2}} \,<\, e^{\e ht/2} + e^{(1 + 3\e/2)ht} + \dotsb\,.
\]
The right hand side limits to zero. Likewise, the reverse inequality is similar. Therefore the claim follows by the squeeze theorem.

By the claim, there is a $t_\mathrm{L2}$ such that $t < t_\mathrm{L2}$ implies $|\Theta(t)| <  e^{h\e t/2}$. Let $\Theta^\pm(t) := \max\{0, \pm \Theta(t)\}$ denote the positive and negative parts of $\Theta(t)$, so that $\Theta(t) = \Theta^+(t) - \Theta^-(t)$. Then $|\Theta(t)| < e^{h\e t/2}$ implies both $\int_{-\infty}^{t_\mathrm{L2}}\dd t\,\Theta^+(t)$ and $\int_{-\infty}^{t_\mathrm{L2}}\dd t\,\Theta^-(t)$ converge to real numbers. Therefore $\int_{-\infty}^{t_\mathrm{L2}}\dd t\,\Theta(t)$ converges to a real number as well  and hence so does $\int_{-\infty}^{t_\mathrm{L}}\dd t\,\Theta(t)$. Then
\[
    \eta(t) = -\int_t^{t_\mathrm{L}}\dd\tilde t\,a(\tilde t)^{-1} = h^{-1}e^{-ht_\mathrm{L}} -h^{-1} e^{-ht} - \int_t^{t_\mathrm{L}}\dd\tilde t\,\Theta(\tilde t)\,.
\]
Therefore for $t < t_\mathrm{L1}$, we have
\[
    \frac{1}{a(t)} +H(t)\eta(t) \,<\, \frac{e^{-ht}}{1 - e^{(1+\e)ht}} + H(t)\left(\frac{e^{-ht_\mathrm{L}}}{h} -  \frac{e^{-ht}}{h} - \int_t^{t_\mathrm{L}}\dd\tilde t\,\Theta(\tilde t)\right)\,.
\]
The right hand side limits to the finite number
\[
    c:= e^{-ht_\mathrm{L}} - \int_{-\infty}^{t_\mathrm{L}}\dd  t\,\Theta( t)
\]
as $t\to-\infty$. Indeed, a quick calculation shows that the hypotheses on $a(t)$ and $\dot{a}(t)$ imply $\frac{e^{-ht}}{1-e^{(1+\e)ht}} -H(t)\frac{e^{-ht}}{h}$ limits to $0$ as $t \to -\infty$. The reverse inequality is similar. Thus $1/a(t) + H(t) \eta(t)$ limits to $c$ by the squeeze theorem.
\qed

\medskip
\medskip

\noindent\emph{Remarks.}
\begin{itemize}
\item[-] It is not hard to see that Lemma \ref{C1 lem} still holds under the weaker set of assumptions ``$a(t) \stackrel{t\to-\infty}{=} e^{ht} + o\big(e^{(2+\e)ht}\big)$ and $\dot{a}(t) \stackrel{t\to-\infty}{=} he^{ht} + o\big(e^{(1+\e)ht}\big)$.'' However, if $a(t) \stackrel{t\to-\infty}{=} e^{ht} + o\big(e^{(2+\e)ht}\big)$ and $\lim_{t \to -\infty}\frac{\dot{a}(t) - he^{ht}}{e^{(2+\e)ht}}$ exists or diverges to $\pm \infty$, then an application of L'H{\^ o}pital's rule shows that the limit must be zero, and hence we must have had $\dot{a}(t) \stackrel{t\to-\infty}{=} he^{ht} + o\big(e^{(2+\e)ht}\big)$ to begin with. Therefore the only time the weaker set of assumptions holds but $\dot{a}(t) \neq he^{ht} + o\big(e^{(2+\e)ht}\big)$ as $t\to-\infty$ is when $\lim_{t \to -\infty} \frac{\dot{a}(t) -he^{ht}}{e^{(1+\e)ht}} = 0$, $\lim_{t \to -\infty} \frac{\dot{a}(t) - he^{ht}}{e^{(2+\e)ht}}$ does not exist, and $\lim_{t \to -\infty}\frac{a(t) - e^{ht}}{e^{(2+\e)ht}} = 0$ --- a situation which seems of scant importance.

\item[-] In fact, using L'H{\^ o}pital's rule again, the hypothesis of Lemma \ref{C1 lem} is equivalent to the set of assumptions ``$a(t) \searrow 0$ as $t \to -\infty$ and $\dot{a}(t) \stackrel{t\to-\infty}{=} he^{ht} + o\big(e^{(2 + \e)ht}\big)$.''

\item[-] If $a(t) = e^{t} + e^{2t}$
or
\[
    a(t)=\frac{e^t}{1+e^t/t}\stackrel{t\to-\infty}{=}e^t-\frac{e^{2t}}{t}+\ldots\,,
\]
then direct evaluation shows that \eqref{C1 condition} does not hold. The latter is a counter-example showing that Lemma \ref{C1 lem} would not hold if the assumption were ``$a(t) \stackrel{t\to-\infty}{=} e^{ht} + o(e^{2ht})$ and $\dot{a}(t) \stackrel{t\to-\infty}{=} he^{ht} + o(e^{2ht})$.'' Hence taking $\e = 0$ in the hypothesis of Lemma \ref{C1 lem} is not sufficient to achieve its conclusion.
\end{itemize}

\medskip
\medskip

\begin{thm}\label{higher reg thm}
Let $(M,g)$ be the smooth spacetime given by 
\[
    M \,=\, (-\infty, t_{\rm max}) \times \R^3 \quad \text{ and } \quad g\,=\,-\dd t^2 + a^2(t)h_{\mathbb{E}}\,,
\]
time oriented by declaring $\pd_t$ to be future directed.
Let $(\l,v)$ denote the coordinates given by \emph{\eqref{l and v coord}}.
\begin{itemize}
\item[\emph{(a)}] Suppose $a(t)=e^{ht} + o\big(e^{(2+\e)ht} \big)$ and $\dot{a}(t)=he^{ht} + o\big(e^{(2+\e)ht} \big)$ as $t \to -\infty$ for some $\e > 0$. Then $(M,g)$ admits a $C^1$ extension through the $\l = 0$ hypersurface.

\item[\emph{(b)}] Suppose the assumptions from part \emph{(a)} and suppose $\lim_{t \to -\infty}\dot{H}(t)/a(t)^2$ exists within $\R$, where $H(t) = \dot{a}(t)/a(t)$. Then $(M,g)$ admits a $C^2$ extension through the $\l = 0$ hypersurface.
\end{itemize}
\end{thm}

\medskip
\medskip

\noindent\emph{Remark.} By ``admits a $C^k$ extension through the $\l = 0$ hypersurface'' we mean there is a neighborhood $V$ of the $\l = 0$ hypersurface, $\{0\} \times \R \times \mathbb{S}^2$, and a $C^k$ Lorentzian metric $g_\ext$ on $V$ such that $g_\ext = g$ on $M \cap V$, and a time orientation on $V$ which agrees with the one on $M \cap V$. Consequently, if $M_\ext = M \cup V$, then $(M_\ext, g_\ext)$ is a $C^k$ extension of $(M,g)$. 
Note that we are certainly not constructing any type of maximal extension. Moreover, the extended metric $g_\ext$ cannot be the same as the one introduced in Theorem \ref{cont ext thm}. To construct the desired $C^k$ extension, it suffices to show that all the partial derivatives (up to order $k$) of the metric components have continuous limits to the $\lambda = 0$ hypersurface. Then the existence of a sufficiently differentiable metric extension follows from  \cite{WhitneyAE}; see also \cite{Seeley}. However, since the metric components are well behaved with respect to $v$-derivatives, it is easy to construct a $C^k$ extension simply by creating a Taylor series in $\lambda$ with coefficients given by the partial derivatives at the past boundary. This is demonstrated for the $C^1$ case in the proof below.

\medskip
\medskip

\proof\:
\begin{itemize}
\item[(a)] Within $M$ the metric takes the form $g = -2\,\dd\lambda\,\dd v + a^2 \dd v^2 + (ar)^2 \dd\Omega^2$ in the $(\l,v)$ coordinates. In this expression, $a$ and $r$ are functions of $\l$ and $v$. Using \eqref{eq:dardlambda}, we have $\pd_\lambda(ar)^2 = 2a(v-\eta)(H(v-\eta)-1/a)$. Since the condition for $C^0$ extendibility is met, we have $a\eta \to -1/h$ as $\lambda \to 0^+$. It thus follows from Lemma \ref{C1 lem} that $\pd_\l(ar)^2$ extends continuously to the $\l = 0$ hypersurface. In fact, its continuous extension is given by $2(v-c/h)$, where $c$ is given by \eqref{C1 condition}. The other partial derivatives, $\pd_v(ar)^2$, $\pd_\l a^2$, $\pd_v a^2$, all limit to zero. Since these derivatives have continuous limits to the $\l = 0$ hypersurface, it follows that $a^2$ and $(ar)^2$ admit $C^1$ extensions on the lower half-plane $\lambda \leq 0$.  Explicitly, a $C^1$ spacetime extension can be defined via 
\[
    g_\ext \,=\, -2\,\dd \lambda\,\dd v + \frac{1}{h^2}\big(1 + 2h(hv-c)\lambda\big)\dd\Omega^2
\]
for $\l \in (-\delta, 0]$. By choosing $\delta > 0$ small enough, we can ensure that the metric remains Lorentzian. Moreover, we can ensure that $\pd_\l +\pd_v$ is timelike throughout $\l \in (-\delta,0]$. Therefore, we time-orient the extension by declaring $\pd_\l + \pd_v$ to be future directed, which agrees with the original time orientation on $M$.

\item[(b)] Under the assumptions of part (a), it is easy to see that each of the second order partial derivatives of all the metric components limit to a finite value on approach to the $\l = 0$ hypersurface except for $\pd_\l^2 a^2$ and $\pd_\l^2(ar)^2$. It is these terms where we need the extra assumption on $\dot{H}/a^2$.
Recalling \eqref{eq:dadlambda}, we find
\begin{equation}\label{eq:d2adlambda2dHdlambda}
    \frac{\partial^2a}{\partial\lambda^2}=\frac{\partial H}{\partial\lambda}=\frac{\dd H}{\dd t}\frac{\dd t}{\dd\lambda}=\frac{\dot H}{a}\,,
\end{equation}
and so we have $\pd_\l^2 a^2 = 2(\dot{H} + H^2)$. Since $H^2 \to h^2$ and $\dot{H} \to 0$ under the extra assumption, $\pd_\l^2 a^2$ limits to $2h^2$ on approach to the $\l = 0$ hypersurface. Using the calculation of $\pd_\l(ar)^2$ from part (a), as well as \eqref{eq:dardlambda} and \eqref{eq:d2adlambda2dHdlambda}, we have
\begin{align*}
    \pd_\l^2(ar)^2 \,&=\, \pd_\l \big(2a(v-\eta)(vH - H\eta -1/a)\big)
    \\
    &=\,2(vH-H\eta -1/a)\pd_\l\big(a(v-\eta)\big) + 2a(v-\eta)\pd_\l(vH - H\eta - 1/a)
    \\
    &=\, 2(vH - H\eta -1/a)^2 + 2\dot{H}(v-\eta)^2\,.
\end{align*}
The first term above limits to a finite value by Lemma \ref{C1 lem}. For the second term, we have $\dot{H}(v-\eta)^2 = \frac{\dot{H}}{a^2}(av - a\eta)^2$. Since $a\searrow 0$, $a\eta \to -1/h$, and $\dot{H}/a^2$ limits to a finite value by assumption, it follows that $\pd_\l^2(ar)^2$ limits to a finite value on approach to the $\l = 0$ hypersurface. The $C^2$ extension can be obtained in the same way as in part (a).
  \qed
\end{itemize}

\medskip
\medskip

\begin{cor}\label{C2 cor}
Suppose that for some numbers $h \in \R_{>0}$ and $b \in \R$, we have
\begin{align*}
a(t) \,&=\, e^{ht} + be^{3ht} + o(e^{3ht})\,,
\\
\dot{a}(t) \,&=\, he^{ht} + 3hbe^{3ht} + o(e^{3ht})\,,
\\
\ddot{a}(t) \,&=\, h^2e^{ht} + 9h^2be^{3ht} + o(e^{3ht})\,.
\end{align*}
Then $(M,g)$ admits a $C^2$ extension through the $\l =0$ hypersurface.
\end{cor}

\proof
A straightforward limit calculation shows that $\dot{H}(t)/a^2(t) \to 4bh^2$ as $t \to -\infty$. 
\qed

\medskip
\medskip

\noindent\emph{Remarks.}
In the previous corollary, note that $a(t)$, $\dot a(t)$, and $\ddot a(t)$ respect the conditions of Lemma \ref{lem:Hto1dotHto0}, hence the spacetime in that example may be called past-asymptotically de Sitter.

Note also that we may equivalently rewrite $a(t)$, $\dot a(t)$, and $\ddot a(t)$ in the previous corollary as $\sim e^{ht}+O(e^{3ht})$ as $t\to-\infty$. This asymptotic condition matches the one already found in \cite{Nomura:2021lzz}. Furthermore, the existence and finiteness of $\lim_{t\to-\infty}\dot H/a^2$ as a condition for spacetime extendibility matches the results \cite{Yoshida:2018ndv,Nomura:2021lzz}.

\subsubsection{Continuously differentiable more than twice (up to smoothness)}\label{sec:Ck}

Let us start by defining some new notation: for any function $\mathcal{F}(\lambda):\mathbb{R}_{>0}\to\mathbb{R}$ and any integer $k\in\mathbb{Z}_{\geq 0}$, let us use the short-hand notation
\begin{equation*}
	l_\lambda^k[\mathcal{F}]:=\lim_{\lambda\to 0^+}\frac{\dd^k\mathcal{F}(\lambda)}{\dd\lambda^k}\,.
\end{equation*}
If $\lim_{\lambda\to 0^+}\frac{\dd^k\mathcal{F}(\lambda)}{\dd\lambda^k}$ exists and is finite, then $l_\lambda^k[\mathcal{F}]$ is a functional (which simply yields a finite real number) and $\mathcal{F}(\lambda)$ is of differentiability class $C^k$ as $\lambda\to 0^+$. Similarly for a function $\mathcal{G}(a):\mathbb{R}_{>0}\to\mathbb{R}$, we write
\begin{equation*}
	l_a^k[\mathcal{G}]:=\lim_{a\to 0^+}\frac{\dd^k\mathcal{G}(a)}{\dd a^k}\,.
\end{equation*}

Let us now make the observation that since our metric is
\begin{equation}
	g=-2\,\dd\lambda\,\dd v+a(\lambda)^2\dd v^2+a(\lambda)^2r(v,\lambda)^2\dd\Omega^2\,,\label{eq:metriclambdavcoordagain}
\end{equation}
where $r(v,\lambda)=v-\eta(\lambda)$, the metric is $C^k$ extendible as $\lambda\to 0^+$ if and only if $l_\lambda^r[a^2]$ and
\begin{equation}
	\lim_{\lambda\to0^+}\frac{\partial^r}{[\partial\lambda\partial v]^r}\Big(a(\lambda)\big(v-\eta(\lambda)\big)\Big)^2=\lim_{\lambda\to 0^+}\frac{\partial^r}{[\partial\lambda\partial v]^r}\left(a(\lambda)^2v^2
	-2a(\lambda)^2\eta(\lambda)v
	+a(\lambda)^2\eta(\lambda)^2\right)
	\label{eq:mixedpartialderiv}
\end{equation}
exist and are finite for all $r\in\{0,\ldots,k\}$. In the above, $[\partial\lambda\partial v]^r$ schematically means any combination of partial derivatives with respect to $\lambda$ and/or $v$ of total order $r$.\footnote{For example, if $r=3$, then we have $\partial_\lambda^3$, $\partial_v\partial_\lambda^2$, $\partial_\lambda\partial_v\partial_\lambda$, $\partial_\lambda^2\partial_v$, $\partial_v^2\partial_\lambda$, $\partial_v\partial_\lambda\partial_v$, $\partial_\lambda\partial_v^2$, and $\partial_v^3$.} Let us note that the existence and finiteness of $l_\lambda^r[a^2]$, $l_\lambda^r[a^2\eta]$, and $l_\lambda^r[a^2\eta^2]$ for all $r\in\{0,\ldots,k\}$ is equivalent to that of \eqref{eq:mixedpartialderiv} by repeated use of Clairaut's theorem on equality of mixed partial derivatives, noting that $\partial_v(a^2v^2)=2a^2v$, $\partial_v^2(a^2v^2)=2a^2$, $\partial_v(a^2\eta v)=a^2\eta$, and any other $v$ derivatives vanish. We recall here that the limit $\lambda\to 0^+$ should always be taken at fixed $v$.
Therefore, the existence and finiteness of $l_\lambda^r[a^2]$, $l_\lambda^r[a^2\eta]$, and $l_\lambda^r[a^2\eta^2]$ for all $r\in\{0,\ldots,k\}$ is equivalent to the metric \eqref{eq:metriclambdavcoordagain} being $C^k$ extendible as $\lambda\to 0^+$. This brings us to the following results:

\medskip

\begin{lem}\label{lem:lalaetakiffCk}
    The existence and finiteness of $l_\lambda^r[a]$ and $l_\lambda^r[a\eta]$ for all $r\in\{0,\ldots,k\}$ is equivalent to the metric given in \eqref{eq:metriclambdavcoordagain} being $C^k$ extendible as $\lambda\to 0^+$. 
\end{lem}

The proof is given in Appendix \ref{app:longCkProofs}.\footnote{Most proofs in this subsection are relegated to an appendix to improve the presentation. The proofs are mostly combinatorial in nature and do not present additional geometrical or physical insight.}

\medskip
\medskip

Next, we note that for $C^2$ extendiblity, the existence and finiteness of $l_\lambda^0[a]$, $l_\lambda^0[a\eta]$, and $l_\lambda^0[\dot H/a^2]$ implied the existence and finiteness of $l_\lambda^2[a]$ and $l_\lambda^2[a\eta]$ since $\partial^2_\lambda a=a(\dot H/a^2)$ and $\partial^2_\lambda(a\eta)=(a\eta)(\dot H/a^2)$. As we see below, one can generalize this relation to arbitrary higher derivatives.

\medskip

\begin{lem}\label{lem:lalaetalhdotoa2km2tok}
    This lemma consists of two similar (but independent) statements:
    \begin{enumerate}[label=(\alph*)]
        \item[\emph{(a)}] the existence and finiteness of $l_\lambda^r[a]$ and $l_\lambda^r[\dot H/a^2]$ for all $r\in\{0,\ldots,k-2\}$ implies the existence and finiteness of $l_\lambda^{k-1}[a]$ and $l_\lambda^k[a]$;
        \item[\emph{(b)}] the existence and finiteness of $l_\lambda^r[a\eta]$ and $l_\lambda^r[\dot H/a^2]$ for all $r\in\{0,\ldots,k-2\}$ implies the existence and finiteness of $l_\lambda^{k-1}[a\eta]$ and $l_\lambda^k[a\eta]$.
    \end{enumerate}
\end{lem}

The proofs are given in Appendix \ref{app:longCkProofs}.

\medskip
\medskip

\begin{cor}\label{cor:aaetdotHa2km2lambdaCk}
    The existence and finiteness of $l_\lambda^r[a]$, $l_\lambda^r[a\eta]$, and $l_\lambda^r[\dot H/a^2]$ for all $r\in\{0,\ldots,k-2\}$ implies the metric \eqref{eq:metriclambdavcoordagain} is $C^k$ extendible as $\lambda\to 0^+$.
\end{cor}

\proof
From Lemma \ref{lem:lalaetalhdotoa2km2tok}, we get the existence and finiteness of $l_\lambda^r[a]$ and $l_\lambda^r[a\eta]$ for all $r\in\{0,\ldots,k\}$, hence from Lemma \ref{lem:lalaetakiffCk} the metric \eqref{eq:metriclambdavcoordagain} is $C^k$ extendible as $\lambda\to 0^+$.
\qed

\medskip
\medskip

For applications, it will turn out to be useful to relate $\lambda$-derivatives of $\dot{H}/a^2$ with $a$-derivatives of $\dot{H}/a^2$ in the limit as $\lambda \to 0^+$. Of course, $a$-derivatives only make sense if we can express $\dot{H}/a^2$ as a function of $a$. This will hold if $a$ and $\lambda$ are invertible on some interval $(0, \lambda_\mathrm{L})$, which will hold, for example, if $\dot{a}(t) > 0$ on some interval $(-\infty, t_\mathrm{L})$ --- this is certainly achieved for a \ref{itm:case2b} inflationary universe satisfying \eqref{eq:defInf}. In the following statements, whenever we make a claim about an $a$-derivative, we are implicitly assuming that this invertibility assumption holds.

\medskip

\begin{lem}\label{lem:llaladothoa2km2}
    The existence and finiteness of $l_\lambda^r[a]$ and $l_a^r[\dot H/a^2]$ for all $r\in\{0,\ldots,k-2\}$ implies the existence and finiteness of $l_\lambda^{r}[\dot H/a^2]$ for all $r\in\{0,\ldots,k-2\}$.
\end{lem}

The proof is given in Appendix \ref{app:longCkProofs}.

\medskip
\medskip

\begin{cor}\label{cor:laaetaldothoa2akm2Ck}
    The existence and finiteness of $l_\lambda^r[a]$, $l_\lambda^r[a\eta]$, and $l_a^r[\dot H/a^2]$ for all $r\in\{0,\ldots,k-2\}$ implies the metric \eqref{eq:metriclambdavcoordagain} is $C^k$ extendible as $\lambda\to 0^+$.
\end{cor}

\proof
From Lemma \ref{lem:llaladothoa2km2}, we get the existence and finiteness of $l_\lambda^r[a]$, $l_\lambda^r[a\eta]$, and $l_\lambda^r[\dot H/a^2]$ for all $r\in\{0,\ldots,k-2\}$, hence from Corollary \ref{cor:aaetdotHa2km2lambdaCk} the metric \eqref{eq:metriclambdavcoordagain} is $C^k$ extendible as $\lambda\to 0^+$.
\qed

\medskip
\medskip

Gathering the above allows us to prove our main result:
\begin{thm}\label{thm:CkdotHoa2ofa}
    If the metric \eqref{eq:metriclambdavcoordagain} is $C^{k-2}$ extendible as $\lambda\to 0^+$ and if $l_a^{r}[\dot H/a^2]$ exists and is finite for all $r\in\{0,\ldots,k-2\}$, then the metric \eqref{eq:metriclambdavcoordagain} is $C^{k}$ extendible as $\lambda\to 0^+$.
\end{thm}

\proof
By the hypothesis, Lemma \ref{lem:lalaetakiffCk} tells us that $l_\lambda^r[a]$ and $l_\lambda^r[a\eta]$ exist and are finite for all $r\in\{0,\ldots,k-2\}$. Thus, we can apply Corollary \ref{cor:laaetaldothoa2akm2Ck}, which completes the proof.
\qed

\medskip
\medskip

\noindent\emph{Remarks and examples.}
We see from this theorem that for the metric to be smoothly extendible, one needs $\dot H/a^2$ seen as function of $a$ to be smooth as $a\to 0^+$. (This proves a result stated without proper proof in the appendix of \cite{Yoshida:2018ndv}.) This is the case, for instance, if $\dot H\propto a^q$ for $q\in\mathbb{Z}_{\geq 2}$ or if $\dot H=\sum_{q=2}^N c_qa^{q}$ for some coefficients $c_q\in\mathbb{R}$, $q$ from 2 up to $N\in\mathbb{Z}_{\geq 2}\cup\{\infty\}$. In the former case, there is a unique solution to when $\dot H=-ba^q$ for some proportionality constant $b\in\mathbb{R}_{>0}$: demanding $H=\dot a/a\to h$ as $t\to-\infty$, it is of the form
\begin{equation}
	a(t)=\left(\frac{h^2q}{2b}\mathrm{sech}^2\left(\frac{hq}{2}(t-t_0)\right)\right)^{1/q}\,,\qquad t_0\in\mathbb{R}\,.
\end{equation}
Letting $b=2q$, $h=1$, and $t_0=0$ for illustrative purposes, the asymptotic expansion of the above is
\begin{equation*}
	a(t)\stackrel{t\to-\infty}{=}e^t-\frac{2}{q}e^{(1+q)t}+\frac{2+q}{q^2}e^{(1+2q)t}+\ldots\,.
\end{equation*}
Recall here $q\in\mathbb{Z}_{\geq 2}$, so the above is asymptotically $a(t)\stackrel{t\to-\infty}{=}e^t+O(e^{3t})$.

Another interesting class of examples is
\begin{equation}
	a(t)=e^t+e^{nt}\,,\qquad n\in\mathbb{R}_{\geq 3}\,.
\end{equation}
We will show $C^\infty$ extendibility for $n = 3$.\footnote{More generally, we conjecture that $C^k$ extendibility follows if $n\in\mathbb{R}_{\geq k+1}$ or if $n\in\{3,4,\ldots,k\}$. Correspondingly, $C^\infty$ would follow if $n\in\mathbb{Z}_{\geq 3}$. We checked several cases that support these conjectures, but a general proof, though achievable, is beyond the scope of this work (and also somewhat tangential).}
When $n=3$, one can analytically invert the scale factor in such a case, yielding
\begin{equation*}
	t(a)=\ln\left(\frac{2^{1/3}\Delta(a)^{2/3}-2\cdot 3^{1/3}}{6^{2/3}\Delta(a)^{1/3}}\right)\,,\qquad\Delta(a):=9a+\sqrt{12+81a^2}\,.
\end{equation*}
Then,
\begin{equation*}
	\frac{\dot H}{a^2}=\frac{4}{(1+e^{2t})^4}=4\left(\frac{1}{3}+\frac{\Delta(a)^{2/3}}{3\cdot 2^{2/3}\cdot 3^{1/3}}+\left(\frac{2}{3}\right)^{2/3}\frac{1}{\Delta(a)^{2/3}}\right)^{-4}\,,
\end{equation*}
which is a smooth function of $a$ as $a\to 0^+$. One way to see this is to notice the series expansion as $a\to 0^+$ is
\begin{equation*}
	\frac{\dot H}{a^2}=4-16a^2+72a^4-352a^6+1820a^8-9792a^{10}+\ldots\,,
\end{equation*}
hence $\dot H/a^2$, as a function of $a$, is real analytic in an open neighborhood about $a=0^+$; consequently, it is smooth in that neighborhood.

More examples, including a realization of slow-roll inflation with a scalar field, can be found in Appendix \ref{app:moreexamples}.

\subsection{Comments on curvature singularities}\label{sec:curv}

To end this section, we discuss the relationship of a $C^2$ extension with curvature singularities. Let $(M,g)$ be a $C^2$ spacetime. We shall call a \emph{scalar curvature invariant} $\mathcal{S}$ a scalar function on $M$ that is a polynomial in the components of the metric $g_{\mu\nu}$, its inverse $g^{\mu\nu}$, and the (Riemann) curvature tensor $R_{\mu\nu\a\b}$. Common scalar curvature invariants include the scalar curvature $R = g^{\mu\nu}R_{\mu\nu} = g^{\mu\nu}g^{\alpha\beta}R_{\alpha\mu\beta\nu}$ to linear order in curvature and, to quadratic order in curvature, the Kretschmann scalar $\mathcal{K}:=R_{\mu\nu\a\b}R^{\mu\nu\a\b}$, the contraction of the Ricci tensor with itself $R_{\mu\nu}R^{\mu\nu}$, and the Gauss-Bonnet invariant $\mathcal{G}:=R^2-4R_{\mu\nu}R^{\mu\nu}+\mathcal{K}$. To any order in curvature, there is always a finite number of linearly independent polynomials that can be written down due to the finite number of contractions of the Riemann tensor (further constrained by its symmetries); see, e.g., \cite{vanNieuwenhuizen:1976vb,Metsaev:1986yb} to cubic order. The following theorem demonstrates the lack of curvature singularities on approach towards the $\l = 0$ hypersurface (i.e., the past boundary) for those spacetimes that admit $C^2$ extensions.

\medskip

\begin{thm}\label{thm:nonscalarsing}
Suppose $(M,g)$, defined as in the premise of Theorem \emph{\ref{higher reg thm}}, admits a $C^2$ extension through the $\l =0$ hypersurface. (This is achieved, e.g., if we assume the hypotheses of part (b) of Theorem \emph{\ref{higher reg thm}} or Corollary \emph{\ref{C2 cor}}.) If $\mathcal{S}$ is a scalar curvature invariant on $M$, then $\mathcal{S}$ extends continuously to the $\l = 0$ hypersurface.
\end{thm}

\proof
The curvature tensor $R_{\mu\nu\a\b}$ is a finite sum of terms involving the metric $g_{\mu\nu}$, the inverse metric $g^{\mu\nu}$, and their first and second derivatives. Taking derivatives of $g_{\mu\a}g^{\a\nu} = \delta_{\mu}^\nu$, it follows that the inverse metric is as regular as the metric. Therefore, by the $C^2$ extendibility of the metric, each of the terms appearing in any curvature invariant $\mathcal{S}$ extends continuously to the $\l = 0$ hypersurface.
\qed

\medskip
\medskip

\noindent\emph{Remark.}
When the previous theorem is realized, one usually says that the resulting spacetime is non-singular in the sense that it has no scalar curvature singularity.

\medskip
\medskip

Theorem \ref{thm:nonscalarsing} may be stronger than necessary though.
Indeed, as mentioned in Section \ref{sec:padS}, the fact that there is no scalar curvature singularity may already be implied by the past-asymptotically dS nature of the $C^2$ extendible spacetime. In fact, we claim that any past-asymptotically dS spacetime (in the sense of Definition \ref{def:padS}) is free of scalar curvature singularities. To be more precise, we assert that, to any order in curvature (say order $n$), any scalar curvature invariant will be a polynomial in $H^2$ and $\dot H$ of the same order, i.e., it will consist of a linear combination of $\{H^{2n},H^{2(n-1)}\dot H,\ldots,H^2\dot H^{n-1},\dot H^n\}$. The reason for this is that any scalar curvature invariant of order $O(\mathrm{curvature}^n)$ is schematically of the form $g^{-n}\mathrm{Riem}^n$. Then, we can make the observation in FLRW spacetime coordinates (and Cartesian spatial coordinates $i,j\in\{x,y,z\}$) that the non-vanishing contractions involve the components $R^j{}_{iji}=a^2H^2$ and $R^t{}_{iti}=a^2(H^2+\dot H)$ of the Riemann tensor (no summation implied) contracted with $g^{ii}=1/a^2$ and the components $R^i{}_{tit}=H^2+\dot H$ of the Riemann tensor contracted with $g^{tt}=-1$. Thus, a scalar polynomial (which is coordinate independent) can only consist of the terms $H^2$ and $\dot H$ raised to the appropriate order. We can see this explicitly for linear and quadratic invariants [recall \eqref{eq:lqci}], and to give a couple of examples to cubic order, we have $R_{\mu\nu}R^{\nu\rho}R_\rho{}^\mu=108H^6+162H^4\dot H+108H^2\dot H^2+30\dot H^3$ and $R_{\mu\nu}{}^{\alpha\beta}R_{\alpha\beta}{}^{\gamma\sigma}R_{\gamma\sigma}{}^{\mu\nu}=48H^6+72H^4\dot H+72H^2\dot H^2+24\dot H^3$.

Any past-asymptotically dS spacetime thus seems to be free of any scalar curvature singularity (as for exact dS). However, as we have encountered throughout this section, a past-asymptotically dS spacetime is not necessarily extendible (within the $(\lambda,v)$ coordinates); conditions for ($C^0$, $C^1$, $C^2$, etc.)~extendibility may be stronger than those defining past-asymptotically dS. Therefore, we may wonder what goes wrong for those spacetimes that might be ($C^0$, $C^1$, $C^2$, etc.)~\emph{in}extendibile, yet free of scalar curvature singularities.

For $C^k$ (in)extendibility, $k\geq 2$, we can provide some explanation. Recalling part (b) of Theorem \ref{higher reg thm} for $C^2$ extendibility, if the quantity $\dot H/a^2$ convergences toward the past boundary, then $C^2$ extendibility is achievable. If the quantity $\dot H/a^2$ diverges, then we do not know whether a $C^2$ extension exists within the $(\lambda,v)$ coordinates. What we can show, though, is that the spacetime possesses a \emph{null parallelly propagated curvature singularity} on the past boundary. Indeed, it is straightforward to verify that some components of the curvature tensor in the $(\lambda,v)$ coordinates will involve the quantity $\dot H/a^2$ and hence diverge. For instance, the $\lambda\lambda$ component of the Ricci tensor is $R_{\lambda\lambda}=-2\dot H/a^2$. Since
the divergence of the curvature tensor is specific to this coordinate system, this is not a scalar curvature singularity, but bears the aforementioned name often abbreviated p.\,p.~singularity.\footnote{The discussion is generalizable to $C^k$ extendibility, $k\geq 2$, recalling Corollary \ref{cor:aaetdotHa2km2lambdaCk}. In this case, the relevant quantity is $\dd^{k-2}(\dot H/a^2)/\dd\lambda^{k-2}$, and this enters in the components of the $(k-2)$th covariant derivative of the curvature tensor in $(\lambda,v)$ coordinates. See the appendix of \cite{Yoshida:2018ndv} for more details.}
To be precise, one must find a tetrad basis $\{E^M\}$ that is parallelly propagated along null geodesics, that is for which the covariant derivative along the tangent vector $K$ to the null geodesics vanishes: $\nabla_KE^M=K^\mu\nabla_\mu E^M=0$. Different bases can be constructed \cite{Yoshida:2018ndv,Nomura:2021lzz}, but they all share the property that $\dot H/a^2$ appears in the components of the Ricci tensor in the given basis. For instance, the null parallelly propagated basis constructed in \cite{Nomura:2021lzz} can be written in terms of our coordinates as $E^0=\dd v/\sqrt{2}$, $E^1=(2\dd\lambda-a^2\dd v)/\sqrt{2}$, $E^2=ar\,\dd\theta$, $E^3=ar\sin\theta\,\dd\phi$, and in that parallelly propagated basis\footnote{At fixed $v$, $\theta$, and $\phi$, this makes it clear that it is the $\lambda$ components that matter along null geodesics.} one finds $R_{11}=-\dot H/a^2$. Whenever this is divergent, we can say that there is a p.\,p.~curvature singularity.

One may say that a p.\,p.~singularity is weaker than a scalar curvature singularity, but it may nevertheless be physically relevant. In fact, the relation between p.\,p.~singularities and spacetime extendibility was the crux of the work by Clarke \cite{Clarke:1973,Clarke:1982,Clarke:1994cw}, which was applied in cosmology in \cite{Yoshida:2018ndv,Numasawa:2019juw,Nomura:2021lzz,Nishii:2021ylb,Nomura:2022vcj,Harada:2021yul}. For instance, the approach of \cite{Yoshida:2018ndv} was precisely to evaluate the curvature tensor in a null parallelly propagated basis within a flat FLRW spacetime and finding the criterion that determined the presence/absence of a p.\,p.~curvature singularity. The result showed that the quantity $\dot H/a^2$ had to converge for the spacetime to be free of p.\,p.~singularities and for there to be no obstruction to spacetime extendibility. We thus see that this approach and ours yield consistent results --- they certainly are complementary.

\section{Conformal embeddings into the Einstein static universe}\label{sec:confEmbedd}

It is well known that the flat de Sitter model conformally embeds into the Einstein static universe \cite{Hawking:1973uf,Carroll:2004st,Mukhanov:2005sc}. In this section, we show that the past-asymptotically de Sitter models also conformally embed into the Einstein static universe and retain similar geometrical properties. These properties will be used in the subsequent section to go beyond the symmetries of FLRW, i.e., beyond exact homogeneity and isotropy. 

First, we give the definition of a conformal extension. Definitions \ref{conformal extension def} and \ref{def: conformal boundaries} parallel Definitions \ref{def: extension} and \ref{def: future and past boundaries} in Section \ref{sec: further prelim}. 

\medskip

\begin{Def}\label{conformal extension def}\emph{Suppose $(M,g)$ is a $C^\infty$ spacetime and $(\wt{M}, \wt{g})$ is a $C^k$ spacetime. We say that $(\wt{M}, \wt{g})$ is a $C^k$ \emph{conformal extension} of $(M,g)$ if there is an embedding $M \hookrightarrow \wt{M}$  and a $C^k$ function $\Omega\colon \ov{M} \to [0,\infty)$ ($\ov{M}$ is defined below) such that }
\begin{itemize}
\item[\emph{(i)}] \emph{$M \subset \wt{M}$ is a proper open subset,}
\item[\emph{(ii)}] \emph{the restriction $\Omega|_M$ is smooth and positive,}
\item[\emph{(iii)}] \emph{$\pd_0M \neq \emptyset$ where $\pd_0M := \{p \in \pd M \mid \Omega(p) = 0\}$,}
\item[\emph{(iv)}] \emph{the pull back of $\wt{g}$ under the embedding equals $\Omega^2 g$.}
\end{itemize}
\emph{
In this setting, we refer to $M \hookrightarrow \wt{M}$ as a $C^k$ \emph{conformal embedding}.}
\end{Def}

\medskip
\medskip

A couple of comments: (1) as in Section \ref{sec: further prelim}, we are identifying $M$ with its image under the embedding; (2) $\ov{M}$ denotes the closure of $M$ within $\wt{M}$, and likewise, $\pd M$ denotes the topological boundary of $M$ within $\wt{M}$, hence $\ov{M} = M \cup \pd M$.

\medskip
\medskip

\begin{Def}\label{def: conformal boundaries}
\emph{Let $(\wt{M}, \wt{g})$ be a $C^k$ conformal extension of a $C^\infty$ spacetime $(M,g)$. A future-directed $\wt{g}$-timelike curve $\g \colon [a,b] \to \wt{M}$ is called a \emph{future-terminating timelike curve} for a point $p \in \pd M$ provided $\g(b) = p$ and $\g\big([a,b)\big) \subset M$. \emph{Past-terminating timelike curves} are defined time-dually. The \emph{future} and \emph{past conformal boundaries} of $M$ within $\wt{M}$ are defined as
\begin{align*}
\pd^+_0M \,&=\, \{p \in \pd_0 M \mid \text{there is a future-terminating timelike curve for $p$}\}\,,\\
 \pd_0^-M \,&=\, \{p \in \pd_0 M \mid \text{there is a past-terminating timelike curve for $p$}\}\,.
\end{align*}
The \emph{future} and \emph{past boundaries} of $M$ within $\wt{M}$ are defined as
\begin{align*}
\pd^+M \,&=\, \{p \in \pd M \setminus \pd_0M \mid  \text{there is a future-terminating timelike curve for $p$}\}\,,\\
 \pd^-M \,&=\, \{p \in \pd M \setminus \pd_0 M \mid \text{there is a past-terminating timelike curve for $p$}\}\,.
\end{align*}}
\end{Def}

\medskip
\medskip

\noindent\emph{Remark.} The definition of future and past boundaries of $M$ within $\wt{M}$ (Definition \ref{def: conformal boundaries}) is similar to the definition of future and past boundaries of $M$ within $M_\ext$ (Definition \ref{def: future and past boundaries}) where $(M_\ext, g_\ext)$ is a continuous extension of $(M,g)$. While they are in general different, there may be cases where they coincide as we will see in Corollary \ref{cor: extension from conformal}. (\emph{Coincide} here means that one can find the same terminating timelike curve in $M$ for the two boundary points.) Under some mild regularity assumptions along with global hyperbolicty of $(M,g)$, there will be a local isometry up to the boundary between such points \cite{Sbierski:2022roz}.

\medskip

\subsection{Review of the conformal embedding of the flat de Sitter model into the Einstein static universe} \label{flat de Sitter into ESU sec}

The Einstein static universe is the spacetime $(\wt{M}, \wt{g})$ given by 
\[
    \wt{M} \,=\, \R \times \mathbb{S}^3 \:\:\:\: \text{ and } \:\:\:\: \wt{g} \,=\, -(\dd T')^2 + h_{\mathbb{S}^3}\,,
\]
where $h_{\mathbb{S}^3}$ is the usual round metric on the unit 3-sphere and $T'$ is the coordinate on $\R$. (The choice of $T'$ instead of $T$ is so that our notation is consistent with below.) In this subsection, we review how the flat de Sitter comformally embeds into the Einstein static universe. Let $(\hat{M}, \hat{g})$ denote the flat de Sitter model, i.e.,
\[
    \hat{M} \,=\, \R \times \R^3 \:\:\:\: \text{ and } \:\:\:\: \hat{g} \,=\, -\dd\hat{t}\,^2 + e^{2 h \hat{t}}h_{\mathbb{E}}\,,
\]
where $h_{\mathbb{E}}$ is the Euclidean metric as before; if $(r, \theta, \phi)$ denote the usual spherical coordinates, then $h_{\mathbb{E}} = \dd r^2 + r^2(\dd\theta^2 + \sin^2\theta\,\dd\phi^2)$. Moreover, we denote the de Sitter Hubble constant by $h \in \R_{>0}$.

We introduce new coordinates $(T, \chi, \theta, \phi)$ for $(\hat{M}, \hat{g})$ via
\begin{subequations}
\begin{align}
    e^{ h \hat{t}} \,&=\, \sinh( h T) + \cosh( h T) \cos \chi\,, \label{global trans 1}
\\
     h re^{ h \hat{t}}\,&=\, \cosh( h T) \sin \chi\,, \label{global trans 2}
\\
    \theta \text{ and }& \phi \text{ are unchanged.} \notag
\end{align}
\end{subequations}
The Jacobian between $(\hat{t},r, \theta, \phi)$ and $(T, \chi, \theta, \phi)$ is nonzero, so the latter are well-defined coordinates. The metric $\hat{g}$ in $(T, \chi, \theta, \phi)$ coordinates is 
\[
    \hat{g} \,=\, -\dd T^2 + \frac{\cosh^2( h T)}{ h^2 }\,\big[\dd\chi^2 + \sin^2\chi\,(\dd\theta^2 + \sin^2\theta\,\dd\phi^2)\big]\,.
\]
The term in square brackets is the round metric on $\mathbb{S}^3$. The coordinates $(\theta, \phi)$ are the usual coordinates on $\mathbb{S}^2$ and hence their ranges are the usual ones. Now we determine the ranges of $T$ and $\chi$. Squaring equations \eqref{global trans 1} and \eqref{global trans 2} and summing them, we obtain
$\sinh( h T) = \frac{1}{2} h^2 r^2 e^{ h \hat{t}} + \sinh( h \hat{t})$.
Therefore the range of $T$ is $\R$. Plugging this into \eqref{global trans 1} and substituting \eqref{global trans 2}, we find $\tan \chi = 2 h r/(1- h^2 r^2+e^{-2 h \hat t})$. Thus the range of $\chi$ is $(0,\pi)$. The $(T,\chi,\theta,\phi)$ coordinates are usually referred to as the closed de Sitter slicing --- or global de Sitter --- in the literature.

Now we investigate the trajectories of the flat de Sitter comoving observers within the $(T, \chi, \theta, \phi)$ coordinates as $\hat{t} \to -\infty$. Recall that a comoving observer in flat de Sitter coordinates is the timelike curve $\hat{t}\mapsto (\hat{t}, x_0, y_0, z_0)$ for some fixed $(x_0, y_0, z_0)$. (From here on, when we refer to comoving observers, we are referring to these observers.) Let $r_0$ denote the corresponding radial point. Since $r_0$ is fixed, the expression for $\sinh( h T)$ in the above paragraph shows that $T \to -\infty$ as $\hat{t} \to -\infty$ along the comoving observer. Then \eqref{global trans 2} implies $\chi$ approaches either $0$ or $\pi$. However, we can rule out $\chi$ approaching $\pi$ from \eqref{global trans 1}. Thus, we have established the following proposition which will play an important role in things to come.

\medskip

\begin{prop}\label{prop: comoving observation 1}
$T \to -\infty$ and $\chi \to 0$ as $\hat{t} \to -\infty$ along each comoving observer. Hence all the comoving observers limit to the north pole ($\chi=0$) on $\mathbb{S}^3$ as $\hat{t} \to -\infty$. 
\end{prop}

\medskip

Define a new coordinate $T'$ via $T' = 2\arctan(e^{ h T}) - \pi/2$ so that $\cosh( h T) = \sec T'$. Hence $T'$ maps $\R$ diffeomorphically onto $(-\frac{\pi}{2}, \frac{\pi}{2})$. From \eqref{global trans 1} and \eqref{global trans 2}, the relationship between $(\hat{t}, r, \theta, \phi)$ and $(T', \chi, \theta, \phi)$ is
\begin{subequations}
\begin{align}
    e^{ h \hat{t}} \,&=\, \frac{\sin T' + \cos \chi}{\cos T'}\,, \label{conformal trans 1}
\\
     h r\,&=\, \frac{\sin \chi}{\sin T' + \cos \chi}\,, \label{conformal trans 2}
\end{align}
\end{subequations}
while $\theta$ and $\phi$ are unchanged.
Writing the metric in $(T', \chi, \theta, \phi)$ coordinates, we readily see that the metric is conformal to the Einstein static universe:
\begin{equation}\label{flat dS in ESU metric}
    \hat{g} \,=\, \frac{1}{ h^2 \cos^2 T'}\big(-(\dd T')^2 + \dd\chi^2 + \sin^2\chi\,(\dd\theta^2 + \sin^2\theta \,\dd\phi^2) \big)\,.
\end{equation}

The values that $(T', \chi)$ on $M$ assume are restricted as we explain now. Since $T' \in (-\frac{\pi}{2}, \frac{\pi}{2})$, from \eqref{conformal trans 1}, we have $\sin T' + \cos \chi > 0$. Hence $\cos(T' - \frac{\pi}{2}) >  \cos(\chi - \pi)$. Since cosine is strictly increasing on the interval $(-\pi, 0)$, we have $T' > \chi - \frac{\pi}{2}$.  In view of Definition \ref{conformal extension def}, the following properties are now readily established. 
  
\medskip
\medskip

\begin{figure}[t]
\begin{center}
\mbox{
\includegraphics[width=3in]{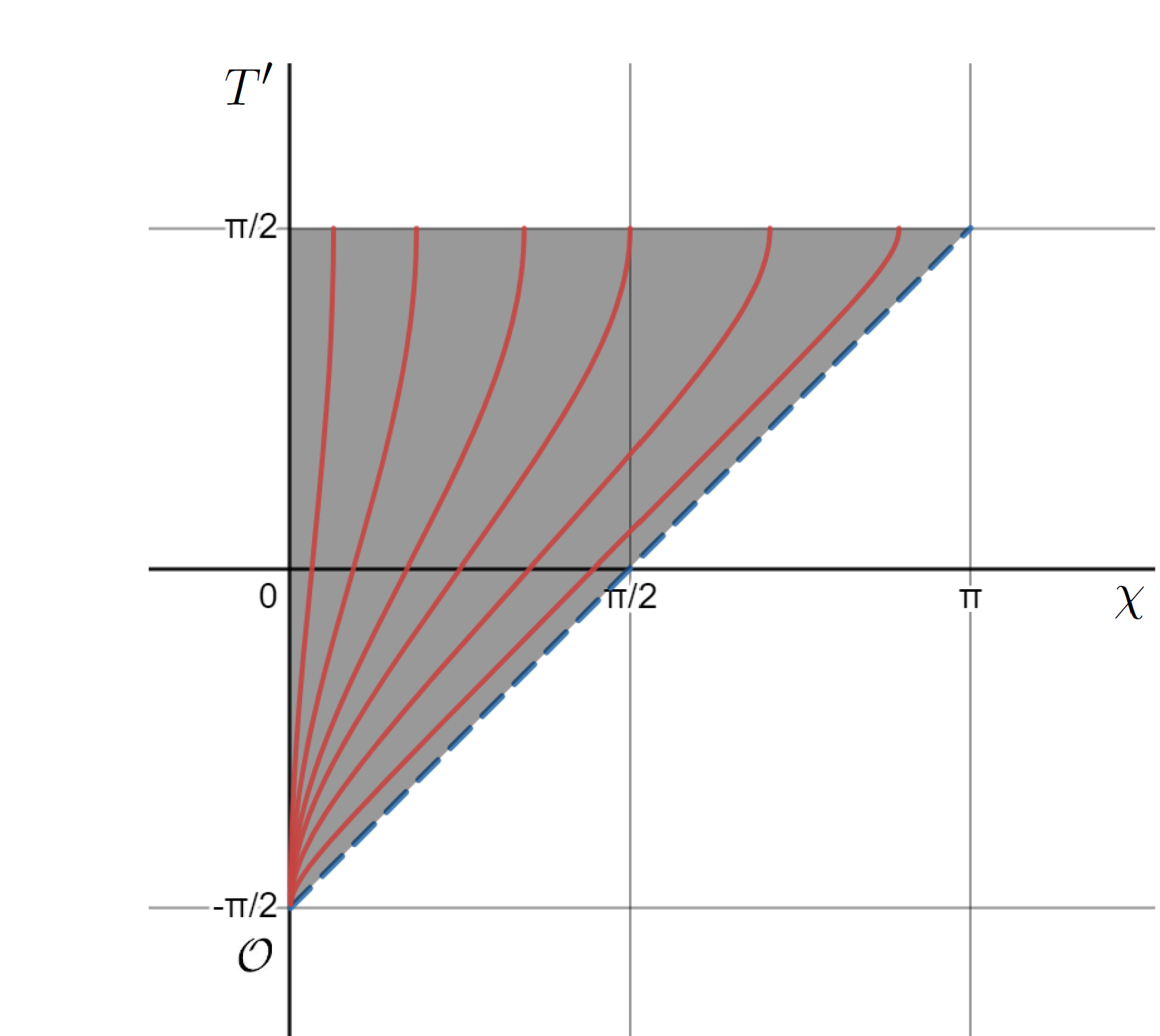}
}
\end{center}
\caption{\small{The shaded region corresponds to the flat de Sitter model $\hat{M}$ within the Einstein static universe $\wt{M}$. The vertical axis is $T'$ and the horizontal axis is $\chi$.  The comoving observers are given by the red lines; they all limit to the point $\mc{O}$ which corresponds to the north pole ($\chi = 0$) at $T' = -\pi/2$. The past conformal boundary is precisely $\pd^-_0\hat{M} = \{\mc{O}\}$. The past boundary $\pd^-\hat{M}$ is given by the lightcone, $T' = \chi-\pi/2$, for values $-\pi/2 < T' < \pi/2$  --- this is depicted by the dashed blue line}.}
\label{flat dS in ESU fig}
\end{figure}

\noindent\underline{Properties of the conformal embedding $\hat{M} \hookrightarrow \wt{M}$}:
\begin{itemize}
    \item[(1)] The Einstein static universe $(\wt{M}, \wt{g})$ is a $C^\infty$ conformal extension of the flat de Sitter model $(\hat{M}, \hat{g})$. See Figure \ref{flat dS in ESU fig}.
    
    \item[(2)] $\hat{M}$ corresponds to the set of points $\{(T',\chi)\in(-\frac{\pi}{2}, \frac{\pi}{2}) \times [0, \pi) \mid T'>\chi-\frac{\pi}{2}\}$.\footnote{We only characterize the $(T',\chi)$ space, i.e., we omit to explicitly write $\times\mathbb{S}^2$ every time.}
    
    \item[(3)] The topological boundary is given by the disjoint union $\pd \hat{M} = \{\mc{O}\} \sqcup \mathcal{B}^- \sqcup \Sigma^+$. Here $\mc{O}$ is the point that corresponds to the north pole (i.e., $\chi = 0$) at $T' = -\frac{\pi}{2}$. Also, $\mathcal{B}^-=\{(T',\chi)\in(-\frac{\pi}{2}, \frac{\pi}{2}) \times (0, \pi) \mid T'=\chi-\frac{\pi}{2}\}$ is the past null boundary of flat de Sitter, and $\Sigma^+=\{T'=\frac{\pi}{2} \textrm{ and } 0\leq\chi\leq\pi\}$ is the future spacelike boundary. In what follows, it shall be useful to also define $\mathcal{P}$, the point at $T'=\frac{\pi}{2}$ that corresponds to the south poles (i.e., $\chi=\pi$).
    
    \item[(4)] The conformal factor $\Omega$ is given by $ h \cos T'$ restricted to the closure of $\hat{M}$ within $\wt{M}$, i.e., restricted to $\hat{M} \cup \pd \hat{M}$.
    
    \item[(5)] The conformal boundary is $\pd_0 \hat{M} = \{\mc{O}\} \sqcup \Sigma^+$, and moreover, $\pd_0^-\hat{M} = \{\mc{O}\}$ and $\pd_0^+\hat{M} = \Sigma^+ \setminus \{\mathcal{P}\}$. Recall $\mc{O}$ and $\mc{P}$ correspond to the north and south pole points at $T' =-\pi/2$ and $T' = \pi/2$, respectively.

    \item[(6)] $\pd^-\hat{M} = \mathcal{B}^-$,\:\:\:\:\:\: $\pd^+\hat{M} = \emptyset$.
    
    \item[(7)] Proposition \ref{prop: comoving observation 1} above implies that $T' \to -\frac{\pi}{2}$ and $\chi \to 0$ as $\hat{t} \to -\infty$ along each comoving observer. Hence all the comoving observers limit to $\mc{O}$ as $\hat{t} \to -\infty$.  

    \item[(8)] There is an open subset $\wt{M}' \subset \wt{M}$ containing $\mc{O}$ such that $\hat{M} = I^+_{\wt{g}}(\mc{O},\wt{M}')$. For example, one can take $\wt{M}' = (-\infty, \frac{\pi}{2}) \times \mathbb{S}^3$. Note that $I^+_{\wt{g}}$ is the timelike future with respect to $\wt{g}$. (The definition of $I^\pm$ is given at the beginning of Section \ref{sec: further prelim}.) 
\end{itemize}

\medskip
\medskip

\noindent\textit{Remark.} We note that $\mathcal{P}$ is absent from the past and future boundaries since it cannot be reached by any timelike curve within $\hat{M}$.

\subsection{Conformal embeddings of quasi-de Sitter spacetimes into the Einstein static universe}

In this section, we show many of the geometrical properties of the previous section hold for spacetimes that approach de Sitter in the asymptotic past. Specifically, we have in mind working under the same premise as at the beginning of Section \ref{sec:C0}.
As before, we fix a flat FLRW spacetime $(M,g)$ given by $M = (-\infty, t_{\rm max}) \times \R^3$ and $g = -\dd t^2 + a(t)^2h_{\mathbb{E}}$
where $a(t)$ is a smooth positive function on $(-\infty, t_{\rm max})$.
We seek functions $\omega(t)>0$ and $\hat{t}(t)$ such that
\begin{equation}
    g=\omega^2\hat g\,,
\end{equation}
i.e.,
\begin{equation}\label{conformal to flat dS metric}
    -\dd t^2 + a(t)^2h_{\mathbb{E}} \,=\, \omega\big(t(\hat t)\big)^2\left(-\dd\hat{t}\,^2 + e^{2 h \hat{t}}h_{\mathbb{E}}\right)\,,
\end{equation}
so that $(M,g)$ is conformal to the flat de Sitter model $(\hat{M},\hat{g})$. This will hold provided
\begin{subequations}
\begin{align}
    \frac{\dd\hat{t}}{\dd t} \,&=\, \frac{1}{\omega(t)}\,, \label{conf to flat 1}
\\
    e^{ h \hat{t}} \,&=\, \frac{a(t)}{\omega(t)}\,. \label{conf to flat 2}
\end{align}
\end{subequations}
Differentiating \eqref{conf to flat 2} with respect $t$ and substituting \eqref{conf to flat 1}, we arrive at the following differential equation for $\omega(t)$:
\begin{equation*}
    \frac{\dot{\omega}}{a} - \omega \frac{\dot{a}}{a^2}\,=\, -\frac{h}{a}\ \iff\ \frac{\dd}{\dd t}\left(\frac{\omega}{a}\right)= -\frac{h}{a}\,;
\end{equation*}
a solution to this differential equation is
\begin{equation}\label{omega def}
    \omega(t) \,=\, h \, a(t) \int_t^{t_0} \frac{\dd\tilde t}{a(\tilde t)}\,,
\end{equation}
where $t_0 \in (-\infty, t_{\rm max})$ is arbitrary. Moreover, comparing with \eqref{def: conformal time} and considering $t_0\leq t_\mathrm{L}<t_\mathrm{max}$, we see that
\begin{equation}\label{eq:omegaaseta}
    \omega(t)= h \, a(t) \left(\eta(t_0)-\eta(t)\right)\,.
\end{equation}
Substituting this solution in \eqref{conf to flat 2}, we can also obtain 
\begin{equation}\label{t-hat def}
    \hat{t}(t) \,=\, \frac{1}{h}\ln\left(\frac{a(t)}{\omega(t)}\right)\,=\,-\frac{1}{h}\ln\left(h\int_t^{t_0}\frac{\dd\tilde t}{a(\tilde t)}\right)\,.
\end{equation}

\medskip
\medskip

\noindent\emph{Remarks.} The function $\omega(t)$ is positive for all $t \in (-\infty, t_0)$. Moreover, $\omega(t) \searrow 0$ as $t \to t_0^-$, and so, $\hat t\to\infty$ as $t\to t_0^-$.

\medskip
\medskip

Recall we are using $\mc{O}$ to denote the point at the north pole $(\chi = 0)$ at $T' = -\pi/2$ within the Einstein static universe $(\wt{M}, \wt{g})$.

\medskip

\begin{prop}\label{conformal embed prop}
Fix any $t_0 \in (-\infty, t_{\rm max})$. Define $M_{t_0} \subset M$ via $M_{t_0} = (-\infty, t_0) \times \R^3$. Suppose the hypothesis of Lemma \ref{lem: aeta limit} is respected, i.e., either $a(t)=e^{ht}+o(e^{ht})$ or $H(t) \to h$ as $t \to -\infty$ for some positive finite number $h \in \R_{>0}$. Then there is a $C^0$ conformal embedding of $(M_{t_0},g)$ into the Einstein static universe $(\wt{M}, \wt{g})$ such that
\begin{itemize}
\item[\emph{(a)}] within $M_{t_0}$, the conformal factor $\Omega$ is $\frac{ h \cos T'}{\omega(t)}$,
\item[\emph{(b)}] $\partial_0M_{t_0}=\{\mc{O}\}\sqcup\{\mathcal{P}\}$, $\pd_0^-M_{t_0}=\{\mc{O}\}$, $\pd^+_0M_{t_0}=\emptyset$,
\item[\emph{(c)}] $\pd^-M_{t_0}=\mathcal{B}^-$, $\pd^+M_{t_0}=\Sigma^+\setminus\{\mathcal{P}\}$,
\item[\emph{(d)}] $\g(t) \to \mc{O}$ as $t \to -\infty$ for each comoving observer $\g(t)$,
\item[\emph{(e)}] there is an open subset $\wt{M}' \subset \wt{M}$ containing $\mc{O}$ such that $M_{t_0} = I^+_{\wt{g}}(\mc{O},\wt{M}')$. For example, one can take $\wt{M}' = (-\infty, \frac{\pi}{2}) \times \mathbb{S}^3$.
\end{itemize} 
\end{prop}

\proof
Let $\omega(t)$ and $\hat{t}(t)$ be given by \eqref{omega def} and \eqref{t-hat def}.
Recall $a(t)\searrow 0$ as $t\to-\infty$. From \eqref{eq:omegaaseta} and the result of Lemma \ref{lem: aeta limit} (which says that $a\eta\to-1/h$ as $t\to-\infty$), we find that $\omega(t)\to 1$ as $t\to-\infty$. Consequently, $\hat t\to-\infty$ as $t\to-\infty$.

Let $(T', \chi, \theta, \phi)$ denote the coordinates for the Einstein static universe introduced in Section \ref{flat de Sitter into ESU sec}. Recall that the flat de Sitter model $(\hat{M}, \hat{g})$ corresponds to the set of points $T' \in (-\frac{\pi}{2}, \frac{\pi}{2})$ and $\chi \in (0, \pi)$ such that $T' > \chi - \frac{\pi}{2}$. From equation \eqref{conformal trans 1}, it follows that the hypersurface $\hat{t} = +\infty$ corresponds to the hypersurface $T'=\pi/2$. Therefore, $M_{t_0}$ corresponds to all of $\hat{M}$ as in Figure \ref{flat dS in ESU fig}.

From \eqref{conformal to flat dS metric} and \eqref{flat dS in ESU metric}, we have
\begin{equation}\label{eq: g conf to ESU}
    g \,=\, \frac{\omega(t)^2}{ h^2 \cos^2 T'}\left(-(\dd T')^2 + \dd\chi^2 + \sin^2\chi\,(\dd\theta^2 + \sin^2\theta\,\dd\phi^2)\right)\,.
\end{equation}
Since $\omega(t)$ is positive and limits to 1 as $t \to -\infty$, it follows that $\hat{t}$ maps $(-\infty, t_0)$ diffeomorphically onto $(-\infty, \infty)$, so it follows that $M_{t_0}$ is diffeomorphic to $\hat{M}$. Therefore, the past conformal properties of the conformal embedding $M_{t_0} \hookrightarrow \wt{M}$ are inherited from the conformal embedding $\hat{M} \hookrightarrow \wt{M}$ from Section \ref{flat de Sitter into ESU sec}. That is, (b)--(e) hold for $M_{t_0}$ since analogous statements hold for $\hat{M}$. The only difference is that now the conformal factor is
\[
    \Omega=\frac{h \cos T'}{\omega(t)}=\frac{h}{a(t)}\left(\sin T'+\cos\chi\right)\,,
\]
where we combined \eqref{conformal trans 1} and \eqref{conf to flat 2}. Hence $\Omega$ vanishes when $T'=\pm\pi/2$ and when $\omega(t)$ is non-zero; this happens only for $T'=-\pi/2$, so at $\{\mathcal{O}\}$, since $\omega\to 0$ as $T'\to\pi/2$ (i.e., as $t\to t_0^-$ or $\hat t\to\infty$). The only other situation where $\Omega$ can vanish is when $T'=\pi/2$ and $\chi=\pi$, so at $\{\mathcal{P}\}$, since then $a(t)$ is non-zero, but $\sin T'+\cos\chi=0$.
Since $h \neq 0$, it follows that $\Omega$ extends as a $C^0$ function to the topological boundary $\pd M_{t_0}$ within $\wt{M}$; hence we can only guarantee that we have a $C^0$ conformal embedding. 
\qed

\medskip
\medskip

From equation \eqref{eq: g conf to ESU}, we have
\[
    g = \frac{\omega(t)^2}{ h^2 \cos^2 T'}\,\wt{g}\,.
\]
Since $\omega(t) \to 1$ as $t \to -\infty$, we have the following corollary.

\medskip

\begin{cor}\label{cor: extension from conformal}
Suppose $a(t)=e^{ht}+o(e^{ht})$ or $H(t) \to h$ as $t \to -\infty$ for some positive finite number $h \in \R_{>0}$. Consider the $C^0$ conformal embedding $M_{t_0} \hookrightarrow \wt{M}$ given in Proposition \emph{\ref{conformal embed prop}}.
Define the $C^0$ spacetime $(M_\ext, g_\ext)$ via
\[
    M_\ext \,=\, M \cup \left\{\wt{M}\,\big|\,T' \leq \chi - \frac{\pi}{2} \:\: \text{ and } \:\: T' \in \left(-\frac{\pi}{2}, \frac{\pi}{2}\right) \right\}\,,
\]
\[
    g_\ext \,=\,  \left\{
    \begin{array}{ll}
      g & \text{ on } M \\
      \frac{1}{h^2\cos^2 T'}\,\wt{g} & \text{ on }  M_\ext \setminus M\,, \\
    \end{array} 
    \right.
\]
with time orientation determined by declaring $\pd_{T'}$ to be future directed.
Then $(M_\ext, g_\ext)$ is a $C^0$ extension of $(M,g)$. The past boundary of $M$ within $M_\ext$ (Definition \emph{\ref{def: future and past boundaries}}) coincides with the past boundary of $M$ within $\wt{M}$ (Definition \emph{\ref{def: conformal boundaries}}). Each is denoted by $\pd^-M$ and corresponds to the lightcone $T'= \chi - \frac{\pi}{2}$ within $M_\ext$. 
\end{cor}

\medskip
\medskip

\noindent\emph{Remark.}
The only assumption appearing in Corollary \ref{cor: extension from conformal} is that either $a(t)=e^{ht}+o(e^{ht})$ or $H(t)$ limits to a positive finite number $h$ as $t \to -\infty$. This is in agreement with the $C^0$-extendibility result from Section \ref{sec:C0}, which provided an alternative proof to why $a(t)=e^{ht}+o(e^{ht})$ or $H\to h>0$ as $t\to-\infty$ yields a $C^0$ extension. Note that the choices of extension are clearly different though: in Theorem \ref{cont ext thm}, the extension was essentially handpicked to be flat spacetime (Minkowski); here the extension is closed (global) de Sitter.

\medskip
\medskip

Now we take a closer look at the comoving observers, i.e., the integral curves of $u := \pd_t$.  In Proposition \ref{conformal embed prop}, we saw that all the comoving observers emanate from the north pole $(\chi = 0)$ at $T' = -\frac{\pi}{2}$. Note that the integral curves of $u$ are parameterized by $g$-proper time. Now consider the vector field $\frac{1}{\Omega}u$ on $M_{t_0}$, where the integral curves of $\frac{1}{\Omega}u$ are parameterized by $\wt{g}$-proper time. Next we prove the following proposition.

\medskip

\begin{prop}\label{prop: comoving observation 2}
For each integral curve of $\frac{1}{\Omega}u$, the vector field along said integral curve extends continuously to $\pd_{T'}$ at $\mc{O}$. 
\end{prop}

\medskip

\noindent\emph{Remark.} In other words, Proposition \ref{prop: comoving observation 2} says the following: let $\wt{\g} \colon (0,b) \to M$ be an integral curve of $\frac{1}{\Omega}u$; hence it is parameterized by $\wt{g}$-proper time. From Proposition \ref{conformal embed prop}, we know that $\wt{\g}(s) \to \mc{O}$ as $s \to 0^+$. Proposition \ref{prop: comoving observation 2} says that its tangent vector, $\wt{\g}'(s)$, converges to $\pd_{T'} \in T_{\mc{O}}\wt{M}$ as $s \to 0^+$ as well. And this is true for each comoving observer. In fact, this behavior can be seen in Figure \ref{flat dS in ESU fig}.

\medskip

\proof
Using \eqref{conformal trans 1} and \eqref{conformal trans 2}, the relationship between $T'$ and $\chi$ of a comoving observer at a fixed $r_0$ is given by 
\begin{equation}\label{eq: comoving T'}
     h r_0(\sin T' + \cos \chi) \,=\, \sin \chi\,.
\end{equation}
Since $T'$ is a time function for $(\wt{M}, \wt{g})$, we can parameterize the curve for the comoving observer via $T'$. With this parameterization, the comoving observer $
t \mapsto (t,r_0, \theta_0, \phi_0)$ with respect to the $(t,r, \theta, \phi)$ coordinates is given by $T' \mapsto (T', \chi(T'), \theta_0,\phi_0)$ with respect to the $(T', \chi, \theta, \phi)$ coordinates. Implicitly differentiating \eqref{eq: comoving T'} with respect to $T'$, one calculates the derivative of $\chi(T')$:
\[
    \frac{\dd\chi}{\dd T'} = \frac{ h r_0\cos T'}{\cos \chi +  h r_0 \sin \chi}\,.
\]
We can relate $ h r_0 \cos T'$ to $\chi$ via \eqref{eq: comoving T'}. Substituting this relationship into the expression for $\dd\chi/\dd T'$ gives a $\chi$-parameterization of the curve for the comoving observer. To obtain a $\wt{g}$-proper time paramterization, one normalizes the vector field along the integral curve with respect to $\wt g$. Doing this yields the following expression for the vector field along the integral curve of $\frac{1}{\Omega}u$ at constant $r_0$:
\[
    \sqrt{\frac{\cos^2\chi + 2 h r_0 \sin \chi +  h^2 r_0^2 \sin^2 \chi}{1 + 2 h r_0\sin\chi(1-\cos\chi)}}\,\pd_{T'} \,+\, \sqrt{\frac{ h^2 r_0^2 - \sin^2 \chi + 2 h r_0\cos \chi \sin \chi -  h^2 r_0^2 \cos^2 \chi}{1 + 2 h r_0\sin \chi (1-\cos \chi)}}\,\pd_\chi\,.
\] 
As $\chi \to 0^+$, the above vector field along the integral curve approaches $\pd_{T'}$. \qed

\medskip
\medskip

The next proposition and corollary will motivate an important hypothesis in Theorem \ref{thm: main nonhomog setting}.

\medskip

\begin{prop}\label{prop: ricci tensor components}
Within $M_{t_0}$, the nonzero components of the $g$-Ricci tensor, ${\rm Ric}_g$, in the $(T', \chi, \theta, \phi)$-coordinates are determined by
\begin{align*}
    \Omega^2 \,{\rm Ric}_g(\pd_{T'}, \pd_{T'}) \,&=\, -3H^2 + \frac{\dot{H}}{a^2}\left(-3\frac{\omega^2}{\cos^2 T'}(1 + \sin T' \cos \chi)^2 + \omega^2 \sin^2 \chi \right)\,,
\\
    \Omega^2 \,{\rm Ric}_g(\pd_\chi, \pd_\chi) \,&=\, 3H^2 + \frac{\dot{H}}{a^2}\left(-3\omega^2 \sin^2\chi + \frac{\omega^2}{\cos^2 T'}(1 + \sin T' \cos \chi)^2 \right)\,,
\\
    \Omega^2\,{\rm Ric}_g(\pd_{T'},\pd_\chi) \,&=\, 2\frac{\dot{H}}{a^2} \frac{\omega^2}{\cos T'}\sin\chi (1+ \sin T' \cos \chi)\,,
\end{align*}
and $\Omega^2\,{\rm Ric}_g(\pd_{\theta},\pd_\theta) = (3H^2 + \dot{H})\sin^2\chi = \Omega^2{\rm Ric}_g(\pd_\phi, \pd_\phi)/\sin^2\theta$.
\end{prop}

\proof
The proof is computational. We sketch the proof for the first equality. The rest are analogous. Set $R_{T'T'} := \text{Ric}_g(\pd_{T'}, \pd_{T'})$ and likewise with $R_{tt}$, etc. Recall that $R_{tt} = -3(H^2 + \dot{H})$, $R_{rr} = a^2(3H^2 + \dot{H})$, and $R_{tr} = 0$. Since $R_{tr} = 0$, the component transformation law gives $R_{T'T'} = R_{tt}(\frac{\pd t}{\pd T'})^2 + R_{rr}(\frac{\pd r}{\pd T'})^2$. From \eqref{conformal trans 1} and \eqref{conformal trans 2}, we have
\[
    \frac{\pd t}{\pd T'} \,=\, \frac{\pd t}{\pd \hat{t}}\frac{\pd \hat{t}}{\pd T'} \,=\, \frac{\omega}{ h \cos T'}\frac{1 + \sin T' \cos \chi}{\sin T' + \cos \chi}
\]
and
\[
    \frac{\pd r}{\pd T'} \,=\, -\frac{\cos T' \sin\chi}{ h (\sin T' + \cos \chi)^2}\,.
\]
Plug these into the component transformation law for $R_{T'T'}$, and use the fact that $a/\omega = e^{ h \hat{t}}$, $\Omega = \frac{ h \cos T'}{\omega}$, and $\frac{\dot{H}}{a^2} = \dot{H}\frac{\cos^2 T'}{\omega^2}\frac{1}{(\sin T' + \cos \chi)^2}$.
\qed

\medskip
\medskip

We say that a function $f\colon M_{t_0} \to \R$ \emph{extends continuously} to $M_{t_0} \cup \{\mc{O}\}$ if there is a continuous function $\wt{f} \colon M_{t_0} \cup \{\mc{O}\} \to \R$ such that the restriction of $\wt{f}$ to $M_{t_0}$ coincides with $f$. The topology on $M_{t_0} \cup \{\mc{O}\}$ is just the subspace topology inherited from $\wt{M}$.

\medskip

\begin{cor}\label{cor: Ric ext}
If $H$ limits to a positive finite number and $\dot{H}/a^2$ limits to a finite number as $t \to -\infty$, then the $(T', \chi, \theta, \phi)$-components of $\Omega^2{\rm Ric}_g$ extend continuously to $M_{t_0} \cup \{\mc{O}\}$.  
\end{cor}

\medskip

\noindent\emph{Remark.} The hypothesis ``$\dot{H}/a^2$ limits to a finite number as $t \to -\infty$" was the hypothesis used to obtain a $C^2$ extension in Theorem \ref{higher reg thm}. 

\medskip

\proof
Since $H$ limits to a positive finite number, $\omega$ limits to unity (see the proof of Proposition \ref{conformal embed prop}). Therefore, from Proposition \ref{prop: ricci tensor components}, it suffices to show that the function $f \colon M_{t_0} \to \R$ given by $f = (1 + \sin T' \cos \chi)/\cos T'$ extends continuously to $M_{t_0} \cup \{\mc{O}\}$. Indeed, we will show that the continuous extension is given by $\wt{f}(\mc{O}) = 0$. 

Recall that $\mc{O}$ corresponds to the point $T' = -\frac{\pi}{2}$ and $\chi = 0$.
Note that we can write
\[
    f=\frac{1+\sin T'}{\cos T'}-(1-\cos\chi)\tan T'\,.
\]
As $T'\to-\pi/2$, one can show by l'H\^opital's rule that $(1+\sin T')/\cos T'$, which is of the form $0/0$, limits to $-\cot(-\pi/2)=0$.
Regarding the second term in $f$ above, we can Taylor-expand $\tan T'$ about $T' = -\pi/2$ and $\cos \chi$ about $\chi = 0$, so for points near $\mc{O}$, we have
\[
    (1-\cos\chi)\tan T' = -\frac{\chi^2}{2(T'+\pi/2)}\left(1-\frac{\chi^2}{12}+O\left(\chi^4\right)\right)\left(1-\frac{1}{3}\left(T'+\frac{\pi}{2}\right)^2+O\left(\left(T'+\frac{\pi}{2}\right)^4\right)\right)\,.
\]
Thus we are left with showing that $\chi^2/(T'+\pi/2)$ extends continuously to $M_{t_0} \cup \{\mc{O}\}$ and takes on the value $0$ at $\mc{O}$. This is indeed the case: since $0<\chi < (T'+\pi/2)$ within $M_{t_0}$, we have $0<\chi^2/(T'+\pi/2) < \chi$ within $M_{t_0}$, which limits to $0$ at $\mc{O}$.
\qed

\section{The cosmological constant appears as an initial condition}\label{sec:inhomoaniso}

So far, we have been able to derive conditions on spacetime extendibility in past-eternal quasi-de Sitter universes, which were restricted to the symmetry assumptions of homogeneity and isotropy (FLRW). While those results are fairly exhaustive and purely geometrical --- they do not depend on the physical input, i.e., they are independent of any gravitational field equations --- they necessarily are of limited physical applicability as soon as one departs from an exact FLRW background. This is particularly relevant since, within the theory of inflationary cosmology as a proposed early universe scenario to generate the initial conditions for structure formation, there has to be primordial quantum fluctuations (hence spacetime inhomogeneities). In this section, inspired by the consequences of the analysis in Section \ref{sec:confEmbedd}, we will derive results that go beyond the isotropy and homogeneity assumptions of FLRW. In particular, in Theorem \ref{thm: main nonhomog setting}, we show that in some sense the cosmological constant appears as an initial condition for spacetimes satisfying geometrical properties described in the previous section. However, to do so we shall specify some physical input; conservatively, let us apply the Einstein field equations of classical general relativity.

\subsection{The cosmological constant appears as an initial condition for past-asymptotically de Sitter spacetimes}

As shown in \cite[Thm.~12.11]{ON}, FLRW spacetimes satisfy the Einstein equations with a perfect fluid $(u, \rho, p)$,
\begin{equation}\label{eq: perfect fluid}
    \text{Ric} - \half Rg\,=\, 8\pi T \,=\, 8\pi\big((\rho + p)u_* \otimes u_* + pg\big)\,,
\end{equation}
where $u_* = g(u,\cdot)$ is the one-form metrically equivalent to the fluid's velocity vector field $u = \pd_t$ (in the comoving frame of the fluid).
Note that we set Newton's gravitational constant to unity.
We emphasize that for FLRW spacetimes, the (total) energy density $\rho$ and pressure $p$ are purely geometrical quantities given by $\rho = \frac{1}{8\pi} G(u,u)$ and $p = \frac{1}{8\pi}G(e,e)$, where $e$ is any unit spacelike vector orthogonal to $u$ (its choice does not matter by spatial isotropy).\footnote{In the physics literature, it is more customary to define $\rho=T(u,u)$ and $p=T(e,e)$, but if the Einstein equations $G=8\pi T$ hold, then the physical and geometrical definitions are equivalent. If gravity is modified, then one has to be more careful as the definitions may deviate. In this paper, we work under the geometrical definition.} Here $G := \text{Ric} - \half Rg$ is the Einstein tensor, which is related to the energy-momentum tensor $T$ via $G = 8\pi T$. To incorporate a positive cosmological constant $\Lambda>0$, we define $T_{\rm m} = T + \frac{\Lambda}{8\pi}g$, so that the Einstein equations become
\begin{equation*}
    \text{Ric} - \half R g + \Lambda g \,=\, 8\pi T_{\rm m}\,.
\end{equation*}
Setting $\rho_{\rm m} = T_{\rm m}(u,u)$ and $p_{\rm m} = T_{\rm m}(e,e)$, we have
\begin{equation*}
    \rho  \,=\, \rho_{\rm m}+ \rho_{\Lambda} \:\:\:\:  \text{ and } \:\:\:\: p  \,=\, p_{\rm m} + p_{\Lambda}\,,
\end{equation*}
where $\rho_{\Lambda} = \frac{\Lambda}{8\pi}$ and $p_{\Lambda} = -\frac{\Lambda}{8\pi}$. Note that 
\begin{equation}\label{cosmo const eq st} 
    p_{\Lambda} \,=\, - \rho_{\Lambda}\,.
\end{equation} 
Equation \eqref{cosmo const eq st} is the \emph{equation of state} for a cosmological constant. 

For a flat ($k = 0)$ FLRW spacetime, the Friedmann equations \cite[Thm.~12.11]{ON} are given by
\begin{equation}\label{Friedmann equations}
    \frac{8\pi}{3}\rho(t) \,=\, H(t)^2 \:\:\:\: \text{ and } \:\:\:\: -8\pi p(t) \,=\, 2\frac{\ddot{a}(t)}{a(t)} + H(t)^2\,.
\end{equation}
These equations imply
\begin{equation}\label{eq: hubble, rho, and p}
    \dot{H}(t) \,=\, -4\pi\big(\rho(t) + p(t)\big)\,.
\end{equation}

\medskip

\begin{prop}\label{prop: cosmo const eq of state}
If a flat FLRW spacetime as defined in Section \ref{sec:firstPreliminaries} is past-asymptotically de Sitter (recall Definition \ref{def:padS}), then $\rho(t)$ and $p(t)$ converge as $t \to -\infty$; furthermore, if $\rho(-\infty)$ and $p(-\infty)$ denote their limits, then $p(-\infty) = -\rho(-\infty) = -\frac{3}{8\pi}H_\Lambda^2$.
Conversely, if $\rho(t)$ and $p(t)$ converge as $t \to -\infty$, with $\rho(-\infty) > 0$, in an expanding flat FLRW spacetime, then the spacetime is past-asymptotically de Sitter with $H_\Lambda$ determined by $\frac{3}{8\pi}H_\Lambda^2 = -p(-\infty) = \rho(-\infty)$.
\end{prop}

\proof
For the first part, suppose a flat FLRW spacetime is past asymptotically de Sitter. Since $\dot{H} = \ddot{a}/a - H^2$, it follows that $\ddot{a}(t)/a(t)$ converges as $t \to -\infty$. And clearly $\rho(t)$ converges from the first Friedmann equation in \eqref{Friedmann equations}. Thus $p(t)$ converges as $t \to -\infty$ as well from the second Friedmann equation in \eqref{Friedmann equations}. Then $p(-\infty) = -\rho(-\infty)$ follows from \eqref{eq: hubble, rho, and p}.

For the second part, suppose for a flat FLRW spacetime $\rho(t)$ and $p(t)$ converge as $t \to -\infty$ with $\rho(-\infty) > 0$. Then \eqref{Friedmann equations} and \eqref{eq: hubble, rho, and p} imply both $H$ and $\dot{H}$ converge as $t \to -\infty$. If we denote the limit of $H$ as $t\to-\infty$ by $H_\Lambda$,
then it must be that $H_\Lambda>0$ by \eqref{Friedmann equations} and the assumption $\rho(-\infty) > 0$. It then also follows from Lemma \ref{lem:HtohdotHtoconst} that $\dot{H}(t) \to 0$ as $t \to -\infty$. Therefore, spacetime is past-asymptotically de Sitter and also $p(-\infty) = -\rho(-\infty)$ according to \eqref{eq: hubble, rho, and p}.
\qed

\medskip
\medskip

\noindent\textit{Remarks.}
Given equation \eqref{cosmo const eq st}, we refer to $p(-\infty) = -\rho(-\infty)$ as \emph{the cosmological constant appearing as an initial condition}. An implication of the above proposition is that
all the past-asymptotically de Sitter spacetime cases discussed in previous sections were equivalent to having
a cosmological constant as an initial condition. For instance, Corollary \ref{cor:padSC0} could now be restated as follows: if $p(-\infty)=-\rho(-\infty)$, then the spacetime is $C^0$ extendible. As another example, if the hypotheses of the $C^2$ extendibility Theorem \ref{higher reg thm}(b) are respected, then certainly it means that a cosmological constant appears as an initial condition for that spacetime.

\medskip
\medskip

We generalize Proposition \ref{prop: cosmo const eq of state} in the next section (see Theorem \ref{thm: main nonhomog setting}).

\subsection{Beyond homogeneity and isotropy}

In this section, we generalize Proposition \ref{prop: cosmo const eq of state} to a nonhomogeneous setting. We show that the reason why the cosmological constant appears as an initial condition has to do with the \emph{geometry} of the conformal embedding in Proposition \ref{conformal embed prop}. It is the fact that the comoving observers all emanate from the origin point $\mc{O}$ (see Proposition \ref{conformal embed prop}) which forces the cosmological constant to appear as an initial condition; it has nothing to do with the symmetry assumptions appearing in flat FLRW spacetimes. As such our theorem has applications to inflationary scenarios in spacetimes that are not isotropic or homogeneous, which we discuss in Section \ref{sec: applications to inflationary theory}. 

We set the stage for our theorem. Let $(M,g)$ be a smooth spacetime. Let $(\wt{M}, \wt{g})$ be a $C^0$ conformal extension of $(M,g)$ with conformal factor $\Omega$ (Definition \ref{conformal extension def}). Assume $M = I^+_{\wt{g}}(\mc{O}, \wt{M})$ for some point $\mc{O} \in \pd^-_0 M$. (Here $\wt{M}$ plays the role of  $\wt{M}'$ in Proposition \ref{conformal embed prop}(e).) Let $f \colon M \to \R$ be a smooth function. We say $f$ \emph{extends continuously} to $M \cup \{\mc{O}\}$ if there is a continuous function $\wt{f} \colon M \cup \{\mc{O}\} \to \R$ such that $\wt{f}|_M = f$. In this case, we call $\wt{f}$ the \emph{continuous extension} of $f$ to $M \cup \{\mc{O}\}$. The topology on $M \cup \{\mc{O}\}$ is the subspace topology inherited from $\wt{M}$. In other words, $\wt{f}$ is continuous at $\mc{O}$ means that for each $\e > 0$ there is a neighborhood $U \subset \wt{M}$ of $\mc{O}$ such that $|\wt{f}(\mc{O}) - \wt{f}(x)| < \e$ for all $x \in U \cap (M \cup \{\mc{O}\})$. 

Likewise, a smooth vector field $X$ on $M$ \emph{extends continuously} to $M \cup \{\mc{O}\}$ provided there is a coordinate neighborhood $U \subset \wt{M}$ of $\mc{O}$ with coordinates $(x^0, \dotsc, x^3)$ such that each of the components $X^\mu$ in $X = X^\mu \pd_\mu$ extends continuously to $(U \cap M) \cup \{\mc{O}\}$. If $\wt{X}^\mu$ are the continuous extensions of $X^\mu$, then the \emph{continuous extension} of $X$ at $\mc{O}$ is simply the vector $\wt{X}^\mu(\mc{O}) \pd_\mu \in T_{\mc{O}}\wt{M}$. A similar definition applies to smooth tensors on $M$ by requiring each of its components to extend continuously. (This definition does not depend on the choice of coordinate system by the usual transformation law for tensor components.) 

For example, the conformal metric $\Omega^2 g$ on $M$ extends continuously to $M \cup \{\mc{O}\}$ by definition. For another example, suppose $T$ is a smooth tensor on $\wt{M}$, then  obviously the restriction, $T|_M$, extends continuously to $M \cup \{\mc{O}\}$ since it extends smoothly. 

The following theorem generalizes Proposition \ref{prop: cosmo const eq of state}. It is an analogue of \cite[Thm.~2.2]{Ling:2022uzs} for the past-asymptotically de Sitter spacetimes we consider in this paper. The main difference between \cite[Thm.~2.2]{Ling:2022uzs} and Theorem \ref{thm: main nonhomog setting} is the use of a conformal boundary.

\medskip

\begin{thm}\label{thm: main nonhomog setting}
Let $(\wt{M}, \wt{g})$ be a $C^0$ conformal extension of a smooth spacetime $(M,g)$ with conformal factor $\Omega$ such that $M = I^+_{\wt{g}}(\mc{O}, \wt{M})$ for some point $\mc{O} \in \pd^-_0 M$. We make the following assumptions:
\begin{itemize}
    \item[\emph{(a)}] $(M,g)$ solves the Einstein equations with a perfect fluid $(u, \rho, p)$ as described in \eqref{eq: perfect fluid}. 

    \item[\emph{(b)}] Each of the integral curves of $u$ have past endpoint $\mc{O}$ within $\wt{M}$. Moreover, the vector field along each integral curve of $\frac{1}{\Omega}u$ extends continuously to a $\wt{g}$-timelike vector at $\mc{O}$.

    \item[\emph{(c)}] The energy density $\rho$ and pressure $p$ extend continuously to $M \cup \{\mc{O}\}$. Moreover, $\Omega^2\rm{Ric}_g$ extends continuously to $M \cup \{\mc{O}\}$, where $\rm{Ric}_g$ is the Ricci tensor for $(M,g)$. 

    \item[\emph{(d)}] $(\wt{M}, \wt{g})$ is strongly causal at $\mc{O}$.
\end{itemize}
Then the continuous extensions of $\rho$ and $p$ satisfy $\wt{p} = -\wt{\rho}$ at $\mc{O}$.
\end{thm}

\medskip
\medskip

Before proving Theorem \ref{thm: main nonhomog setting}, we make some remarks.

\medskip

\noindent\emph{Remarks.}
\begin{itemize}
    \item[-]  The conclusion of Theorem \ref{thm: main nonhomog setting} is that $\wt{p}(\mc{O}) = -\wt{\rho}(\mc{O})$. This generalizes Proposition \ref{prop: cosmo const eq of state}, where we found $p(-\infty) = -\rho(-\infty)$.

    \item[-] Note that $M = I^+_{\wt{g}}(\mc{O}, \wt{M})$ holds for $M_{t_0}$ in Proposition \ref{conformal embed prop} where $\wt{M}$ in Theorem \ref{thm: main nonhomog setting} plays the role of $\wt{M}'$ in Proposition \ref{conformal embed prop}(e).
    
    \item[-] Assumption (b) in Theorem \ref{thm: main nonhomog setting} holds for $M_{t_0}$ in Proposition \ref{conformal embed prop} as well. Assumption (b) means that if $\g \colon [0, b] \to M \cup \{\mc{O}\}$ is an integral curve of $\frac{1}{\Omega}u$ on $(0,b]$ with past endpoint $\g(0) = \mc{O}$, then $\g'(0)$ is a future-directed $\wt{g}$-timelike vector and hence $\g$ is a future-directed timelike curve within $\wt{M}$. Therefore Proposition \ref{conformal embed prop} and Proposition \ref{prop: comoving observation 2} imply that assumption (b) holds for $M_{t_0}$. Using a more relaxed definition of timelike curves than the one we are using in this paper, assumption (b) in Theorem \ref{thm: main nonhomog setting} can be relaxed. For example, see assumption (b) in \cite[Thm.~2.2]{Ling:2022uzs} and the remark after it.

    \item[-] Regarding assumption (c), we saw in Proposition \ref{prop: cosmo const eq of state} that if $H$ converges and $\dot{H}$ converges to zero, then $\rho$ and $p$ converge as $t \to -\infty$. From Corollary \ref{cor: Ric ext}, we see that $\Omega^2 \text{Ric}$ extends continuously to $M \cup \{\mc{O}\}$ provided $H$ limits to a positive finite number and $\dot{H}/a^2$ limits to a finite number as $t \to -\infty$. 

    \item[-] Regarding assumption (d), recall that $(\wt{M}, \wt{g})$ is \emph{strongly causal} at $\mc{O}$ means that for any neighborhood $U$ of $\mc{O}$ there is a neighborhood $V \subset U$ of $\mc{O}$ such that for any future-directed causal curve $\g \colon [a,b] \to \wt{M}$, if $\g(a),\g(b) \in V$, then $\g\big([a,b]\big) \subset U.$ This assumption holds for any globally hyperbolic spacetime and hence holds for the Einstein static universe appearing in Proposition \ref{conformal embed prop}.
\end{itemize}

\medskip
\medskip

\newpage

\noindent\underline{\emph{Proof of Theorem \emph{\ref{thm: main nonhomog setting}}}}.

\smallskip

The proof only requires some modifications to the proof of \cite[Thm.~2.2]{Ling:2022uzs}. However, for completeness, we provide all the details.

Since $(M,g)$ solves the Einstein equations with a perfect fluid $(u, \rho, p)$, the Einstein tensor satisfies equation \eqref{eq: perfect fluid}. Recall that $u_*$ is the one-form which is $g$-metrically equivalent to the vector field $u$ (i.e., in coordinates, $u_* = g_{\mu\nu}u^\mu\dd x^\nu$, and recall $g_{\mu\nu}u^\mu u^\nu=-1$). Contracting this equation gives an expression for the scalar curvature $R_g$ in terms of $\rho$ and $p$. Plugging this expression back into \eqref{eq: perfect fluid}, we have 
\begin{equation*}
    \text{Ric}_g \,=\, 8\pi\big((\rho + p)u_* \otimes u_* + pg\big) + 4\pi(\rho - 3p)g\,.
\end{equation*}
Multiplying by $\Omega^2$ gives
\begin{equation}\label{eq: Omega squared Ricci}
    \Omega^2 \text{Ric}_g \,=\, 8\pi\big((\rho + p)(\Omega u_*) \otimes (\Omega u_*) + p\Omega^2g\big) + 4\pi(\rho - 3p)\Omega^2g\,.
\end{equation}

Seeking a contradiction, suppose $\wt{p}(\mc{O}) \neq -\wt{\rho}(\mc{O})$. Then there is a coordinate neighborhood $U \subset \wt{M}$ of $\mc{O}$ such that $\wt{\rho} + \wt{p} \neq 0$ for points in  $U \cap (M \cup \{\mc{O}\})$. Therefore, within this set, we can solve for $(\Omega u_*) \otimes (\Omega u_*)$ in the above equality, which yields
\[
    (\Omega u_*) \otimes (\Omega u_*) \,=\, \frac{\frac{1}{8\pi}\Omega^2 \text{Ric}_g - \frac{1}{2}(\rho - p)\Omega^2 g}{\rho + p}\,.
\]
Since $\wt{g} = \Omega^2 g$, and since $\wt{g}$ and $\Omega^2\mathrm{Ric}_g$ extend continuously to $M \cup \{\mc{O}\}$, it follows that the right-hand side of the above equality extends continuously to $M \cup \{\mc{O}\}$. Therefore so does the left-hand side. 

Let $S$ denote the continuous extension of $(\Omega u_*) \otimes (\Omega u_*)$ to $M \cup \{\mc{O}\}$. Let $(x^0, \dotsc, x^3)$ denote the coordinates on $U$. Let $S_{\mu\nu} =  S(\pd_\mu, \pd_\nu)$. Then $S_{\mu\nu} = \Omega^2 u_\mu u_\nu$ within $U \cap M$, where $u_\mu = g_{\mu\nu}u^\nu$ are the components of $u_*$. Define a one form $\alpha$ on $M \cup \{\mc{O}\}$ via  $\alpha|_M = \Omega u_*$ and the extension 
\[
\alpha_\mu(\mc{O}) \,=\, \left\{
\begin{array}{ll}
      +\sqrt{S_{\mu\mu}(\mc{O})} & \text{ if } S_{\mu\mu}(\mc{O}) \neq 0 \text{ and } \Omega u_\mu > 0 \text{ near } \mc{O}\,; \\
      -\sqrt{S_{\mu\mu}(\mc{O})} & \text{ if } S_{\mu\mu}(\mc{O}) \neq 0 \text{ and } \Omega u_\mu < 0 \text{ near } \mc{O}\,;
      \\
      0 & \text{ if } S_{\mu\mu}(\mc{O}) = 0\,.
\end{array} 
\right.
\]
Then $\alpha$ is a continuous extension of $\Omega u_*$ to $M \cup \{\mc{O}\}$. Now consider the vector field $X$, defined on $M \cup \{\mc{O}\}$, which is $\wt{g}$-metrically equivalent to the continuous extension of $\Omega u_*$.  Using components via a coordinate system in $M$, it is easy to see that, within $M$, $X$ is just the vector field $\frac{1}{\Omega}u$: 
\[
    X=\wt{g}^{\a\nu}(\Omega g_{\mu\nu}u^\mu)\pd_\a \,=\, \Omega^{-2}g^{\a\nu}(\Omega g_{\mu\nu}u^\mu)\pd_\a \,=\, \Omega^{-1}u^\a \pd_\a\,. 
\]
Therefore $X$ is a continuous extension of $\frac{1}{\Omega}u$ to $M \cup \{\mc{O}\}$. 

Since $g(u, u) = -1$ (by definition of a perfect fluid), continuity implies $\wt{g}(X, X) = -1$ at $\mc{O}$. Using \cite[Lem.~2.9]{Ling:2019cac} and applying the Gram-Schmidt orthogonalization process appropriately, for any $0 < \e <1$, we can assume that the coordinates $(x^0, \dotsc, x^3)$ on $U$ satisfy assumptions (1)--(6) below:
\begin{itemize}
    \item[(1)] $\pd_0|_{\mc{O}} = X|_{\mc{O}}$,
    \item[(2)] $x^0$ is a time function on $U$,
    \item[(3)] $\wt{g}_{\mu\nu}(\mc{O}) = \eta_{\mu\nu}$ and $|\wt{g}_{\mu\nu}(x) - \eta_{\mu\nu}| < \e$ for all $x \in U$ where $\wt{g}_{\mu\nu} = \wt{g}(\pd_\mu, \pd_\nu)$.
\end{itemize}
Here $\eta_{\mu\nu}$ are the usual components of the Minkowski metric with respect to the coordinates $(x^0, \dotsc, x^3)$, that is,
\[
    \eta \,=\, \eta_{\mu\nu}\dd x^\mu \dd x^\nu \,=\, -(\dd x^0)^2 + h_{\mathbb{E}}\,,
\]
where as before $h_{\mathbb{E}}=\delta_{ij}\dd x^i\dd x^j$.
By choosing $U$ even smaller, similar arguments as in \cite[Lem.~2.9]{Ling:2019cac} show that we can also assume that
\begin{itemize}
    \item[(4)] $\eta^\e(Y,Y) \leq 0 \,\Longrightarrow\, \wt{g}(Y,Y) < 0$ for all nonzero $Y \in T_p\wt{M}$ whenever $p \in U$,
\end{itemize}
where $\eta^\e$ is the narrow Minkowskian metric on $U$ given by
\[
    \eta^\e \,=\, -\frac{1-\e}{1+\e}(\dd x^0)^2 + h_{\mathbb{E}} \,=\, \eta + \frac{2\e}{1+\e}(\dd x^0)^2\,.
\]
Moreover, since $X$ is a continuous extension of $\frac{1}{\Omega}u$ to $M \cup \{\mc{O}\}$, we can also assume that
\begin{itemize}
    \item[(5)] $\left|X^\mu (x) - X^\mu(\mc{O})\right| < \frac{\e}{2}$ for all $x \in U \cap (M \cup \{\mc{O}\})$.
\end{itemize}
Lastly, if $\phi \colon U \to \R^{3+1}$ denotes the coordinate map [i.e., $\phi = (x^0, \dotsc, x^3)$], then, by restricting the domain of $\phi$, we can assume that 
\begin{itemize}
\item[(6)] $\phi(U) = B_{2r}$ where $B_{2r} \subset \R^{3+1}$ is an open ball with radius $2r > 0$ (as measured by the Euclidean metric $\delta = \delta_{\mu\nu}\dd x^\mu \dd x^\nu$ on $U$) centered at the origin: $\phi(\mc{O}) = (0, 0,0, 0)$.  
\end{itemize}

Choose $\e = \frac{3}{5}$. Then $\eta^\e$ has lightcones with `slope' $2$  in these coordinates. Consider the curve $c \colon [0, r] \to B_{2r}$ lying on this narrow lightcone defined by $c(s) = (s, \frac{s}{2}, 0, 0)$. By (4), the curve $\phi^{-1}\circ c(s)$ is future-directed timelike. Let $q = \phi^{-1} \big(c(r)\big)$. Since $M = I^+_{\wt{g}}(\mc{O}, \wt{M})$, it follows that $q \in M$. Let $\g\colon [0,b] \to M \cup \{\mc{O}\}$ denote the integral curve of $\frac{1}{\Omega}u$, i.e., $\g'(t) = X|_{\g(t)}$ on $(0,b]$, with future endpoint $\g(b) = q$ and past endpoint $\g(0) = \mc{O}$. Note that $t$ is the $\wt{g}$-proper time of $\g$.

We can assume $\g\big([0,b]\big) \subset U$. This follows by strong causality of $(\wt{M}, \wt{g})$ at $\mc{O}$. To see this, note that strong causality implies that there is a neighborhood $V \subset U$ of $\mc{O}$ such that if $\g$ has endpoints in $V$, then the image of $\g$ is contained in $U$. Let $V' \subset V$ denote a neighborhood of $\mc{O}$ satisfying assumption (6) above. Then we work in $V'$ to construct the curve $\g$ in exactly the same way as in the paragraph before this one. Then strong causality implies that the image of $\g$ is contained in $U$. 

Since $\g\big([0,b]\big) \subset U$ it follows by (2) that we can reparameterize $\g$ by $x^0$. Let $\bar{\gamma} \colon [0,r] \to M \cup \{\mc{O}\}$ be the reparameterization of $\g$ by $x^0$. Then
\[
    \bar{\g}(s) \,=\, \g \circ (x^0 \circ \g)^{-1}(s)\,, \:\:\:\: \text{ where } \:\:\:\: x^0 \circ \g(t) \,=\, \int_0^t \dd\tilde t\,\frac{\dd(x^0 \circ \g)}{\dd \tilde t}\,.
\]
Note that  $\bar{\gamma}(0) = \mc{O}$ and $\bar{\gamma}(r) = q$. Since $\phi(q) = (r, \frac{r}{2}, 0, 0)$, the mean value theorem implies that there exists a $s_* \in (0,r)$ such that $(x^1 \circ \bar{\gamma})'(s_*) = \frac{1}{2}$. Set  $\gamma^\mu = x^\mu \circ \gamma$ and $\bar{\gamma}^\mu = x^\mu \circ \bar{\gamma}$. Using the fact that $t$ and $s = x^0 \circ \g$ are inverses of each other, the chain rule gives
\[
    \frac{1}{2} \,=\, \frac{\dd \bar{\gamma}^{1}}{\dd s}(s_*) \,=\, \frac{\dd\g^1}{\dd t}\big(t(s_*)\big)\frac{\dd t}{\dd s}(s_*)\,=\, \frac{\dd\gamma^1/\dd t}{\dd\g^0 /\dd t}\big(t(s_*)\big) \,=\, \frac{X^1}{X^0}\big(\bar{\g}(s_*)\big)\,.
\]
However, by (1) and (5), we have
\[
    \sup_{x \in U \cap (M \cup \{\mc{O}\})}\,\frac{X^1}{X^0}(x) \,\leq\, \frac{0 + \e/2}{1 - \e/2} \,=\, \frac{3}{7} \,<\, \frac{1}{2}\,,
\]
which is a contradiction.
\qed

\medskip
\medskip

\begin{cor}\label{cor: Ricci is Einstein}
Assume the hypotheses of Theorem \emph{\ref{thm: main nonhomog setting}}. Then the continuous extension of $\Omega^2 {\rm Ric}_g$ at $\mc{O}$ is given by $8\pi \wt{\rho}\,\wt{g}$ at $\mc{O}$.
\end{cor}

\proof
The proof is analogous to \cite[Cor.~2.5]{Ling:2022uzs}. Essentially, it follows from ``plugging in" $\wt{\rho}(\mc{O}) +\wt{p}(\mc{O}) = 0$ into \eqref{eq: Omega squared Ricci}. However, one needs to be careful to control $(\Omega u_*) \otimes (\Omega u_*)$. To be rigorous, consider a coordinate neighborhood $U$ of $\mc{O}$. Within $U \cap M$, we have
\begin{equation}\label{eq: Ricci raised}
    \wt{g}^{\a\mu}\wt{g}^{\beta \nu}\Omega^2 R_{\mu\nu} \,=\, 8\pi\left((\rho+ p)\left(\frac{1}{\Omega}u^\a\right)\left(\frac{1}{\Omega}u^\beta\right) + p \wt{g}^{\a\b}\right) + 4\pi\left(\rho - 3p \right)\wt{g}^{\a\b}\,,
\end{equation}
where $u^\a = g^{\a\mu}u_\mu$. Fix an integral curve $\g$ of $\frac{1}{\Omega}u$ and evaluate \eqref{eq: Ricci raised} along $\g$. (It is sufficient to consider the approach along a single curve $\g$ since we are already assuming that $\Omega^2 \text{Ric}_g$ extends continuously to $M \cup \{\mc{O}\}$.) By assumption (b) of Theorem \ref{thm: main nonhomog setting}, it follows that the components $\frac{1}{\Omega}u^\a$ are bounded on approach to $\mc{O}$ along $\g$. Therefore $(\rho + p)(\frac{1}{\Omega}u^\a)(\frac{1}{\Omega}u^\b)$ vanishes as one approaches $\mc{O}$ along $\g$. The result follows.
\qed

\subsection{Applications to inflationary theory}\label{sec: applications to inflationary theory}

The results in the previous section have applications to inflationary theory in a nonhomogeneous setting. We will show that the geometrical assumptions appearing in Theorem \ref{thm: main nonhomog setting} will imply inflationary scenarios under favorable conditions. This section is analogous to \cite[Sec.~3]{Ling:2022uzs}, so we will be brief with the details. 

Let $M \hookrightarrow \wt{M}$ be a $C^0$ conformal embedding satisfying the hypotheses of Theorem \ref{thm: main nonhomog setting}. Set $\mathscr{H} = \frac{1}{3}\text{div}_g(u)$. This $\mathscr{H}$ is a generalization of the Hubble parameter, $H$, in FLRW spacetimes. (Recall $u$ is the four-velocity of the perfect fluid.)  Following \cite{Ellis_repub}, we define an \emph{average length scale} $\mathfrak{a}(t)$ along the flow lines of $u$ via $\dot{\mathfrak{a}}/\mathfrak{a} = \mathscr{H}$, where a dot denotes a derivative with respect to the proper time $t$ of the flow lines, i.e., generally $\dot{\mathfrak{a}}:=\nabla_u\mathfrak{a}$. The average length scale, $\mathfrak{a}(t)$, is a generalization of the scale factor, $a(t)$, for FLRW spacetimes. The generalization of Friedmann's second equation is the \emph{Raychaudhuri equation},
\begin{equation}\label{eq: Raychaudhuri}
    3\frac{\ddot{\mathfrak{a}}}{\mathfrak{a}} \,=\, -\text{Ric}_g(u,u) + 2\omega^2 - 2\sigma^2 +\mathrm{div}_g \dot{u}\,,
\end{equation}
where $\omega$ and $\sigma$ are the \emph{vorticity} and \emph{shear} scalars, respectively, and $\dot{u} = \nabla_u u$ is the \emph{acceleration} of the flow.
If $u$ were hypersurface orthogonal (as in FLRW spacetimes), then we would have $\omega = 0$ and $\mathscr{H}$ would be the mean curvature of the corresponding spatial slices. Also, if $u$ were a geodesic vector field (as in FLRW spacetimes), then the acceleration would vanish and so would the divergence term (i.e., $\mathrm{div}_g\dot u=0$).

From Corollary \ref{cor: Ricci is Einstein}, it follows by continuity that for points in $M$ that are sufficiently close to $\mc{O}$, we have $\text{Ric}_g(u,u) \approx 8\pi \wt{\rho}(\mc{O}) g(u,u) = - 8\pi\wt{\rho}(\mc{O})$. Plugging this into
\eqref{eq: Raychaudhuri} and not making the assumptions of hypersurface orthogonality and vanishing acceleration gives
\begin{equation}\label{eq: approx Raych}
    3\frac{\ddot{\mathfrak{a}}}{\mathfrak{a}} \,\approx\, 8\pi\wt{\rho}(\mc{O}) + 2\omega^2 - 2\sigma^2 + \mathrm{div}_g\dot u\,.
\end{equation}
We thus see that if the shear is sufficiently small and the divergence of the acceleration is not too negative, so that
\begin{equation}
    8\pi \wt{\rho}(\mc{O}) + 2\omega^2 \gtrsim 2\sigma^2-\mathrm{div}_g\dot u\,,\label{eq:inflationApproxCond}
\end{equation}
then \eqref{eq: approx Raych} yields an accelerated expansion, $\ddot{\mathfrak{a}} > 0$, for points in $M$ that are sufficiently close to $\mc{O}$. This is a generalization of an inflationary era. 

Let us discuss the potential sources on the right-hand side of \eqref{eq:inflationApproxCond}.
A consequence of the spatial isotropy for FLRW spacetimes is that its shear vanishes: $\s^2 = 0$. In fact, the shear is zero for any flow with uniform expansion. In this sense, assuming $2\sigma^2$ is small in \eqref{eq:inflationApproxCond} can be thought of as an approximation to the spatial isotropy associated with FLRW spacetimes.
In other words, ignoring inhomogeneities, the background should be closer to flat FLRW than, e.g., Bianchi I for anisotropies and thus shear to be sufficiently small. Indeed, Bianchi I can be seen as an average FLRW background with the addition of shear anisotropies, and the FLRW limit is precisely the one where the matter energy density dominates over shear; see, e.g., \cite{Ganguly:2021pke} and references therein for such a discussion.

We see that the vorticity $\omega^2$ in \eqref{eq:inflationApproxCond} generally `helps' successfully starting inflation. It is a well-known fact that vorticity in fact generally mitigates spacetime singularities, but the absence of evidence for vorticity in our universe today indicates that the presence of vorticity in the very early universe is unlikely (see, e.g., \cite{ellis_maartens_maccallum_2012}).

Finally, the divergence of the acceleration in \eqref{eq:inflationApproxCond} can be sourced by the presence of (not necessarily linear) inhomogeneities. To see this, it is useful to express the divergence of the acceleration as
\begin{align*}
    \mathrm{div}_g\dot u=-\vec{\nabla}\cdot\left(\frac{\vec{\nabla}p}{\rho+p}\right)&=-\frac{1}{p+\rho}\left(\nabla^2\delta p-\frac{|\vec{\nabla}\delta p|^2+\vec{\nabla}\delta p\cdot\vec{\nabla}\delta\rho}{p+\rho}\right)\\
    &=-\frac{c_\mathrm{s}^2}{p+\rho}\left(\nabla^2\delta\rho-\frac{1+c_\mathrm{s}^2}{p+\rho}\big|\vec{\nabla}\delta\rho\big|^2\right)\,,
\end{align*}
where $\vec{\nabla}$ and $\nabla^2$ denote the spatial derivative and Laplacian, respectively. The first relation above is straightforward to derive from the momentum conservation equation for a perfect fluid (see, e.g., \cite{ellis_maartens_maccallum_2012}). In the second equality above, we are expressing the pressure and energy density into homogeneous and inhomogeneous parts, i.e., $p(t,\vec{x})=\bar p(t)+\delta p(t,\vec{x})$ and $\rho(t,\vec{x})=\bar\rho(t)+\delta\rho(t,\vec{x})$, without necessarily assuming $\delta p$ and $\delta\rho$ to be smaller than their homogeneous parts.
In the last equality, we assume the inhomogeneous parts are related by a spatially constant parameter as $\delta p=c_\mathrm{s}^2\delta\rho$, which would hold if $\delta p$ and $\delta\rho$ were small linear perturbations --- in this case the linearized Einstein equations would show that $\delta\rho$ satisfies a modified wave equation, where $c_\mathrm{s}$ can be interpretated as the fluid's sound speed. Inhomogeneities are typically assumed to be much smaller than the homogeneous background (to meet observational constraints), but without this assumption at early times, the divergence of the acceleration above could be large\footnote{Note that the term proportional to $|\vec{\nabla}\delta\rho|^2$ generally makes the divergence of the acceleration positive, so it would favor inflation in light of \eqref{eq:inflationApproxCond}. The term proportional to $\nabla^2\delta\rho$ may oscillate though.}, especially since it involves $1/(p+\rho)$ and $p\to-\rho$ at the origin. However, precisely because the equation of state approaches a cosmological constant at the origin, one may expect inhomogeneities to also vanish in that limit. Yet, Theorem \ref{thm: main nonhomog setting} does not tell us anything about the limit of $\vec{\nabla}p$ and $\vec{\nabla}\rho$ --- it would require knowledge of the derivatives, i.e., a higher regularity result, which we do not have at this point.

\section{Conclusions}\label{sec:conclusions}

\subsection{Summary}

In this paper, we investigated different aspects of initial spacetime singularities in the context of inflationary cosmology. In particular, in Sections \ref{sec:metricext} and \ref{sec:confEmbedd}, we obtained several interesting conditions on spacetime extendibility in flat past-eternal quasi-de Sitter universes where homogeneity and isotropy assumptions are valid (i.e., flat FLRW spacetimes). These results are fairly exhaustive and purely geometrical --- they do not depend on the physical input. While the central theme of our work focused on flat FLRW spacetimes, in Section \ref{sec:inhomoaniso} we derived, without the isotropy and homogeneity assumptions of FLRW, that in some sense a cosmological constant appears as an initial condition. Here we provide a detailed summary of our results.

Within flat FLRW, we showed that if inflation were to be past eternal, i.e., if the universe had been expanding at an accelerating rate from its inception, then the only situation that might be free of scalar curvature singularities is when the universe extends to infinite proper time in the past (Classification \ref{itm:case2b}). When spatial sections reach zero size in finite proper time in the past, a scalar curvature singularity is necessarily reached, whether the scale factor is decelerating (Classification \ref{itm:case1a}; see Theorem \ref{thm:pre-inf-sing}) or accelerating (Classification \ref{itm:case2a}; see Proposition \ref{prop:cases2a}) initially. The prototypical example of an inflationary universe within Classification \ref{itm:case2b} is the flat foliation of de Sitter spacetime, where the scale factor follows an exponential function of time, whence the Hubble rate reaches a constant. This motivated us to consider situations that are close to de Sitter (quasi-de Sitter), or often more specifically, flat past-asymptotically de Sitter spacetimes according to our Definition \ref{def:padS}. Such spacetimes are expected to be always free of scalar curvature singularities, but whether they are extendible beyond the past boundary at $t=-\infty$ is one of the main research questions of this work. This paper investigated low regularity extensions. If the extended metric is $C^2$ (or just $C^{1,1}$), existence and uniqueness of geodesics emanating from the past boundary into the original spacetime ensures geodesic incompleteness of that patch. However, the geodesics extend beyond the boundary, and they could be complete in the extended spacetime.

Building on the fact that flat dS in FLRW coordinates can be extended beyond the past boundary using different coordinates [dubbed the $(\lambda,v)$ coordinates], we derived sufficient asymptotic conditions on the extendibility of quasi-de Sitter metrics within $(\lambda,v)$ coordinates. In particular, $a(t)=e^{ht}+o(e^{ht})$ or $H\to h$ as $t\to-\infty$ for some $h>0$ ensures continuous ($C^0$) extendibility (see Theorem \ref{cont ext thm}). For higher regularity, $a(t)=e^{ht}+o(e^{(2+\e)ht})$ and $a(t)=he^{ht}+o(e^{(2+\e)ht})$ as $t\to-\infty$ for some $\e>0$, guarantees $C^1$ extendibility, while the convergence of $\dot H/a^2$ yields $C^2$ extendibility (see Theorem \ref{higher reg thm}). Furthermore, we discussed how these conditions can be generalized to $C^k$ extendibility, $k\in\mathbb{Z}_{\geq 3}\cup\{\infty\}$. In particular, when the quantity $\dot H/a^2$ can be reexpressed as a function of the scale factor, smooth extensions follow if, for instance, $\dot H/a^2$ is real analytic in a neighborhood of the past boundary (see Theorem \ref{thm:CkdotHoa2ofa}). An example includes $a(t)=e^t+e^{3t}$.

Let us stress again at this point that the above results did not presuppose any gravitational theory, i.e., no field equations were ever solved, so the results were purely geometrical. In fact, an alternative geometrical method was used to prove that past-asymptotically dS spacetimes are continuously extendible, by showing that they could be conformally embedded into the Einstein static universe.
Indeed, a spacetime close enough to dS in the asymptotic past can be conformally embedded into the Einstein static universe just like exact dS, and as such, their past conformal boundaries share similar geometrical and topological properties (see Proposition \ref{conformal embed prop} and Corollary \ref{cor: extension from conformal}). A fundamental geometric property for these spacetimes is that the comoving observers (i.e., the integral curves of $u = \pd_t$ in comoving coordinates) all emanate from a single origin point $\mc{O}$ in the past conformal boundary.

As mentioned before, if one assumes that the Einstein field equations are respected with a perfect fluid, then we were able to show a result that did not rely on the strong symmetry assumptions of FLRW, i.e., it applied for metrics that may be inhomogeneous and anisotropic. Specifically, the statement (Theorem \ref{thm: main nonhomog setting}) stipulated that if the spacetime admits a conformal extension satisfying the fundamental geometric property alluded to in the above paragraph, then the equation of state at the origin must be that of a cosmological constant.
Applications of Theorem \ref{thm: main nonhomog setting} to inflationary theory in a nonhomogeneous setting were discussed in Section \ref{sec: applications to inflationary theory}.

\subsection{Discussion and future directions}\label{sec:discussion}

Although Theorem \ref{thm: main nonhomog setting} is applicable to spacetimes which are not homogneeous, the only examples of spacetimes that we are aware of which satisfy its hypotheses are past-asymptotically de Sitter spacetimes, i.e., ones that are already isotropic. Therefore, an important mathematical question is to find examples of spacetimes satisfying the hypotheses of Theorem \ref{thm: main nonhomog setting}, which do not belong to the class of past-asymptotically de Sitter spacetimes. The same question remains in the Milne-like case as well \cite{Ling:2022uzs}.

Physical insight can be gained by examining the contrapositive of Theorem \ref{thm: main nonhomog setting}: if the `initial' equation of state at the origin is not that of a cosmological constant, then under the premise at least one of the assumptions (a)--(d) must break down. If we enforce the Einstein equations (more on this below), reasonable causality, and that the integral curves of $u$ all emanate from the origin, this leaves us with conditions pertaining to the extendibility of physically meaningful (dimensionful) quantities such as the energy density, the pressure, and the Ricci tensor.\footnote{Finding a deeper meaning to why it is the quantity $\Omega^2\mathrm{Ric}_g$ (and not just, e.g., $\mathrm{Ric}_g$ or the conformally transformed Ricci tensor) that enters in the hypothesis of the theorem eludes us. This could be investigated in future work.} Consequently, a possible conclusion in such a case is that such quantities diverge at the origin. While this is not an \emph{inextendibility} result on its own, it certainly suggests that any universe with a pointlike past conformal boundary other than those `beginning' with a cosmological constant might well be inextendible (at least in a conformal sense). If this proves to be true, it may be interpreted from different viewpoints: if general relativity always holds and if inflation is fundamentally past eternal, then it either means that a cosmological constant is `selected' as the only viable initial condition (and thus the spacetime is extendible), or it means that an initial singularity is truly unavoidable.

The latter possibility prevails from a dynamical point of view. Indeed, within general relativity, the regime of small spatial volume and large spacetime curvature (and ultimately the approach to a singularity) is generally expected to be highly anisotropic. This is known as the Belinski-Khalatnikov-Lifshitz (BKL) chaotic mixmaster approach to a singularity \cite{Belinsky:1970ew,Misner:1969hg,belinski_henneaux_2017}. In a similar fashion to why we may expect a cosmological constant to dominate at late times (cf.~the cosmic no-hair conjecture; see, e.g., \cite{Wald:1983ky,Barrow:1984zz,Kitada:1992uh,Maleknejad:2012as,Andreasson:2013pga,Kleban:2016sqm,Mirbabayi:2018suw,Creminelli:2019pdh,Creminelli:2020zvc,Wang:2021hzv,Azhar:2022yip}), we may expect exactly the opposite at early times. Let us consider a Bianchi I metric in general relativity with a perfect fluid for instance. The Hamilton constraint equation implies that the average expansion rate squared, $\mathscr{H}^2$, is proportional to the sum of the shear anisotropies, where $\sigma^2\propto \mathfrak{a}^{-6}$, and of the matter energy densities, where $\rho\propto \mathfrak{a}^{-4}, \mathfrak{a}^{-3}, \mathfrak{a}^{0}$ for radiation, dust, and a cosmological constant, respectively; $\mathfrak{a}$ is the averaged scale factor. Therefore, while the energy density of a cosmological constant dominates as $\mathfrak{a}\to\infty$, it is easily overtaken by other matter components --- and eventually anisotropies --- as $\mathfrak{a}\searrow 0$. In fact, it might be that, from a dynamical point of view (and within general relativity), the slightest perturbations preclude the existence of realistic spacetimes satisfying the conditions of a flat past-asymptotically dS metric. In other words, inflation happening all the way to $a\searrow 0$ as $t\to-\infty$ is quite special, `infinitely tuned' in the sense that the energy density of any matter component (i.e., the slightest perturbation) would become dominant (and ultimately blow up) over the energy density of the inflaton (approximately constant as for a cosmological constant) in that limit. As soon as these matter components dominate (standard matter in the sense that it respects the strong energy condition for instance), then we necessarily end up in Case \ref{itm:case1a}, with a decelerating pre-inflationary phase and an initial big bang-like scalar curvature singularity.\footnote{Assuming the Bousso bound on entropy \cite{Bousso:1999xy} (which is motivated by holography and quantum gravity \cite{Bousso:2002ju}), a general (semi-classical) singularity theorem has been proved \cite{Bousso:2022cun}, from which the following corollary follows: the slightest radiation added to global de Sitter sufficiently far into the future or past results in a null geodesically incomplete spacetime. As expected, the energy density in radiation will get to dominate over the cosmological constant in the approach to the throat and disrupt the global de Sitter bounce.} This can be interpreted as saying that inflation is unlikely to `start' the universe\footnote{How likely it is that the universe becomes inflationary is somewhat of a related issue; see, e.g., \cite{Goldwirth:1991rj,Brandenberger:2016uzh} for standard reviews, as well as some of the more recent developments from a numerical perspective \cite{Easther:2014zga,East:2015ggf,Clough:2016ymm,Clough:2017efm,Aurrekoetxea:2019fhr,Joana:2020rxm,Corman:2022alv,Garfinkle:2023vzf} or from a mathematical perspective \cite{Kleban:2016sqm,Mirbabayi:2018suw,Creminelli:2019pdh,Creminelli:2020zvc,Wang:2021hzv}.}; alternatively, it might mean that some fundamental principle needs to `select inflation' as the `initial condition' (such as finite quantum amplitudes; see \cite{Lehners:2019ibe,Jonas:2021xkx}).

It is worth noting that our analysis in this paper is at the level of classical gravity. However, in the approach to a singularity, classical physics might break down due to large quantum effects, or at the very least from an effective field theory point of view, higher-curvature terms can start becoming important. This is particularly relevant in the high-curvature regime that might correspond to the past of inflation.
Yet, most of our results in Sections \ref{sec:3cases} to \ref{sec:confEmbedd} solely relied on having a real, Lorentzian metrical theory of gravity, so they are not limited to GR and could be readily applicable to effective theories of gravity with higher derivatives too. Even in Section \ref{sec:inhomoaniso}, where we explicitly relied on the Einstein field equations, we may conjecture that the results could be generalized to theories with higher curvature terms. For instance, if hypothesis (a) of Theorem \ref{thm: main nonhomog setting} is replaced by the field equations of whatever higher-curvature theory (e.g., involving the Riemann tensor, contractions thereof, and derivatives thereof, all raised to some higher powers), then as long as the appropriate higher-curvature terms are assumed to be continuously (or potentially smoothly) extendible in hypothesis (c), then we may conjecture the same result will follow, namely $\wt p(\mathcal{O})=-\wt\rho(\mathcal{O})$. This is interesting since novel solutions exist in higher-curvature theories. For instance, in quadratic gravity one can have solutions with constant Hubble parameter $H$ and constant shear $\sigma^2$, so the solution is de Sitter-like, but anisotropic \cite{Barrow:2006xb,Barrow:2005qv,Barrow:2009gx,Middleton:2010bv}, though such solutions are unlikely to be past dynamical attractors \cite{Barrow:2006xb,Muller:2017nxg}.

Certainly, the issues of extendibility, geodesic incompleteness, and spacetime singularity change if the notion of spacetime continuum as we know it completely breaks down or, perhaps less drastically, simply if it is no longer real and Lorentzian.\footnote{Even most conservatively, at the level of quantum field theory on curved spacetime, the notions of extendibility and completeness change. For aspects of quantum completeness in the context of inflation, see \cite{Hofmann:2015xga,Hofmann:2019dqu}.}$^{,}$\footnote{There still are singularity theorems within non-smooth, low-regularity geometry constructions of spacetime (see, e.g., \cite{Kunzinger:2014fka,Kunzinger:2015gwa,Graf:2017tli,Grant:2018jtw,Alexander:2019qcd,Graf:2019wuk,Kunzinger:2021das,Steinbauer:2022hvq,Cavalletti:2022phz,McCann:2023egg} and references therein), but it is currently unknown whether such geometries could find applications in quantum gravity.} For example, the Hartle-Hawking no-boundary proposal is a well-known example of quantum prelude to inflation, in which the `pre-inflationary phase' is conjectured to be Riemannian \cite{Hartle:1983ai} (or at least, the metric may have a complex transition from Riemannian to Lorentzian --- see, e.g., \cite{Jonas:2022uqb} and references therein for how such complex metrics might be theoretically constrained). Nevertheless, classical physics is still relevant in the context of quantum cosmology: if we imagine performing a semi-classical path integral of gravity (as a proxy for the unknown theory of quantum gravity), the leading-order `tree-level' contribution is often the classical on-shell action, i.e., the action evaluated on classical solutions to the equations of motion. Determining whether the classical solutions are past extendible or inextendible is therefore crucial since the conclusions may depend significantly on the nature of the classical solutions. For instance, it was found in \cite{Lehners:2019ibe,Jonas:2021xkx} that non-singular bouncing or cyclic universes that are geodesically complete often have divergent classical on-shell actions (thus such solutions would have to be disregarded from a quantum viewpoint), while geodesically incomplete accelerating solutions emanating of an initial singularity in higher-curvature gravitational theories have finite classical on-shell actions (thus they are well defined).\footnote{Alternatively, \cite{Jia:2022gzo} claims that classically singular solutions should not be included in gravitational path integrals. (See \cite{Cotler:2023eza} for a concrete de Sitter quantum gravity analysis.) Either way, it still remains crucial to correctly determine whether or not a given classical solution is singular and/or (in)extendible.} In this work, we certainly confirm that an accelerating scale factor that reaches $0$ in finite time has an initial scalar curvature singularity, but we have shown when it reaches $0$ in infinite time some asymptotic conditions (e.g., corrections to $e^{ht}$) lead to extendible spacetimes. The issue of quantum transition amplitudes in such universes could be an interesting follow-up subject.

An aspect of our work that we have touched upon but have not explored in great detail is the possible forms of the extensions when the spacetime is extendible. As we mentioned several times, if the spacetime is extendible beyond its past boundary, the choice of extension is often not unique. As an example, the spacetime extension chosen in Theorem \ref{cont ext thm} (which applies, e.g., for the continuous extension of past-asymptotically de Sitter spacetimes) corresponds to Minkowski; however, many different choices could have been taken. A Minkowski extension is also what we showed for the warm-up in exact de Sitter (Section \ref{flat de Sitter sec}). It is quite clear, though, that for exact de Sitter we could have taken the extended metric and manifold to be copies of the original metric and manifold (so the extension is a sort of spacetime `reflection'). The continuous extension would follow just as well, and in fact, as already alluded to, the resulting extension would be the smooth extension corresponding to global de Sitter (this is shown in \cite{Yoshida:2018ndv}). For quasi-de Sitter spacetimes that admit extensions, we conjecture that a similar `reflected' extension may work in general; proving this and analyzing the geometrical and topological properties of the resulting extended spacetime could be the subject of a follow-up work.\footnote{First steps in that direction may be found in \cite{Nishii:2021ylb}.} Such extensions might find physical applications to symmetric `mirror' universes, such as in the works of \cite{Boyle:2018rgh,Boyle:2018tzc,Boyle:2021jej,Boyle:2022lyw}. Another physical aspect of the resulting `reflected' global spacetime that would be interesting to investigate is whether it would generally manifest a (non-singular) bounce, as closed de Sitter does.

%\vskip23pt
\vskip30pt
\subsection*{Acknowledgments}
We thank Eric Woolgar for stimulating discussions in the initial stages of this project. We further thank the stimulating atmosphere at the Fields Institute throughout the Thematic Program on Nonsmooth Riemannian and Lorentzian Geometry and especially during the Low Regularity Physics and Geometry Seminar and the Workshop on Mathematical Relativity, Scalar Curvature and Synthetic Lorentzian Geometry.
This publication was supported by the Fields Institute for Research in Mathematical Sciences. Its contents are solely the responsibility of the authors and do not necessarily represent the official views of the Institute. E.~Ling was supported by Carlsberg Foundation CF21-0680 and Danmarks Grundforskningsfond CPH-GEOTOP-DNRF151.
This research was also supported by a Discovery Grant from the Natural Science and Engineering Research Council of Canada (NSERC) and partly by the Perimeter Institute for Theoretical Physics. Research at the Perimeter Institute is supported by the Government of Canada through the Department of Innovation, Science and Economic Development and by the Province of Ontario through the Ministry of Colleges and Universities.
%\vskip23pt
\vskip30pt

%%%%%%%%%%%%%%%%%%%%%%%%%%%%%%%%%%%%%%%%%%%%%%%% APPENDIX

\appendix

\section{Proofs of Section \ref{sec:Ck}}\label{app:longCkProofs}

We have gathered the missing proofs of Section \ref{sec:Ck} in this appendix.

\medskip
\medskip

\noindent\textbf{Lemma \ref{lem:lalaetakiffCk}.}
\textit{The existence and finiteness of $l_\lambda^r[a]$ and $l_\lambda^r[a\eta]$ for all $r\in\{0,\ldots,k\}$ is equivalent to the metric \eqref{eq:metriclambdavcoordagain} being $C^k$ extendible as $\lambda\to 0^+$.}

\proof
($\Rightarrow$) If we show that the existence and finiteness of $l_\lambda^r[a]$ and $l_\lambda^r[a\eta]$ for all $r\in\{0,\ldots,k\}$ implies the existence and finiteness of $l_\lambda^r[a^2]$, $l_\lambda^r[a^2\eta]$, and $l_\lambda^r[a^2\eta^2]$ for all $r\in\{0,\ldots,k\}$, then we are done (recall the discussion at the beginning of Section \ref{sec:Ck}).
Applying the general Leibniz rule at any order $k$, we find
\begin{align}
	\frac{\dd^k(a^2\eta^2)}{\dd\lambda^k}&=\frac{\dd^{k-1}}{\dd\lambda^{k-1}}\left(\frac{\dd(a\eta)^2}{\dd\lambda}\right)=\frac{\dd^{k-1}}{\dd\lambda^{k-1}}\left(2a\eta\cdot\frac{\dd(a\eta)}{\dd\lambda}\right)\nonumber\\
	&=2\sum_{m=0}^{k-1}\frac{(k-1)!}{m!(k-1-m)!}\frac{\dd^{k-1-m}(a\eta)}{\dd\lambda^{k-1-m}}\frac{\dd^{m+1}(a\eta)}{\dd\lambda^{m+1}}\,,\nonumber
\end{align}
\begin{align}
	\frac{\dd^k(a^2)}{\dd\lambda^k}&=\frac{\dd^{k-1}}{\dd\lambda^{k-1}}\left(\frac{\dd(a^2)}{\dd\lambda}\right)=\frac{\dd^{k-1}}{\dd\lambda^{k-1}}\left(2a\cdot\frac{\dd a}{\dd\lambda}\right)\nonumber\\
	&=2\sum_{m=0}^{k-1}\frac{(k-1)!}{m!(k-1-m)!}\frac{\dd^{k-1-m}a}{\dd\lambda^{k-1-m}}\frac{\dd^{m+1}a}{\dd\lambda^{m+1}}\,,\nonumber
\end{align}
\begin{align}
	\frac{\dd^k(a^2\eta)}{\dd\lambda^k}&=\frac{\dd^{k-1}}{\dd\lambda^{k-1}}\left(\frac{\dd(a\cdot a\eta)}{\dd\lambda}\right)=\frac{\dd^{k-1}}{\dd\lambda^{k-1}}\left(\frac{\dd a}{\dd\lambda}\cdot a\eta+a\cdot\frac{\dd(a\eta)}{\dd\lambda}\right)\nonumber\\
	&=\sum_{m=0}^{k-1}\frac{(k-1)!}{m!(k-1-m)!}\left(\frac{\dd^{k-m}a}{\dd\lambda^{k-m}}\frac{\dd^{m}(a\eta)}{\dd\lambda^{m}}+\frac{\dd^{k-1-m}a}{\dd\lambda^{k-1-m}}\frac{\dd^{m+1}(a\eta)}{\dd\lambda^{m+1}}\right)\,.\nonumber
\end{align}
Taking the limit as $\lambda\to 0^+$, we thus have
\begin{subequations}\label{eq:squaretopower1limits}
\begin{align}
    l_\lambda^k[a^2\eta^2]&=2\sum_{r=1}^{k}\frac{(k-1)!}{(r-1)!(k-r)!}l_\lambda^{k-r}[a\eta]l_\lambda^{r}[a\eta]\,,\label{eq:a2eta2kexp}\\
    l_\lambda^k[a^2]&=2\sum_{r=1}^{k}\frac{(k-1)!}{(r-1)!(k-r)!}l_\lambda^{k-r}[a]l_\lambda^{r}[a]\,,\\
    l_\lambda^k[a^2\eta]&=\sum_{r=0}^{k-1}\frac{(k-1)!}{r!(k-1-r)!}l_\lambda^{k-r}[a]l_\lambda^{r}[a\eta]+\sum_{r=1}^{k}\frac{(k-1)!}{(r-1)!(k-r)!}l_\lambda^{k-r}[a]l_\lambda^{r}[a\eta]\nonumber\\
    &=l_\lambda^k[a]l_\lambda^0[a\eta]+\sum_{r=1}^{k-1}\frac{k!}{r!(k-r)!}l_\lambda^{k-r}[a]l_\lambda^r[a\eta]+l_\lambda^0[a]l_\lambda^k[a\eta]\,,\label{eq:a2etakexp}
\end{align}
\end{subequations}
where some combinatorical simplifications are performed in the last line.
These expressions confirm that the existence and finiteness of $l_\lambda^r[a]$ and $l_\lambda^r[a\eta]$ for all $r\in\{0,\ldots,k\}$ implies the existence and finiteness of $l_\lambda^r[a^2]$, $l_\lambda^r[a^2\eta]$, and $l_\lambda^r[a^2\eta^2]$ for all $r\in\{0,\ldots,k\}$, therefore completing this part of the proof.

($\Leftarrow$) If the metric \eqref{eq:metriclambdavcoordagain} is $C^k$ extendible as $\lambda\to 0^+$, then $l_\lambda^r[a^2]$, $l_\lambda^r[a^2\eta]$, and $l_\lambda^r[a^2\eta^2]$ must exist and be finite for all $r\in\{0,\ldots,k\}$. Therefore, we wish to show that this implies the existence and finiteness of $l_\lambda^r[a]$ and $l_\lambda^r[a\eta]$ for all $r\in\{0,\ldots,k\}$. Let us proceed by induction. For $k=0$, it is obvious that the existence and finiteness of $l_\lambda^0[a^2]$ and $l_\lambda^0[a^2\eta^2]$ imply the existence and finiteness of $l_\lambda^0[a]$ and $l_\lambda^0[a\eta]$. Now, let us assume the result at order $k-1$, so $l_\lambda^r[a]$ and $l_\lambda^r[a\eta]$ exist and are finite for all $r\in\{0,\ldots,k-1\}$. Then, given $l_\lambda^k[a^2\eta^2]$ exists and is finite, we can derive a similar expression to \eqref{eq:a2eta2kexp}:
\[
    l_\lambda^k[a^2\eta^2]-2\sum_{r=1}^{k-1}\frac{(k-1)!}{(r-1)!(k-r)!}l_\lambda^{k-r}[a\eta]l_\lambda^r[a\eta]=2\lim_{\lambda\to 0^+}\left(a\eta\frac{\dd^k(a\eta)}{\dd\lambda^k}\right)\,.
\]
Everything on the left-hand side of this expression exists and is finite, so must the right-hand side, but we actually know $l_\lambda^0[a\eta]=-1/h\neq 0$, hence $l_\lambda^k[a\eta]$ must also exist and be finite.
Similarly, given $l_\lambda^k[a^2\eta]$ exists and is finite, we can derive a similar expression to \eqref{eq:a2etakexp}:
\[
    l_\lambda^k[a^2\eta]-\sum_{r=1}^{k-1}\frac{k!}{r!(k-r)!}l_\lambda^{k-r}[a]l_\lambda^r[a\eta]-l_\lambda^0[a]l_\lambda^k[a\eta]=\lim_{\lambda\to 0^+}\left(\frac{\dd^ka}{\dd\lambda^k}a\eta\right)\,.
\]
Everything on the left-hand side of this expression exists and is finite (including $l_\lambda^k[a\eta]$ now), so must the right-hand side, but we actually know $l_\lambda^0[a\eta]=-1/h\neq 0$, hence $l_\lambda^k[a]$ must also exist and be finite.
\qed

\medskip
\medskip

\noindent\textbf{Lemma \ref{lem:lalaetalhdotoa2km2tok}.}
\textit{This lemma consists of two similar (but independent) statements:}
\begin{enumerate}[label=(\alph*)]
    \item \textit{the existence and finiteness of $l_\lambda^r[a]$ and $l_\lambda^r[\dot H/a^2]$ for all $r\in\{0,\ldots,k-2\}$ implies the existence and finiteness of $l_\lambda^{k-1}[a]$ and $l_\lambda^k[a]$;}
    \item \textit{the existence and finiteness of $l_\lambda^r[a\eta]$ and $l_\lambda^r[\dot H/a^2]$ for all $r\in\{0,\ldots,k-2\}$ implies the existence and finiteness of $l_\lambda^{k-1}[a\eta]$ and $l_\lambda^k[a\eta]$.}
\end{enumerate}

\proof
Let us prove the statements separately:
\begin{enumerate}[label=(\alph*)]
    \item Recalling \eqref{eq:d2adlambda2dHdlambda}, the general Leibniz rule gives
    \begin{align}
	\frac{\dd^ka}{\dd\lambda^k}&=\frac{\dd^{k-2}}{\dd\lambda^{k-2}}\left(\frac{\dd^2a}{\dd\lambda^2}\right)=\frac{\dd^{k-2}}{\dd\lambda^{k-2}}\left(a\cdot\frac{\dot H}{a^2}\right)\nonumber\\
	&=\sum_{m=0}^{k-2}\frac{(k-2)!}{m!(k-2-m)!}\frac{\dd^{k-2-m}a}{\dd\lambda^{k-2-m}}\frac{\dd^m}{\dd\lambda^m}\left(\frac{\dot H}{a^2}\right)\,.\nonumber
    \end{align}
    Therefore, in the limit $\lambda\to 0^+$,
    \begin{equation}
	l_\lambda^k[a]=\sum_{r=0}^{k-2}\frac{(k-2)!}{r!(k-2-r)!}l_\lambda^{k-2-r}[a]l_\lambda^r[\dot H/a^2]\,.\nonumber
    \end{equation}
    This relation shows that the existence and finiteness of $l_\lambda^r[a]$ and $l_\lambda^r[\dot H/a^2]$ for all $r\in\{0,\ldots,k-2\}$ implies the existence and finiteness of $l_\lambda^k[a]$. Then, in particular, the existence and finiteness of $l_\lambda^r[a]$ and $l_\lambda^r[\dot H/a^2]$ for all $r\in\{0,\ldots,k-3\}$ implies the existence and finiteness of $l_\lambda^{k-1}[a]$.

    \item Recalling $\dd\eta=a^{-1}\dd t$ and $\dd\lambda=a\,\dd t$, we have
    \begin{equation}
        \frac{\dd(a\eta)}{\dd\lambda}=\frac{1}{a}\frac{\dd(a\eta)}{\dd t}=\frac{1}{a}\left(\frac{\dd a}{\dd t}\eta+a\frac{\dd\eta}{\dd t}\right)=H\eta+\frac{1}{a}\,,\nonumber
    \end{equation}
    and then,
    \begin{equation}
        \frac{\dd^2(a\eta)}{\dd\lambda^2}=\frac{1}{a}\frac{\dd}{\dd t}\left(H\eta+\frac{1}{a}\right)=\frac{\dot H\eta}{a}\,.\nonumber
    \end{equation}
    To arbitrary order $k$, the general Leibniz rule then gives
    \begin{align}
	\frac{\dd^k(a\eta)}{\dd\lambda^k}&=\frac{\dd^{k-2}}{\dd\lambda^{k-2}}\left(\frac{\dd^2(a\eta)}{\dd\lambda^2}\right)=\frac{\dd^{k-2}}{\dd\lambda^{k-2}}\left(a\eta\cdot\frac{\dot H}{a^2}\right)\nonumber\\
	&=\sum_{m=0}^{k-2}\frac{(k-2)!}{m!(k-2-m)!}\frac{\dd^{k-2-m}(a\eta)}{\dd\lambda^{k-2-m}}\frac{\dd^m}{\dd\lambda^m}\left(\frac{\dot H}{a^2}\right)\,,\nonumber
    \end{align}
    so in the limit to $\lambda\to 0^+$,
    \begin{equation}
        l_\lambda^k[a\eta]=\sum_{r=0}^{k-2}\frac{(k-2)!}{r!(k-2-r)!}l_\lambda^{k-2-r}[a\eta]l_\lambda^r[\dot H/a^2]\,.\nonumber
    \end{equation}
    This relation shows that the existence and finiteness of $l_\lambda^r[a\eta]$ and $l_\lambda^r[\dot H/a^2]$ for all $r\in\{0,\ldots,k-2\}$ implies the existence and finiteness of $l_\lambda^k[a\eta]$. Then, in particular, the existence and finiteness of $l_\lambda^r[a\eta]$ and $l_\lambda^r[\dot H/a^2]$ for all $r\in\{0,\ldots,k-3\}$ implies the existence and finiteness of $l_\lambda^{k-1}[a\eta]$.
\end{enumerate}
\qed

\medskip
\medskip

The following is a new lemma, absent in Section \ref{sec:Ck}, but it is needed as an intermediate step to subsequently show Lemma \ref{lem:llaladothoa2km2}.

\medskip

\begin{lem}\label{lem:dlkm2dotHoa2dadotHoa2}
    The $\lambda$- and $a$-derivatives of $\dot H/a^2$, up to order $k-2$, are related as follows:
   for $k=2$, we trivially have $\dot H/a^2$ as a function of $\lambda$ equal to $\dot H/a^2$ as a function of $a$; then for any $k\geq 3$,
    \begin{equation}
	\frac{\dd^{k-2}}{\dd\lambda^{k-2}}\left(\frac{\dot H}{a^2}\right)=\sum_{m=1}^{k-2}\mathcal{H}_m^{(k)}\frac{\dd^{m}}{\dd a^{m}}\left(\frac{\dot H}{a^2}\right)\,,\label{eq:dotHoa2derivrel}
    \end{equation}
    where the functions $\mathcal{H}_m^{(k)}$ are non-negative-integer polynomials in $H$ and its $\lambda$-derivatives (up to order $k-3$). Specifically,
    \begin{equation}
	\mathcal{H}_1^{(k)}=\frac{\dd^{k-3}H}{\dd\lambda^{k-3}}\,,\qquad\mathcal{H}_{k-2}^{(k)}=H^{k-2}\,,\label{eq:dotHoa2derivrelcoeff}
    \end{equation}
    and, for $m\in\{2,\ldots,k-3\}$ when $k\geq 5$, there exists $c_{n}^{(m,k)}\in\mathbb{Z}_{\geq 0}$ such that
    \begin{equation}
        \mathcal{H}_m^{(k)}=\sum_{n\in\mathcal{N}_m^{(k)}}c_n^{(m,k)}\prod_{i=1}^m\frac{\dd^{n_i}H}{\dd\lambda^{n_i}}\,,
    \end{equation}
    where
    \begin{equation}
	\mathcal{N}_m^{(k)}:=\left\{\{n_i\}_{i=1}^m~\bigg|~n_i\in\{0,\ldots,k-2-m\}\ \mathrm{and}\ \sum_{i=1}^{m}n_i=k-2-m\right\}\,.\label{eq:calRrangeris}
    \end{equation}
\end{lem}

\medskip

\noindent\textit{Remark.} For our purposes, the exact form of the coefficients $c_n^{(m,k)}$ is not needed, but it would be interesting to see if they exhibit a pattern.

\medskip

\proof
The proof for $k\geq 3$ goes by induction. For $k=3$, we have
\begin{equation}
	\frac{\dd}{\dd\lambda}\left(\frac{\dot H}{a^2}\right)=\frac{\dd t}{\dd\lambda}\frac{\dd}{\dd t}\left(\frac{\dot H}{a^2}\right)=\frac{1}{a}\frac{\dd a}{\dd t}\frac{\dd}{\dd a}\left(\frac{\dot H}{a^2}\right)=H\frac{\dd}{\dd a}\left(\frac{\dot H}{a^2}\right)\,, \nonumber
\end{equation}
which is indeed the $k=3$ case of \eqref{eq:dotHoa2derivrel}--\eqref{eq:dotHoa2derivrelcoeff}.

Now, let us assume the lemma holds at order $k-3$, so we have
\begin{equation}
	\frac{\dd^{k-3}}{\dd\lambda^{k-3}}\left(\frac{\dot H}{a^2}\right)=\sum_{m=1}^{k-3}\mathcal{H}_m^{(k-1)}\frac{\dd^{m}}{\dd a^{m}}\left(\frac{\dot H}{a^2}\right)\,,\qquad\mathcal{H}_1^{(k-1)}=\frac{\dd^{k-4}H}{\dd\lambda^{k-4}}\,,\qquad\mathcal{H}_{k-3}^{(k-1)}=H^{k-3}\,,\nonumber
\end{equation}
and for $m\in\{2,\ldots,k-4\}$ there exists $c_{n}^{(m,k-1)}\in\mathbb{Z}_{\geq 0}$ such that
\begin{equation}
	\mathcal{H}_m^{(k-1)}=\sum_{n\in\mathcal{N}_m^{(k-1)}}c_n^{(m,k-1)}\prod_{i=1}^m\frac{\dd^{n_i}H}{\dd\lambda^{n_i}}\,,\nonumber
\end{equation}
with
\begin{equation}
	\mathcal{N}_m^{(k-1)}=\left\{\{n_i\}_{i=1}^m~\bigg|~n_i\in\{0,\ldots,k-3-m\}\ \textrm{and}\ \sum_{i=1}^{m}n_i=k-3-m\right\}\,.\nonumber
\end{equation}
Then, trying to prove the result at order $k-2$, we have
\begin{subequations}
\begin{align}
	\frac{\dd^{k-2}}{\dd\lambda^{k-2}}\left(\frac{\dot H}{a^2}\right)&=\frac{\dd}{\dd\lambda}\left(\frac{\dd^{k-3}}{\dd\lambda^{k-3}}\left(\frac{\dot H}{a^2}\right)\right)\nonumber\\
    &=\frac{\dd}{\dd\lambda}\left(\sum_{m=1}^{k-3}\mathcal{H}_m^{(k-1)}\frac{\dd^{m}}{\dd a^{m}}\left(\frac{\dot H}{a^2}\right)\right)\nonumber\\
	&=\sum_{m=1}^{k-3}\frac{\dd\mathcal{H}_m^{(k-1)}}{\dd\lambda}\frac{\dd^{m}}{\dd a^{m}}\left(\frac{\dot H}{a^2}\right)+\sum_{m=1}^{k-3}\mathcal{H}_m^{(k-1)}H\frac{\dd^{m+1}}{\dd a^{m+1}}\left(\frac{\dot H}{a^2}\right)\nonumber\\
	&=\sum_{m=1}^{k-3}\frac{\dd\mathcal{H}_m^{(k-1)}}{\dd\lambda}\frac{\dd^{m}}{\dd a^{m}}\left(\frac{\dot H}{a^2}\right)+\sum_{m=2}^{k-2}H\mathcal{H}_{m-1}^{(k-1)}\frac{\dd^{m}}{\dd a^{m}}\left(\frac{\dot H}{a^2}\right)\label{eq:deftildeH1}\\
	&=:\sum_{m=1}^{k-2}\mathcal{\wt H}_m^{(k)}\frac{\dd^{m}}{\dd a^{m}}\left(\frac{\dot H}{a^2}\right)\,,\label{eq:deftildeH2}
\end{align}
\end{subequations}
where last equality defines $\mathcal{\wt H}_m^{(k)}$. Comparing \eqref{eq:deftildeH1} and \eqref{eq:deftildeH2}, we notice
\begin{equation}
	\mathcal{\wt H}_1^{(k)}=\frac{\dd\mathcal{H}_1^{(k-1)}}{\dd\lambda}=\frac{\dd}{\dd\lambda}\left(\frac{\dd^{k-4}H}{\dd\lambda^{k-4}}\right)=\frac{\dd^{k-3}H}{\dd\lambda^{k-3}}=\mathcal{H}_1^{(k)}\,,\nonumber
\end{equation}
\begin{equation}
	\mathcal{\wt H}_{k-2}^{(k)}=H\mathcal{H}_{k-3}^{(k-1)}=HH^{k-3}=H^{k-2}=\mathcal{H}_{k-2}^{(k)}\,,\nonumber
\end{equation}
and, for $m\in\{2,\ldots,k-3\}$,
\begin{align}
	\mathcal{\wt H}_m^{(k)}=&~\frac{\dd\mathcal{H}_m^{(k-1)}}{\dd\lambda}+H\mathcal{H}_{m-1}^{(k-1)}\nonumber\\
	=&~\sum_{n\in\mathcal{N}_m^{(k-1)}}c_n^{(m,k-1)}\frac{\dd}{\dd\lambda}\left(\prod_{i=1}^m\frac{\dd^{n_i}H}{\dd\lambda^{n_i}}\right)+\sum_{n\in\mathcal{N}_{m-1}^{(k-1)}}c_n^{(m-1,k-1)}H\prod_{i=1}^{m-1}\frac{\dd^{n_i}H}{\dd\lambda^{n_i}}\nonumber\\
	=&~\sum_{n\in\mathcal{N}_m^{(k-1)}}c_n^{(m,k-1)}\sum_{i=1}^m\frac{\dd^{n_i+1}H}{\dd\lambda^{n_i+1}}\prod_{\substack{j=1\\ j\neq i}}^m\frac{\dd^{n_j}H}{\dd\lambda^{n_j}}+\sum_{n\in\mathcal{N}_{m-1}^{(k-1)}}c_n^{(m-1,k-1)}\underbrace{\prod_{i=1}^{m}\frac{\dd^{n_i}H}{\dd\lambda^{n_i}}}_{\textrm{with}~n_m=0}\,.\label{eq:Hmkp}
\end{align}
Note that the first summation term of \eqref{eq:Hmkp} can be rewritten as\footnote{Indeed, a summation of
\[
	\frac{\dd^{n_i+1}H}{\dd\lambda^{n_i+1}}\prod_{\substack{j=1\\ j\neq i}}^m\frac{\dd^{n_j}H}{\dd\lambda^{n_j}}
\]
over all combinations of $n_i$ such that $\sum_{i=1}^{m}n_i=k-3-m$ is the same as a summation of
\[
	\prod_{i=1}^m\frac{\dd^{n_i}H}{\dd\lambda^{n_i}}
\]
over all combinations of $n_i$ such that $\sum_{i=1}^{m}n_i=k-2-m$.}
\begin{equation}
	\sum_{n\in\mathcal{N}_m^{(k-1)}}c_n^{(m,k-1)}\sum_{i=1}^m\frac{\dd^{n_i+1}H}{\dd\lambda^{n_i+1}}\prod_{\substack{j=1\\ j\neq i}}^m\frac{\dd^{n_j}H}{\dd\lambda^{n_j}}=\sum_{n\in\mathcal{N}_m^{(k)}}\wt c_n^{(m,k)}\prod_{i=1}^m\frac{\dd^{n_i}H}{\dd\lambda^{n_i}}\,,\nonumber
\end{equation}
where the new coefficients $\wt c_n^{(m,k)}$ come from sums of different $c_n^{(m,k-1)}\in\mathbb{Z}_{\geq 0}$, hence $\wt c_n^{(m,k)}\in\mathbb{Z}_{\geq 0}$. Similarly, note that the second summation of \eqref{eq:Hmkp} can be rewritten as
\begin{equation}
	\sum_{n\in\mathcal{N}_{m-1}^{(k-1)}}c_n^{(m-1,k-1)}\underbrace{\prod_{i=1}^{m}\frac{\dd^{n_i}H}{\dd\lambda^{n_i}}}_{\textrm{with}~n_m=0}=\sum_{n\in\mathcal{N}_m^{(k)}}c_n^{(m-1,k-1)}\prod_{i=1}^{m}\frac{\dd^{n_i}H}{\dd\lambda^{n_i}}\nonumber
\end{equation}
since a summation over
\begin{equation}
	n\in\mathcal{N}_{m-1}^{(k-1)}=\left\{\{n_i\}_{i=1}^{m-1}~\bigg|~n_i\in\{0,\ldots,k-2-m\}\ \textrm{and}\ \sum_{i=1}^{m-1}n_i=k-2-m\right\}\nonumber
\end{equation}
becomes equivalent to a summation over
\begin{equation}
	n\in\mathcal{N}_m^{(k)}=\left\{\{n_i\}_{i=1}^{m}~\bigg|~n_i\in\{0,\ldots,k-2-m\}\ \textrm{and}\ \sum_{i=1}^{m}n_i=k-2-m\right\}\nonumber
\end{equation}
when $n_m=0$.
Therefore, \eqref{eq:Hmkp} becomes
\begin{equation}
	\mathcal{\wt H}_m^{(k)}=\sum_{n\in\mathcal{N}_m^{(k)}}c_n^{(m,k)}\prod_{i=1}^m\frac{\dd^{n_i}H}{\dd\lambda^{n_i}}=\mathcal{H}_m^{(k)}\,,\nonumber
\end{equation}
with $c_n^{(m,k)}=\wt c_n^{(m,k)}+c_n^{(m-1,k-1)}\in\mathbb{Z}_{\geq 0}$.
\qed

\medskip
\medskip

\noindent\textbf{Lemma \ref{lem:llaladothoa2km2}.}
\textit{The existence and finiteness of $l_\lambda^r[a]$ and $l_a^r[\dot H/a^2]$ for all $r\in\{0,\ldots,k-2\}$ implies the existence and finiteness of $l_\lambda^{r}[\dot H/a^2]$ for all $r\in\{0,\ldots,k-2\}$.}

\proof
If $k=2$, obviously $l_\lambda^0[\dot H/a^2]=l_a^0[\dot H/a^2]$. Then for $k\geq 3$, taking the limit as $\lambda\to 0^+$ (equivalently $a\to 0^+$) of equation \eqref{eq:dotHoa2derivrel} in Lemma \ref{lem:dlkm2dotHoa2dadotHoa2}, we find
\begin{align}
	l_\lambda^{k-2}[\dot H/a^2]=&~l_\lambda^{k-3}[H]l_a^1\left[\frac{\dot H}{a^2}\right]+\sum_{m=2}^{k-3}l_a^m\left[\frac{\dot H}{a^2}\right]\sum_{n\in\mathcal{N}_m^{(k)}}c_n^{(m,k)}\prod_{i=1}^ml_\lambda^{n_i}[H]+l_\lambda^0[H^{k-2}]l_a^{k-2}\left[\frac{\dot H}{a^2}\right]\nonumber\\
	=&~l_\lambda^{k-2}[a]l_a^1\left[\frac{\dot H}{a^2}\right]+\sum_{m=2}^{k-3}l_a^m\left[\frac{\dot H}{a^2}\right]\sum_{n\in\mathcal{N}_m^{(k)}}c_n^{(m,k)}\prod_{i=1}^ml_\lambda^{n_i+1}[a]+h^{k-2}l_a^{k-2}\left[\frac{\dot H}{a^2}\right]\,,\nonumber
\end{align}
which completes the proof. Note, in particular, that the $n_i$s in the middle term range from $0$ to at most $k-3$ according to \eqref{eq:calRrangeris}. In going to the second line above, we used the fact that, recalling $\dd a/\dd\lambda=(1/a)(\dd a/\dd t)=H$,
\begin{equation}
    \frac{\dd^{m+1}a}{\dd\lambda^{m+1}}=\frac{\dd^{m}H}{\dd\lambda^{m}}\,,\nonumber
\end{equation}
which implies $l_\lambda^{m}[H]=l_\lambda^{m+1}[a]$. We also used the knowledge that $l_\lambda^0[H]$ is a constant (denoted $h$ or $H_\Lambda$ previously), so $l_\lambda^0[H^{k-2}]=l_\lambda^0[H]^{k-2}=h^{k-2}$ is just a constant as well.
\qed

\section{More smoothly extendible examples}\label{app:moreexamples}

We provide more examples of smoothly extendible spacetimes, continuing the discussion present at the end of Section \ref{sec:Ck}.

An interesting class of examples is
\begin{equation}
    a(t)=\frac{1}{e^{-t}+e^{nt}}=e^t-e^{(2+n)t}+e^{(3+2n)t}-e^{(4+3n)t}+\ldots\,,\qquad n\in\mathbb{R}_{\geq 1}\,.
\end{equation}
Here again, we conjecture that $C^k$ extendibility, $k\geq 2$, follows if $n\in\mathbb{R}_{\geq k-1}$ or if $n\in\{1,2,3,4,\ldots,k-2\}$; $C^\infty$ would thus follow if $n\in\mathbb{Z}_{\geq 1}$. (A proof would be tedious, but the conjecture has been checked to hold in many cases.) For the special case $n=1$ (already discussed in \cite{Yoshida:2018ndv}), we can prove this explicitly since a direct computation shows that $\dot H/a^2$ is constant.

Another example (which can be found in \cite{Nomura:2022vcj}) is
\begin{equation}
    a(t)=\frac{e^t}{\sqrt{1+e^{2t}}}\,.
\end{equation}
This can be analytically inverted to find
\begin{equation*}
    t(a)=\ln\left(\frac{a}{\sqrt{1-a^2}}\right)\,,\qquad \frac{\dot H}{a^2}=-\frac{2}{1+e^{2t}}=-2(1-a^2)\,.
\end{equation*}
This is a smooth function of $a$, so a smooth extension is possible in this case.

As a somewhat more physically motivated example, let us consider a proper field theory realization of inflation. An enormous number of examples could be explored, but let us only revisit a particular case of small-field inflation, for which \cite{Yoshida:2018ndv} found the absence of obstructions to $C^2$ metric extendibility. The potential of the scalar field $\varphi$ is given by $V(\phi)=V_0(1-(\frac{\varphi}{2m})^2)^2$, where the field asymptotically begins at the unstable point $\varphi=0$ with no velocity and slowly rolls down the potential toward one of its minima. The slow-roll solution was derived in \cite{Yoshida:2018ndv} as a function of $\varphi$, but let us repeat it as a function of $t$. Under the slow-roll approximation, the scalar field's velocity follows $\dot\varphi\simeq-\frac{1}{3H}\frac{\dd V}{\dd\varphi}$, and the Hubble parameter satisfies the constraint $3H^2\simeq V(\varphi)$. We are using units where Newton's gravitational constant is equal to $1/(8\pi)$ (or where the reduced Planck mass is set to unity). The resulting equation, given the above potential, is $\dot\varphi\simeq(\varphi/m^2)\sqrt{V_0/3}$, whose solution is $\varphi(t)\simeq\varphi_\mathrm{e}\exp(\sqrt{\frac{V_0}{3}}\frac{t}{m^2})$ for some integration constant $\varphi_\mathrm{e}$. Then, upon solving $\dot a(t)/a(t)\simeq\sqrt{V(\varphi(t))/3}$, we find
\begin{equation}
    a(t)\simeq a_\mathrm{e}\exp[-\frac{\varphi_\mathrm{e}^2}{8}\exp(\frac{2}{m^2}\sqrt{\frac{V_0}{3}}t)]\exp(\sqrt{\frac{V_0}{3}}t)\,,\nonumber
\end{equation}
for some integration constant $a_\mathrm{e}$.
As $t\to-\infty$, this is not asymptotically equal to $e^{ht}+o(e^{ht})$, but it does respect $H\to\sqrt{V_0/3}$, so this is $C^0$ extendible. In fact, it is $C^2$ (as shown in \cite{Yoshida:2018ndv}). Moreover, it can be shown to be $C^\infty$ extendible: upon inverting the above scale factor, one can compute $\dot H/a^2$ and express it as a function of $a$ as
\begin{equation}
    \frac{\dot H}{a^2}\simeq\frac{2V_0}{3m^2a^2}W(-x(a)^2)\,,\qquad \textrm{with}\ x(a)=\frac{\varphi_\mathrm{e}}{2m}\left(\frac{a}{a_\mathrm{e}}\right)^{1/m^2}\,,\nonumber
\end{equation}
and where $W$ is the Lambert $W$ function.
Near $a=0^+$ asymptotically, the above can be expanded as
\begin{equation}
	\frac{\dot H}{a^2}\simeq-\frac{2V_0}{3m^2}\left[\left(\frac{\varphi_\mathrm{e}}{2ma_\mathrm{e}^{1/m^2}}\right)^2a^{\frac{2}{m^2}-2}+\left(\frac{\varphi_\mathrm{e}}{2ma_\mathrm{e}^{1/m^2}}\right)^4a^{\frac{4}{m^2}-2}+\cdots\right]\,.\nonumber
\end{equation}
Provided $m<1$ (it must be so for the field's mass to be sub-Planckian), the above converges as $a\searrow 0$, thus confirming smoothness. However, as is discussed in Section \ref{sec:discussion}, this example is one of infinitely tuned initial conditions for the field to sit at $\varphi=0$ asymptotically on the unstable point of the potential. Therefore, it is an unlikely scenario from a dynamical perspective. This is a well-known initial conditions problem in small-field inflation \cite{Goldwirth:1991rj,Brandenberger:2016uzh}, but the comment applies more broadly as we argue in Section \ref{sec:discussion}.

%%% BIBLIOGRAPHY %%%

%\newpage

\addcontentsline{toc}{section}{References}

\let\oldbibliography\thebibliography
\renewcommand{\thebibliography}[1]{
  \oldbibliography{#1}
  \setlength{\itemsep}{0pt}
  \footnotesize
}

\bibliographystyle{JHEP2}
\bibliography{refs}

\end{document}